\documentclass[aps,prd,preprintnumbers,showpacs,showkeys,nofootinbib,
superscriptaddress,fleqn,floatfix,tightenlines,10pt]{revtex4-1}
\usepackage{amsmath,amsfonts,amssymb,amscd,amsxtra,amsthm}
\usepackage{graphicx}  
\usepackage{epstopdf}
\usepackage{dcolumn}  
\usepackage{bm}        
\usepackage{slashed}
\usepackage{cancel}
\usepackage{float}
\usepackage{mathtools}
\usepackage{amsbsy}
\usepackage{amstext}

\usepackage[utf8]{inputenc} 
\usepackage{booktabs}
\usepackage[normalem]{ulem} 
\usepackage[dvipsnames]{xcolor} 
\usepackage{tabularx}
\usepackage{enumitem}  
\usepackage{array} 
\usepackage{slashed}
\usepackage{tikz}
\usepackage{float}
\usepackage{multirow}
\renewcommand\sout{\bgroup \color{red} \ULdepth=-.5ex \ULset}
\newcommand{\com}[1]{{\sf\color[rgb]{0,0,1}{#1}}}

\makeatletter

\begin{document}  
\preprint{INHA-NTG-01/2020}
\title{Electromagnetic transition form factors, $E2/M1$ and $C2/M1$
  ratios of the baryon decuplet}
\author{June-Young Kim}
\email[E-mail: ]{Jun-Young.Kim@ruhr-uni-bochum.de}
\affiliation{Institut f\"ur Theoretische Physik II, Ruhr-Universit\"at
  Bochum, D-44780 Bochum, Germany}
\affiliation{Department of Physics, Inha University, Incheon 22212,
Republic of Korea}

\author{Hyun-Chul Kim}
\email[E-mail: ]{hchkim@inha.ac.kr}
\affiliation{Department of Physics, Inha University, Incheon 22212,
Republic of Korea}
\affiliation{School of Physics, Korea Institute for Advanced Study 
  (KIAS), Seoul 02455, Republic of Korea}

\date{\today}
\begin{abstract}
We investigate the electromagnetic transition form factors of the
baryon decuplet to the baryon octet, based on the self-consistent
SU(3) chiral quark-soliton model, taking into account the effects of
explicit breaking of flavor SU(3) symmetry. We emphasize the $Q^2$
dependence of the electromagnetic $N\to \Delta$ transition form
factors and the ratios of $E2/M1$ and $C2/M1$ in comparison with the
experimental and empirical data. In order to compare the present
results of the electromagnetic transition form factors of the $N\to
\Delta$ with those from lattice QCD, we evaluate the form factors with
the pion mass deviated from its physical value. The results of the
$E2/M1$ and $C2/M1$ ratios are in good agreement with the lattice
data. We also present the results of the electromagnetic transition
form factors for the decuplet to the octet transitions. 
\end{abstract}
\maketitle
\section{Introduction}
Understanding how a baryon is shaped electromagnetically has been one
of the most important issues in hadronic physics. The EM structure of
the baryon decuplet, which contains the first excitations of the
nucleon and hyperons, is far from complete understanding. The reason
is that it is very difficult to get access to their structures
experimentally on account of their ephemeral nature. On the other
hand, the EM transition $N\gamma^*\to \Delta$ can be examined
experimentally by using the electroproduction of the
pion~\cite{Aznauryan:2009mx, Villano:2009sn, Stave:2008aa,  
  Sparveris:2004jn, Kelly:2005jy, Beck:1999ge, Stave:2006ea,
  Sparveris:2006uk, Pospischil:2000ad, Blomberg:2015zma}. The EM
transitions from the baryon octet to the decuplet are also 
experimentally accessible~\cite{Molchanov:2004iq, Antipov:2004qp, 
  Keller:2011aw,  Keller:2011nt}. Thus, the structure of the baryon
decuplet can be investigated by using the EM transition from the baryon
octet to the decuplet. Theoretically, the EM transitions 
of the $N\gamma^*\to \Delta$ have been extensively 
studied within various frameworks over decades: for example, 
the linear sigma model~\cite{Fiolhais:1996bp}, the Skyrme 
models~\cite{Walliser:1996ps,Wirzba:1986sc, Abada:1995db,
  Haberichter:1996cp}, the chiral bag model~\cite{Lu:1996rj}, the
Dyson-Schwinger approaches~\cite{Segovia:2014aza, Eichmann:2011aa,
  Eichmann:2016yit}, constituent quark models~\cite{Buchmann:1996bd},
relativistic quark models~\cite{Ramalho:2013uza}, QCD sum
rules~\cite{Aliev:2004ju,   Braun:2005be, Rohrwild:2007iz,
  Wang:2009ru}, a $\pi N$ dynamical model~\cite{JuliaDiaz:2006xt},
lattice QCD~\cite{Leinweber:1992pv, Alexandrou:2004xn,
  Alexandrou:2007dt,Alexandrou:2010uk},
AdS/QCD~\cite{Grigoryan:2009pp}, chiral perturbation
theory~\cite{Li:2017vmq}, chiral effective field
theory($\chi$EFT)~\cite{Pascalutsa:2005ts} and so on.    

Since the $\Delta$ isobar decays mostly into the nucleon and pion, it
has been known that the pion clouds play an essential role in
describing the structure of the $\Delta$. For example, the
naive nonrelativistic quark model (NRQM) underestimates the $N\to \Delta$
transition magnetic moment by approximately $30~\%$ and predicts
wrongly the electric quadrupole transition moment $Q_{p\to
  \Delta^+}=0$, which indicates that the $\Delta$ isobar from the NRQM
has a completely spherically symmetric shape. However, experimental
results indicate that the $N\to \Delta$ transition electric quadrupole
moment is small but still finite. Other members of the baryon decuplet
show similar decay modes except for the $\Omega^-$. The first excited 
$\Sigma^*$ and $\Xi^*$ hyperons decay into the corresponding octet
baryons and the pions. This implies at least two important points
about the structure of the baryon decuplet. Firstly, any approach in
describing the baryon decuplet should consider the pions
seriously. Secondly, chiral symmetry and its spontaneous breakdown
should come into play, since the pions are regarded as the
pseudo-Nambu-Goldstone (pNG) bosons. Thus, any theoretical works on
the baryon decuplet should include the pions as the pNG bosons.  

The chiral quark-soliton model ($\chi$QSM) has been developed as a
pion mean-field approach for describing the structure of the
nucleon~\cite{Diakonov:1987ty}. In the large $N_c$
limit~\cite{Witten:1979kh, Witten:1983tw}, the nucleon can be viewed
as a state of the $N_c$ valence quarks bound by the pion mean fields
that are produced by the presence of the $N_c$ valence quarks
self-consistently. The model was extended to the description of the
lowest-lying SU(3) baryons~\cite{Blotz:1992pw} and was successfully
applied to explaining various properties of the SU(3) baryons. For
details, we refer to the reviews~\cite{Christov:1995vm,
  Alkofer:1994ph} and references therein. The $\chi$QSM was also
extended to the description of singly heavy
baryons~\cite{Yang:2016qdz, Kim:2017jpx, Kim:2018cxv, Kim:2019rcx,
  Yang:2020klp}.  Thus, the model provides a unified pion mean-field
approach for both the light and singly heavy baryons. 
In the present work, we want to investigate the EM transition form
factors and related observables from the baryon octet to the decuplet 
within the framework of the self-consistent SU(3) chiral quark-soliton
model ($\chi$QSM). The EM transition $N\gamma \to \Delta$  was already
studied in the $\chi$QSM~\cite{Watabe:1995xy, Silva:1999nz,
  Urbano:2000xg,  Kim:2005gz, Ledwig:2008es}. 
In Ref.~\cite{Ledwig:2008es} the EM transition form factors for
the $N\gamma^*\to \Delta$ were computed without the effects of the
flavor SU(3) symmetry. Moreover, the large $N_c$ argument was used to
improve quantitatively the magnetic dipole and electric quadrupole transition
moments. However, it is known that the absolute magnitudes of the
helicity amplitudes for the $N\gamma\to \Delta$ excitation based on
the $\chi$QSM are underestimated, compared with the experimental and
empirical data. The lattice data are also known to be quite smaller
than the experimental data. This indicates that we need to study the
EM transitions of the baryon decuplet to the octet more in
detail. While the $\chi$QSM model underestimates the absolute values
for the helicity amplitudes, it provides very interesting results,
compared with the lattice data. 

In the present work, we will compute all possible EM
transitions from the baryon octet to the decuplet with the effects of
flavor SU(3) symmetry breaking taken into account. Since radiative
decays of the negative decuplet baryons vanish in the exact flavor
SU(3) symmetric case due to the $U$-spin symmetry, it is of great
importance to consider the explicit breaking of flavor SU(3) symmetry. 
For example, the transition form factors for the $\Sigma^-\gamma \to
\Sigma^{*-}$ and $\Xi^-\gamma\to \Xi^{*-}$ radiative decays have null
results without flavor SU(3) symmetry breaking. This means that the
linear $m_{\mathrm{s}}$ contributions play the leading role in
describing these EM transition form factors. 
We first examine the $N\gamma^*\to \Delta$ transition, because the
lattice data as well as the experimental data exist. The existing
lattice calculation~\cite{Alexandrou:2007dt} still uses the values of
the unphysical pion mass, so that we employ the corresponding values
of the pion mass to compare the numerical results with the lattice
data.  We then present the numerical results of all possible EM
transition form factors from the baryon octet to the decuplet,
focusing on the explicit breaking of flavor SU(3) symmetry. The
results of the $E2/M1$ and $C2/M1$ ratios are compared with the
experimental and lattice data. 

The present work is organized as follows:
In Section II, we briefly explain the definition of the EM transition
form factors and helicity amplitudes. In Section III, we describe
shortly how the expressions of the EM transition form factors are
obtained. In Section IV, the numerical results are presented and are
compared with the experimental data. We also discuss them in
comparison with the lattice data. In the final Section, summary and
conclusions are given. 

\section{EM transition form factors
  for radiative excitations $B_8\gamma^*\to B_{10}$ }
\begin{figure}[htp]
\includegraphics[scale=0.75]{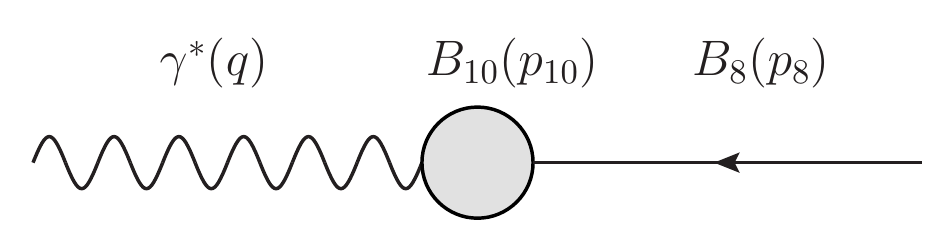}
\caption{Schematic diagram for the $\gamma^{*} B_{8}\to B_{10}$ transition}
\label{fig:1}
\end{figure}

The radiative excitation from an octet baryon to a decuplet baryon,
$B_8 \gamma^*\to B_{10}$, is schematically shown in Fig.~\ref{fig:1}
in the rest frame of a decuplet baryon.  
In this rest frame, $p_{10}$, $p_{8}$, and $q$ denote respectively
the momenta of a decuplet baryon and an octet one, and the momentum 
transfer carried by the photon. They are expressed respectively as  
\begin{align}
{p}_{{10}} = ( M_{10}, \bm{0}), \ \  {p}_{{8}} = ( E_{{8}},
  -\bm{q}), \ \  {q} = ( \omega_q, \bm{q}) , 
\end{align}
where ${\bm q}$ and $\omega_q$ stand for the three-momentum and energy
of the virtual photon, respectively. In this rest frame, the
energy-momentum relations are given as $E_{{8}}^{2} =  M_{{8}}^{2} +{
  |\bm{q}|}^{2}$ and $E_{{10}}^{2} = M_{{10}}^{2}$. Thus, the momentum
and energy of the virtual photon are written by
\begin{align}
|\bm{q}|^{2} =\left (\frac{M_{{10}}^{2} + M_{{8}}^{2}+ Q^{2}}{2
  M_{{10}}}\right )^{2} - M_{{8}}^{2}, \ \ \ 
\omega_q =\left (\frac{M_{{10}}^{2} - M_{{8}}^{2} - Q^{2}}{2 M_{{10}}}\right ),
\label{eq:kinematics}
\end{align}
where $Q^{2} = -q^{2} > 0$.

To explain the radiative excitation $B_8\gamma^*\to B_{10}$, we need
to evaluate the matrix element of the EM current  $V^\mu$ between 
$B_{10}$ and $B_{8}$ as follows
as~\cite{Jones:1972ky}    
\begin{align}
\langle B_{10}(p_{10},\lambda') | V^{\mu}(0) | B_{8}(p_{8},\lambda) \rangle = i
  \sqrt{\frac{2}{3}} \overline{u}_{\beta}(p_{10},1/2) \Gamma^{\beta
  \mu} {u}(p_{8},1/2), 
\label{eq:matel1}
\end{align}
where the EM current $V^\mu$ is defined as
\begin{align}
  \label{eq:EMCurrent1}
V^\mu(0) = \bar{\psi}(0) \gamma^\mu \hat{Q} \psi(0)\com{,}
\end{align}
with the charge operator given by 
$\hat{Q} = \mathrm{diag}(2/3,-1/3,-1/3)$. $\lambda \, (\lambda')$
denotes the helicity of the initial (final) baryon
state. $u^{\beta}(p,\lambda)$ and
$u(p,\lambda)$ stand for the Rarita-Schwinger and Dirac spinors,
respectively. The baryon states and Dirac spinors are normalized by
$\langle B(p',\lambda') | 
B(p,\lambda) \rangle=
(p^{0}/M)\delta_{\lambda'\lambda}\delta^{(3)}(\bm{p}'-\bm{p})$ and
$\overline{u}(p,\lambda')u(p,\lambda) = \delta_{\lambda'\lambda}$,
respectively.
$\Gamma^{\beta \mu}$ in Eq.~\eqref{eq:matel1} can be expressed in
terms of the multipole form factors~\cite{Ernst:1960zza, Jones:1972ky} 
\begin{align}
\Gamma^{\beta \mu} = G^{*}_{M1}(Q^{2}) \mathcal{K}_{M1}^{ \beta \mu} +
  G _{E2}^{*}(Q^{2}) \mathcal{K}_{E2}^{ \beta \mu} + G _{C2}^{*}(Q^{2})
  \mathcal{K}_{C2}^{ \beta \mu}, 
\end{align}
where $G^{*}_{M1}$, $G^{*}_{E2}$, and $G^{*}_{C2}$ are called 
respectively as the magnetic dipole ($M1$) transition form factor, the
electric quadrupole ($E2$) one, and the Coulomb quadrupole ($C2$)
one. The corresponding Lorentz tensors $\mathcal{K}^{ \beta \mu}_{M1,
  E2, C2}$
are written as 
\begin{align}
  {\mathcal{K}}_{M1}^{\beta \mu} &=
   \frac{-3(M_{10}+M_{8})}{2M_{8}[(M_{10}+M_{8})^{2}+Q^{2}]}
\varepsilon^{\beta \mu \sigma \tau} P_{\sigma} q_{\tau}, \cr 
{\mathcal{K}}_{E2}^{\beta \mu} &= -{\mathcal{K}}_{M}^{\beta \mu}  -
                                 \frac{6}{4M^{2}_{10} |\bm{q}|^{2}}
                                 \frac{M_{10}+M_{8}}{M_{8}}
                                 \varepsilon^{\beta \sigma \nu \gamma}
                                 P_{\nu}q_{\gamma}
                                 {{\varepsilon^{\mu}}_{\sigma}}^{\alpha  \delta}
                                 p_{10\alpha}q_{\delta} i\gamma^{5}, \cr
 {\mathcal{K}}_{C2}^{\beta \mu} &=-\frac{3}{4M^{2}_{10} |\bm{q}|^{2}}
                                  \frac{M_{10}+M_{8}}{M_{8}} q^{\beta} [q^{2} P^{\mu} -
                                  q\cdot P q^{\mu}]i\gamma^{5}.
\end{align}
The Lorentz tensors are required to satisfy the identities $q_{\mu}
{\mathcal{K}}_{M1,E2,C2}^{\beta \mu} = 0$ by conservation of the
EM current.  

The transition magnetic moment\footnote{Note that the
  definition of the magnetic transition moment in the present work is 
  different from that in Refs.~\cite{Ledwig:2008es}, where
  the following approximation was used $\sqrt{M_{10}/M_{8}} \approx 1
  + \mathcal{O}(N^{-2}_{c})$. In the present work, we strictly follow
  the definition used in experiments.} and the transition electric
quadrupole moment are defined respectively
by~\cite{Tiator:2003xr,Pascalutsa:2006up}:  
\begin{align} 
 \mu_{B_{8}B_{10}} = \frac{M_{N}}{M_{8}}\sqrt{\frac{M_{10}}{M_{8}}}
  G^{*}_{M1} (0) \mu_{N}, \;\;\;
 \mathcal{Q}_{B_{8}B_{10}} =-\frac{6}{M_{8}} \frac{2
  M_{10}}{M_{10}^{2}-M_{8}^{2}}  \sqrt{\frac{M_{10}}{M_{8}}} G^{*}_{E2} (0),
\label{eq:Static}
\end{align}
where $\mu_N$ denotes the nuclear magneton defined by $\mu_N=e/2M_N$.  

It is also of great interest to examine the helicity amplitudes, since
they can be extracted from experimental data. The transverse 
and Coulomb helicity amplitudes are defined respectively in terms of
the spatial and temporal components of the EM current
\begin{align}
  A_{\lambda} &= -\frac{e}{\sqrt{2\omega_q}}
  \int d^{3} r e^{i {\bm{q}} \cdot {\bm{r}}} \bm{\epsilon}_{+1}
  \cdot \langle B_{10} (3/2 , \lambda) | \overline{\psi} (\bm{r})
  \hat{\mathcal{Q}} \bm{\gamma} \psi (\bm{r}) |
  B_{8}(1/2,\lambda-1) \rangle, \cr
 S_{1/2} &= -\frac{e}{\sqrt{2\omega_q}} \frac{1}{\sqrt{2}}
   \int d^{3} r e^{i {\bm{q}} \cdot {\bm{r}}}  \langle B_{10} (3/2,1/2)
           | \overline{\psi} (\bm{r}) \hat{\mathcal{Q}} \gamma^{0}
           \psi (\bm{r}) | B_{8}(1/2,1/2) \rangle,
\label{eq:MatrixEl1}
\end{align}
where $\lambda$ is the corresponding value of the helicity of the
decuplet baryon $B_{10}$, i.e. $\lambda=3/2$ or $1/2$.  
The transverse photon polarization vector is defined as
$\hat{\bm{\epsilon}} =  -1/\sqrt{2} (1,i,0)$. The helicity amplitudes
can be expressed in terms of the multipole form
factors~\cite{Jones:1972ky,Pascalutsa:2006up}    
\begin{align}
  A_{1/2} &= - \frac{e}{\sqrt{2\omega_{q}}}\frac{1}{4c_{\Delta}}
      ( G^{*}_{M1} - 3 G^{*}_{E2}), \ \ 
       A_{3/2} = - \frac{e}{\sqrt{2\omega_{q}}}\frac{\sqrt{3}}{4c_{\Delta}}
       ( G^{*}_{M1} + G^{*}_{E2}), \ \ S_{1/2}=  \frac{e}{\sqrt{2\omega_{q}}}
      \frac{|\bm{q}|}{4c_{\Delta}M_{10}}G^{*}_{C2}.
\label{eq:Amplitude_decom}
\end{align}
where $c_{\Delta} = \sqrt{\frac{M^{3}_{8} }{2 M_{10} |\bm{q}|^{2}}}
\sqrt{1 + \frac{Q^{2}}{(M_{10}+M_{8})^{2}}}$.
The multipole form factors can be explicitly expressed as follows:
\begin{align}
G^{*}_{M1}(Q^{2})& =  - 2c_{\Delta}  \int d^{3} r 3
                   j_{1}(|\bm{q}||\bm{r}|)
                   \langle B_{10} (3/2 , 1/2) |   [\hat{\bm r} \times
                   {\bm V}]_{11}
                   | B_{8}(1/2,-1/2) \rangle, \cr
G^{*}_{E2}(Q^{2})& \simeq -2c_{\Delta}  \int d^{3} r
                   \sqrt{\frac{20\pi}{27}}  \frac{\omega_{q}}{|{\bm
                   q}|} \left( \frac{\partial}{\partial r} r
                   j_{2}(|{\bm q}||{\bm r}|)\right)  \langle B_{10}
                   (3/2 , 1/2) | Y_{21} (\hat{\bm r}) V_{0} |
                   B_{8}(1/2,-1/2) \rangle, \cr 
G^{*}_{E2}(Q^{2})& = 4c_{\Delta}\frac{M_{10}}{|\bm{q}|}  \int d^{3} r
                   \sqrt{10\pi} j_{2}(|\bm{q}||\bm{r}|)   \langle
                   B_{10} (3/2 , 1/2) | Y_{20} (\hat{\bm r}) V_{0} |
                   B_{8}(1/2,1/2) \rangle. 
 \end{align}
Note that we neglect a term that gives a tiny correction to $E2$ at
low-energy region and implements the current conservation on that
multipole form factor~\cite{Fiolhais:1996bp,Amoreira:2000dn}. Once we
have evaluated the form factors, the well-known ratios $R_{EM}$  and
$R_{SM}$, which are defined respectively as 
\begin{align}
R_{EM}(Q^{2})  = -\frac{G^{*}_{E2}(Q^{2})}{G^{*}_{M1}(Q^{2})}, \;\;\;
R_{SM}(Q^{2})  =
  -\frac{|\bm{q}|}{2M_{10}}\frac{G^{*}_{C2}(Q^{2})}{G^{*}_{M1}(Q^{2})}, 
\end{align}
can be obtained. The decay width in terms of the helicity amplitudes
is expressed as~\cite{Tanabashi:2018oca}: 
\begin{align}
\Gamma(B_{10} \to B_{8} \gamma) &= \frac{\omega^{2}_{q}}{\pi}
                                  \frac{M_{8}}{2M_{10}} \left(
                                  |A_{1/2} |^{2}+ |A_{3/2}
                                  |^{2}\right). 
\label{eq:decay_width}
\end{align}
\section{Expressions of the EM transition form factors} 
We start from the effective chiral action $S_{\mathrm{eff}}[U]$ given
by 
\begin{align}
S_{\mathrm{eff}}[U]= - N_{c} \mathrm{Tr \, ln } \, D(U),
\end{align}
where $N_{c}$ is the number of colors and $D(U)$ represents the Dirac
operator defined by 
\begin{align}
D(U)= i \slashed{\partial}+iMU^{\gamma_{5}}+i\hat{m}.
\label{eq:DiracOp}
\end{align}
The $U^{\gamma_{5}}$ denotes the chiral field, which is written as
\begin{align}
U^{\gamma_{5}}=U \frac{1+\gamma_{5}}{2} + U^{\dagger}
  \frac{1-\gamma_{5}}{2}. 
\end{align}
The $U$ field is represented by 
\begin{align}
  \label{eq:1}
U(x) = \exp[i\pi^a(x) \lambda^a], 
\end{align}
where $\pi^a(x)$ stand for the pNG fields and $\lambda^a$ designate
the Gell-Mann matrices in flavor SU(3). Since the classical pion
fields have a hedgehog symmetry, we keep only the pNG fields for
$a=1,\,2,\,3$ whereas all other fields are set equal to zero. Then the
static SU(3) soliton field can be constructed by Witten's trivial
embedding  
\begin{align}
U(\bm{r}) = \left( \begin{array}{c c} U_{\mathrm{SU(2)}}(\bm{r}) & 0
                     \\ 0 &
                            1
                   \end{array} \right), 
\end{align}
where the $U_{\mathrm{SU(2)}}$ is expressed by incorporating the
hedgehog symmetry as 
\begin{align}
U_{\mathrm{SU(2)}} (\bm{r}) = \mathrm{exp}[i \bm{\pi} \cdot \bm{\tau}]
  =\mathrm{exp}[i  P(r) \hat{\bm{r}} \cdot \bm{\tau}]. 
\end{align}
The soliton profile function $P(r)$ is derived by minimizing the
energy of the soliton. 
$\hat{m}$ in Eq.~\eqref{eq:DiracOp} stands for the current-quark mass
matrix. Assuming isospin symmetry, i.e., $m_{u}=m_{d}=\overline{m}$,
we can express it as $\hat{m}=\mathrm{diag}(\overline{m}, \,
\overline{m}, \, m_{s})=\overline{m}+\delta m$. The $\delta m$ is
defined by 
\begin{align}
\delta m =\frac{-\overline{m}+m_{s}}{3}\bm{1}_{3} +
  \frac{\overline{m}-m_{s}}{\sqrt{3}}\lambda^{8} = m_{1} \bm{1}_{3} +
  m_{8}\lambda^{8}, 
\end{align}
and will be treated as a small perturbation. $m_1$ and $m_8$ are
respectively written as  $m_1 = (-\overline{m}+m_{\mathrm{s}})/3, \,
\,  m_8 = (\overline{m}-m_{\mathrm{s}})/\sqrt{3}$.  To find the
single-quark energies and states, we can define the Dirac 
Hamiltonian $h(U)$ as 
\begin{align}
h(U) = i\gamma_{4} \gamma_{i} \partial_{i} - \gamma_{4}
  MU^{\gamma_{5}} - \gamma_{4}\overline{m}. 
\end{align}

In the large $N_c$ limit, the $1/N_c$ pion-loop fluctuations 
can be neglected. Thus, the classical equation of motion gives rise to
the pion mean fields, which is derived by $\delta
S_{\mathrm{eff}}/\delta P(r)=0$. However, we have to consider the zero
modes of the soliton completely, since they are not at all small. This
is known as the zero-mode collective quantization, which can be
achieved by rotating and translating the soliton as follows: 
\begin{align}
A(t)U_{c}(\bm{r}-\bm{z}(t))A^{\dagger}(t).
\end{align}
Here, $A(t)$ is a time-dependent SU(3) matrix in the flavor space and
$\bm{z}(t)$ represents a time-dependent translation of the
soliton. The rotational velocity is defined as 
\begin{align}
\Omega =
  \frac{1}{2i}\mathrm{Tr}(A^{\dagger}\dot{A}\lambda^{\alpha})\lambda^{\alpha}
  = \frac{1}{2}\Omega_{\alpha}\lambda^{\alpha} ,
\end{align}
which is of order $1/N_c$. Assuming that the soliton rotates slowly,
we can deal with $\Omega_\alpha$ as a perturbative parameter. The 
translational corrections are taken to the zeroth order. Note that the
translational zero modes naturally provides the Fourier transform for
the EM transition form factors. Having performed the zero-mode
quantization, we arrive at the collective Hamiltonian, which 
consists of the flavor SU(3) symmetric and symmetry-breaking terms, 
given as 
\begin{align}
H_{\mathrm{coll}}=H_{\mathrm{sym}} +H_{\mathrm{sb}}, 
\end{align}
with
\begin{align}
H_{\mathrm{sym}}&= M_{\mathrm{sol}} +
                  \frac{1}{2I_{1}}\sum_{i=1}^{3}\hat{J}_{i}\hat{J}_{i}
                  +
                  \frac{1}{2I_{2}}\sum_{a=4}^{7}\hat{J}_{a}\hat{J}_{a}
                  \cr 
H_{\mathrm{sb}}&= \alpha D^{(8)}_{88}(A) + \beta \hat{Y} +
                 \frac{\gamma}{\sqrt{3}}D^{(8)}_{8i}(A)\hat{J}_{i} .
\end{align}
$M_{\mathrm{sol}}$ stands for the classical soliton
mass that is obtained by the minimization procedure.
$D^{(\mathcal{R})}_{ab}(A)$ and $\hat{J}$ denote the SU(3)
Wigner D functions and the angular momentum operators. The explicit
expressions for the moments of inertia ($I_{1}, I_{2}$) and the 
inertial parameters ($\alpha, \beta, \gamma$) are given in
Appendix~\ref{app:b}. 

In the present work, we will present here only the final
expressions of the EM transition form factors, since the general 
formalism can be found in previous works. For a detailed calculation, we refer to a
recent work on the EM form factors of the baryon
decuplet~\cite{Kim:2019gka} and a review~\cite{Christov:1995vm}. 
Having taking into account the rotational $1/N_{c}$ and linear $m_{s}$
corrections, we obtain the magnetic dipole form factor $G_{M1}^{*}$
as 
\begin{align}
  G_{M1}^{B_{8}\to B_{10} *}(Q^{2}) &=-c_{\Delta}  \int d^{3} r \frac{6}{\sqrt{2}}
                      j_{1}(|\bm{q}||\bm{r}|)\mathcal{G}^{B_{8}\to
                      B_{10}}_{M1}(\bm{r}), 
\label{eq:M1}
\end{align}
where the corresponding magnetic dipole density
$\mathcal{G}^{B_{8}\to B_{10}}_{M1}(\bm{r})$ is defined as  
\begin{align}
  \mathcal{G}^{B_{8}\to B_{10}}_{M1}(\bm{r}) &=
\left(  \mathcal{Q}_{0} (\bm{r})  +
\frac{1}{I_{1}}  \mathcal{Q}_{1} (\bm{r}) \right)
\langle B_{10} | D^{(8)}_{Q3 } | B_{8} \rangle
-  \frac{1}{\sqrt{3}} \frac{1}{I_{1}}  \mathcal{X}_{1} (\bm{r})
\langle B_{10} | D^{(8)}_{Q 8}J_{3} | B_{8} \rangle \cr
&-   \frac{1}{I_{2}}  \mathcal{X}_{2} (\bm{r}) \langle B_{10} |
 d_{pq3} D^{(8)}_{Qp} J_{q} | B_{8} \rangle   
 + \frac{2}{\sqrt{3}} m_{8} \left(\frac{K_{1}}{I_{1}} \mathcal{X}_{1}
  (\bm{r}) -   \mathcal{M}_{1} (\bm{r})\right)  \langle B_{10} | D^{(8)}_{83}
 D^{(8)}_{Q8} |B_{8} \rangle \cr
&+2 m_{8} \left(\frac{K_{2}}{I_{2}} \mathcal{X}_{2} (\bm{r})
     -    \mathcal{M}_{2} (\bm{r})\right) \langle B_{10} | d_{pq3}  D^{(8)}_{8p}
  D^{(8)}_{Qq} | B_{8} \rangle  \cr 
  & - 2   m_{1} \mathcal{M}_{0} (\bm{r})
    \langle B_{10} | D^{(8)}_{Q3} | B_{8} \rangle -
    \frac{2}{\sqrt{3}} m_{8} \mathcal{M}_{0} (\bm{r})
    \langle B_{10} | D^{(8)}_{88} D^{(8)}_{Q3} | B_{8}\rangle. 
\label{eq:M1den}
\end{align}
The explicit expressions for the densities $\mathcal{Q}_i$,
$\mathcal{X}$, and $\mathcal{M}_i$ can be found in Appendix~\ref{app:a}.
The results on the matrix elements of collective operators $\langle
B_{10} |...| B_{8}  \rangle$ are given in Appendix~\ref{app:b}.
The expression of the electric quadrupole form factor is given as
\begin{align}
G_{E2}^{B_{8}\to B_{10} *} (Q^{2}) &=c_{\Delta} \int d^{3} r \sqrt{\frac{10}{9}}  
\frac{\omega}{|\bm{q}|} \left( \frac{\partial}{\partial r} r  
j_{2}(|\bm{q}||\bm{r}|)\right)     \mathcal{G}^{B_{8}\to B_{10}}_{E2} (\bm{r}), 
\label{eq:E2}
\end{align}
with the corresponding density $\mathcal{G}^{B_{8}\to B_{10}}_{E2} (\bm{r})$
\begin{align}
\mathcal{G}^{B_{8}\to B_{10}}_{E2}(\bm{r}) =& - \frac{2}{I_{1}}
\mathcal{I}_{1E2} (\bm{r})\left(3
    \langle B_{10} | D^{(8)}_{Q 3} J_{3} | B_{8} \rangle
    - \langle B_{10} | D^{(8)}_{Q i} J_{i} |B_{8} \rangle  \right) \cr
 & + 4 m_{8}\left( \frac{K_{1}}{I_{1}} \mathcal{I}_{1E2}(\bm{r}) -
    \mathcal{K}_{1E2}(\bm{r})\right) \left( 3\langle B_{10} | D^{(8)}_{8 3}
    D^{(8)}_{Q 3} | B_{8}\rangle -\langle B_{10} | D^{(8)}_{8 i} D^{(8)}_{Q i}
   |B_{8} \rangle \right). 
\label{eq:E2den}
\end{align}
The explicit expressions for $\mathcal{I}_{1E2}(\bm{r})$ and $
\mathcal{K}_{1E2}(\bm{r})$ can be found in Appendix~\ref{app:a}.
The Coulomb quadrupole form factor $G_{C2}^{B_8\to B_{10}*}$ is
written as  
\begin{align}
  G_{C2}^{B_8\to B_{10}*}(Q^{2}) &=c_{\Delta} \sqrt{40}\int d^{3} r  
                      \, \frac{M_{10}}{|\bm{q}|} j_{2}(|{\bm q}||{\bm r}|)
                      \mathcal{G}^{B_8\to B_{10}}_{C2} (\bm{r}), 
\label{eq:C2}
\end{align}
where  ${\cal G}^{B_8\to B_{10} }_{C2} (\bm{r})$ is simply the same as
${\cal G}^{B_8\to B_{10} }_{E2} (\bm{r})$.

It is more convenient to decompose the densities into three different terms
\begin{align}
  \mathcal{G}^{B_8\to B_{10} }_{(M1,E2,C2)}(\bm{r}) =
     \mathcal{G}^{B_8\to B_{10}(0)}_{(M1,E2,C2)}(\bm{r})
  + \mathcal{G}^{B_8\to B_{10} (\mathrm{op})}_{(M1,E2,C2)}(\bm{r})
  + \mathcal{G}^{B_8\to B_{10} (\mathrm{wf})}_{(M1,E2,C2)}(\bm{r}).
  \label{eq:decom}
\end{align}
The first term represents the SU(3)-symmetric ones including both the
leading and rotational $1/N_c$ terms, the second one denotes the
linear $m_{\mathrm{s}}$ corrections arising from the current-quark
mass term of the effective chiral action. The last term 
is originated from the collective wave functions. If the effects of
the flavor SU(3) symmetry breaking are considered, a collective baryon
wave function is not any longer in a pure state but a state mixed with
higher representations, as shown in Eq.~\eqref{eq:mixedWF1}. Thus,
there are two different terms that provide the effects of flavor SU(3)
symmetry breaking. The explicit expressions of these three terms are
given in Appendix~\ref{app:c}. 

Finally, we want to mention that while the leading-order contributions
in the rotational $1/N_c$ expansion vanish for the $E2$ transition form factor
because of the hedgehog ansatz in the present approach. This means 
that the rotational $1/N_c$ corrections take a role of the
leading-order contributions. Moreover, we have only the single
rotational $1/N_c$ term that contains the density
$\mathcal{I}_{1E2}(\bm{r})$. The corresponding expression can be found
in Eq.~\eqref{eq:E2den} in Appendix~\ref{app:a}.  Similarly, the $C2$ 
form factors does not get any contribution from the leading-order term.   

\section{Results and discussion}
The parameters in the $\chi$QSM except for the dynamical quark mass
were already fixed by reproducing properties of the pion. Since the
contributions of the sea quarks need to be regularized, 
the cutoff mass $\Lambda$ should be introduced. This is fixed by
reproducing the pion decay constant, $f_\pi=93$ MeV. the average mass
of the up and down current quarks $\overline{m}$ is determined by the
physical pion mass $m_\pi=139$ MeV. While the dynamical quark mass $M$
can be considered as a free parameter, it is also fixed by describing
the electric form factor of the proton. Once we fix all these
parameters, we compute various observables including both the
lowest-lying light and singly heavy baryons. Therefore, we do not have
any room to fit the parameters in the present calculation. 

It is already known from previous investigations~\cite{Silva:1999nz, 
  Urbano:2000xg, Ledwig:2008es} that the magnitudes of the EM
transition form factors of the $\Delta$ are rather underestimated,
compared with the experimental data while the $E2/M1$ and $C2/M1$
ratios are well described. There are several reasons why it is so. In
fact, the pion-loop effects come into essential play in explaining 
the nature of the $\Delta$ isobar, since it decays strongly into the
$\pi N$. Moreover, the $\Delta$ isobar has a rather broad width, so
that the corresponding wavefunction should contain such information 
arising from this broad width. In Ref.~\cite{Yang:2018idi} the strong
decay widths of the baryon decuplet were scrutinized based on the
$\chi$QSM in a model-independent approach, where all the dynamical
parameters were fixed by experimental data without calculating them
self-consistently. While the strong decay width of the $\Delta$ is still
underestimated, the widths of all the other members of the baryon
decuplet are in good agreement with the experimental data. It implies
that one should go beyond the pion mean-field approximation to
describe the strong properties of the $\Delta$ baryon
quantitatively. Since it is rather difficult to take into account the
pion-loop corrections beyond the pion mean-field approximation in the
present framework, we will consider the $Q^2$ dependence of the form
factors and the ratios of $E2/M1$ and $C2/M1$.  These effects beyond
the mean fields may be cancelled in the calculation of these ratios. 
Note that the lattice calculations also have similar problems in
reproducing the experimental data.   

\begin{figure}[htp]
\centering
\includegraphics[scale=0.235]{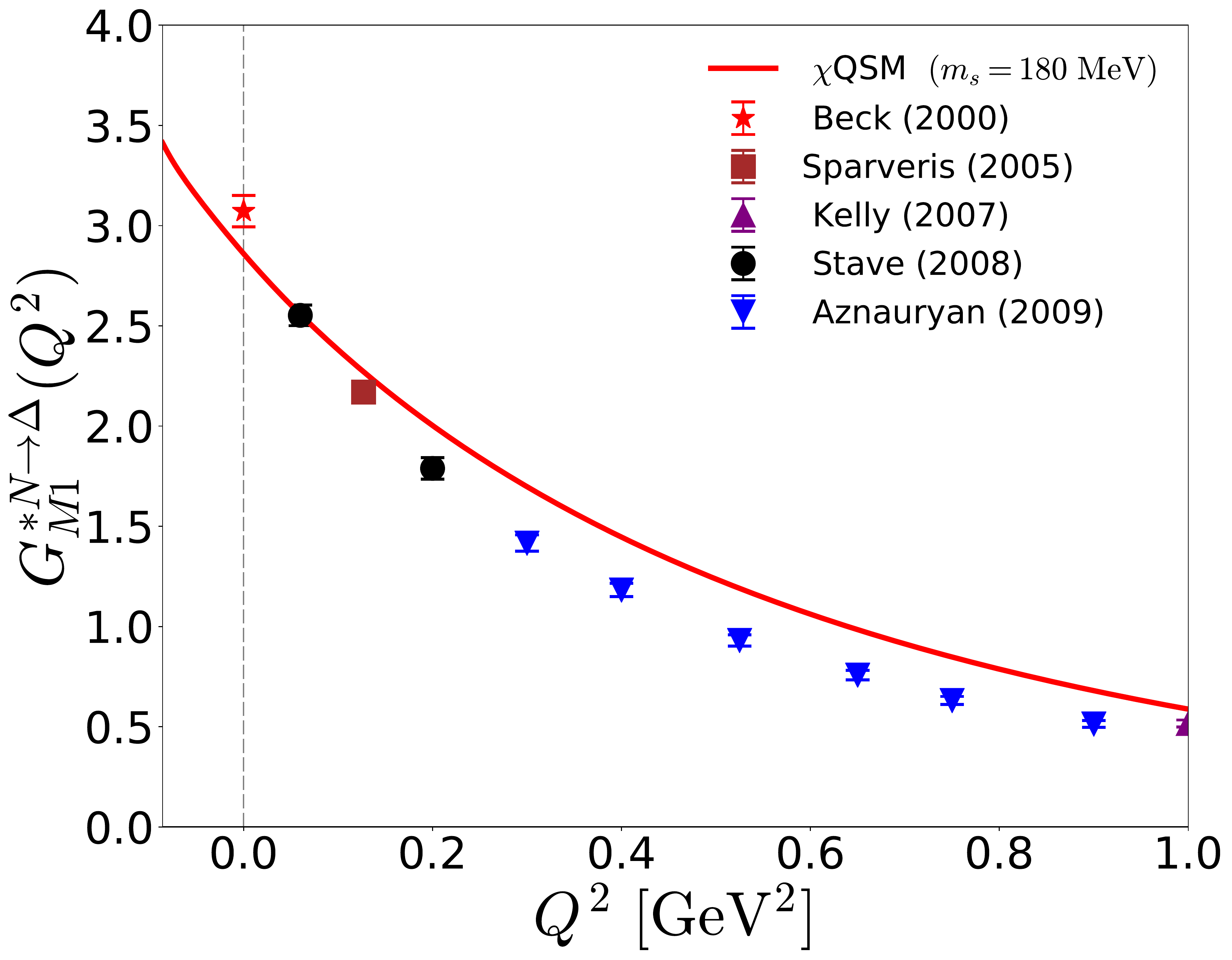}
\includegraphics[scale=0.235]{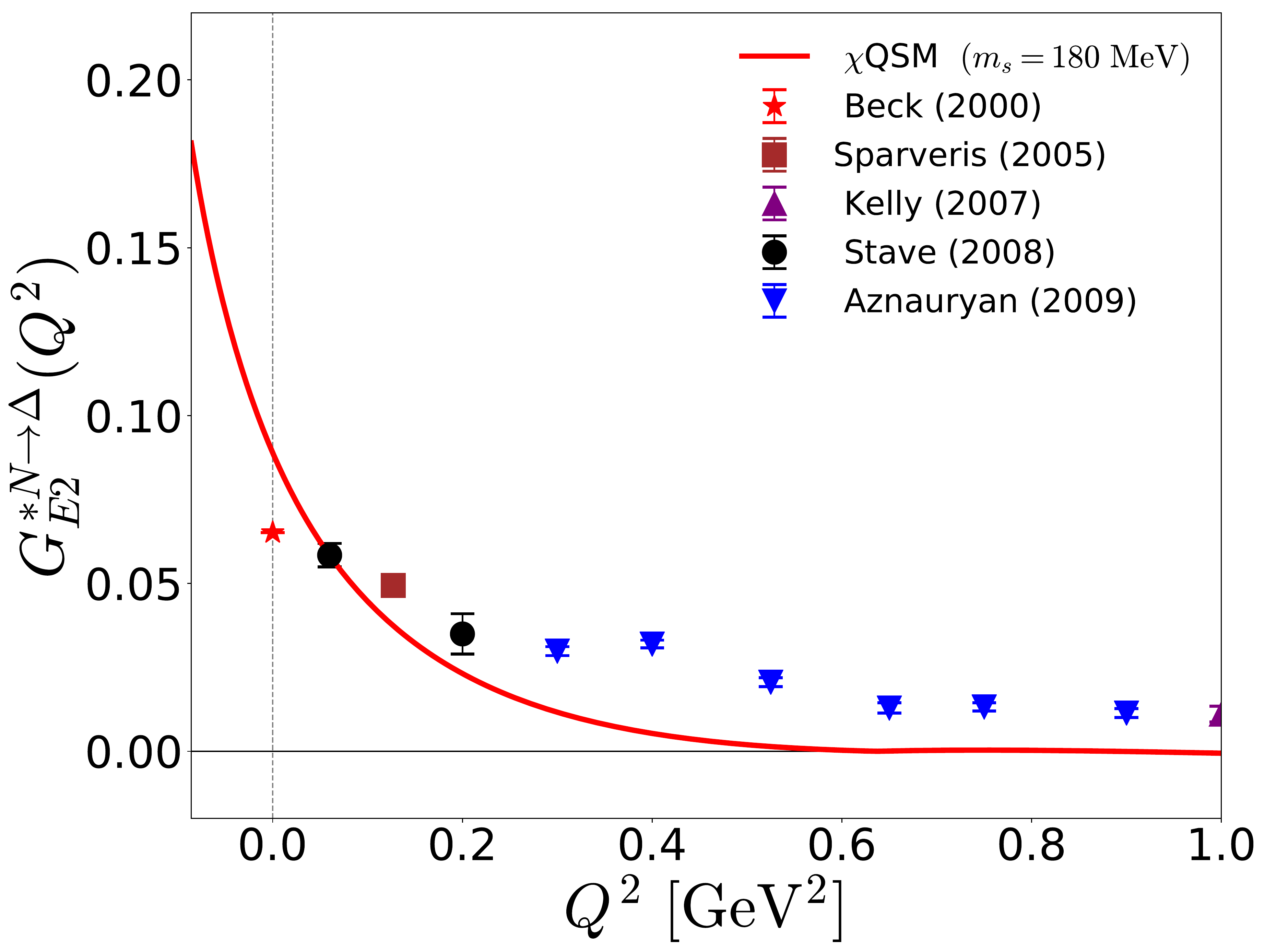}
\includegraphics[scale=0.235]{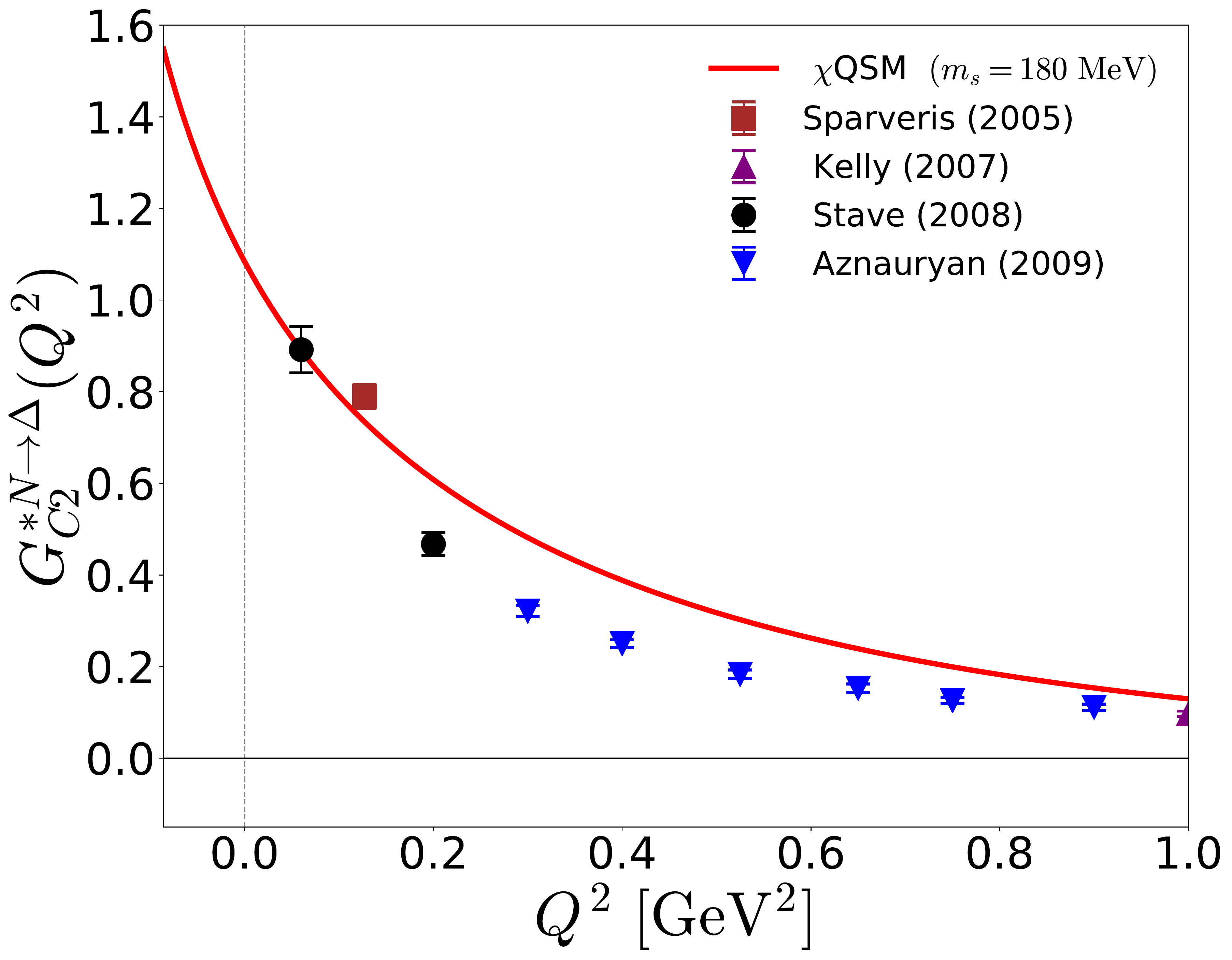}
\caption{The normalized results for the magnetic dipole transition form
  factor for the $N\to \Delta$ EM transition as functions of $Q^2$
  with the strange current quark mass $m_{\mathrm{s}}$ taken to be
  180 MeV. In the upper left and right panels, we draw the results of the
  magnetic dipole and electric quadrupole transition form factors
  respectively, whereas in the lower panel, we depict those of
  the Coulomb quadrupole form factor.  They are compared with the
  experimental and empirical data. We normalize
  the numerical results by the experimental data at
  $Q^2=0.06\,\mathrm{GeV}^2$, i.e. the $M1$ form factors by 1.82, the 
  $E2$ ones by 3.13, and $C2$ ones by 3.18. The solid curve
  illustrates the result of the EM transition form factors with
  $m_{\mathrm{s}}=180$ MeV. The result is normalized by the   data
  from Ref.~\cite{Stave:2008aa}, i.e. by the factor of 1.82.  
  The red triangle denotes the data taken from 
  Ref.~\cite{Beck:1999ge}, the black circles from
  Ref.~\cite{Stave:2008aa}, the brown square from
  Ref.~\cite{Sparveris:2004jn}, the blue squares from
  Ref.~\cite{Aznauryan:2009mx}, and the violet triangle from
  Ref.~\cite{Kelly:2005jy}.   
}  
\label{fig:2}
\end{figure}
We first discuss the results of the $N\to \Delta$ EM transition form
factors, focusing on the $Q^2$ dependence of the form factors.
In order to compare the $Q^2$ dependence of the present results with
the experimental and empirical data, we normalize the values of the
form factors at $Q^2=0.06\,\mathrm{GeV}^2$, using the experimental
data on the helicity amplitudes by the A1
Collaboration~\cite{Stave:2008aa}.  We explicitly multiply the present
values of the $M1$, $E2$, and $C2$ form factors by 1.82, 3.13, and
3.18, respectively. 
The upper left panel of Fig.~\ref{fig:2} draws the result of the
magnetic dipole transition form factor as a function of $Q^2$ with the
strange current quark mass taken to be $m_{\mathrm{s}}=180$ MeV. We
take the experimental and empirical data from
Refs.~\cite{Beck:1999ge, Sparveris:2004jn, Kelly:2005jy, Stave:2008aa,
  Aznauryan:2009mx}. The present result seems to fall off more slowly
than those of the empirical and experimental data, as $Q^2$ 
increases. However, the result is in agreement
with the data. In the upper right panel of Fig.~\ref{fig:2} we show
the result of the electric quadrupole form factor for the $N\to
\Delta$ EM transition as a function of $Q^2$. The result exhibits 
different $Q^2$ dependence from that of the $M1$ form
factor.  It falls off faster than the empirical and experimental
data. As we will discuss explicitly later, the present results of the
$E2/M1$ ratio deviate from the experimental data because of this
$Q^2$ dependence of the $E2$ form factor. Actually, one can understand
this $Q^2$ behavior of the present results. Since the $E2$ transition
form factor is proportional to $\omega_{q}$ within this model
expression, it is strongly suppressed when $\omega_{q}(Q^{2}) = 0$,
which corresponds approximately to $Q^{2}\simeq0.6~\mathrm{GeV}^2$ for
the $N\gamma^{*}\to\Delta$ excitation. That explains why the $E2$ form 
  factor decreases drastically as $Q^2$ increases.  On the other
hand, The result of the Coulomb form factor describes relatively well
the experimental data, as shown in the lower panel of Fig.~\ref{fig:2}.  

\begin{figure}[htp]
\includegraphics[scale=0.235]{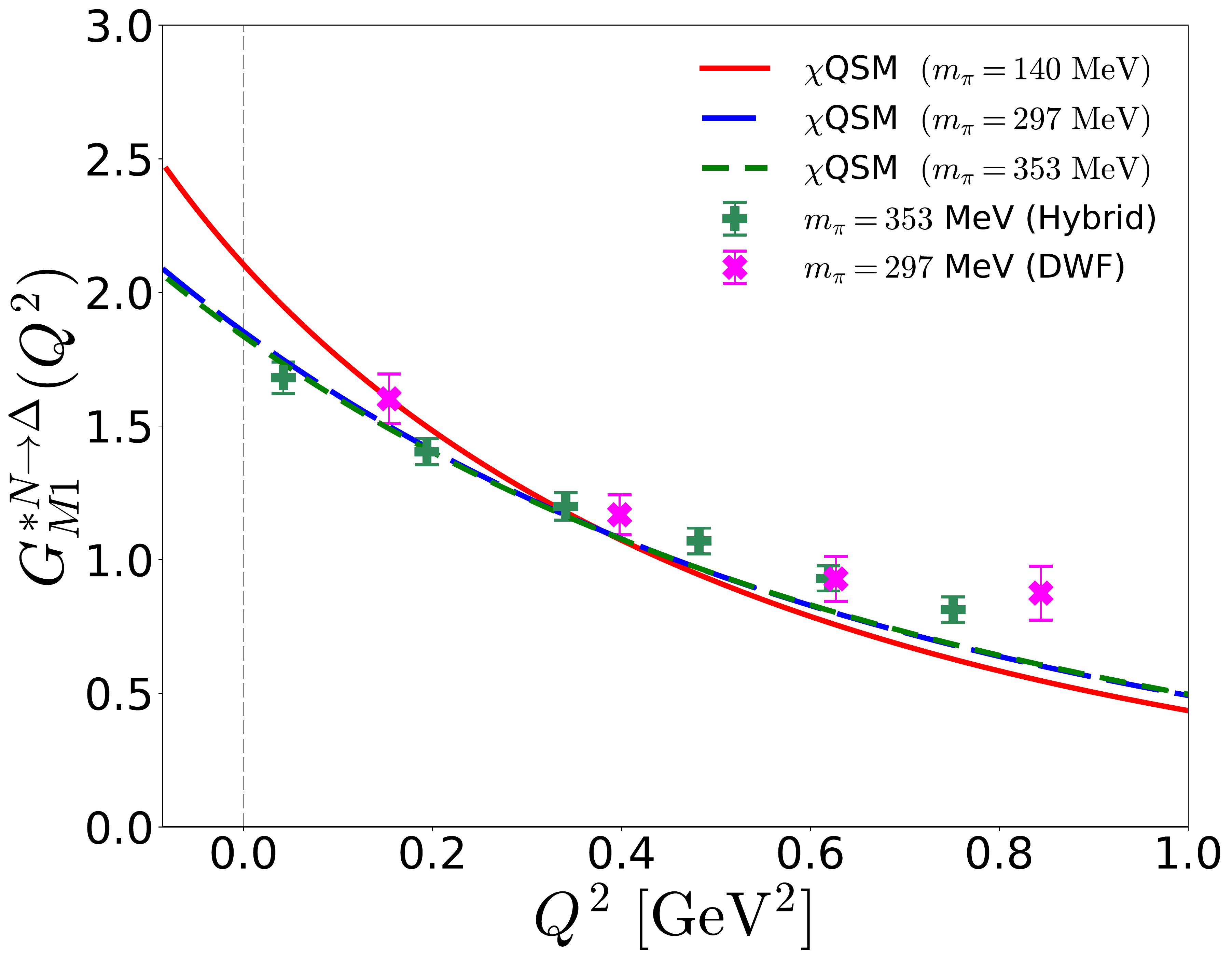}
\includegraphics[scale=0.235]{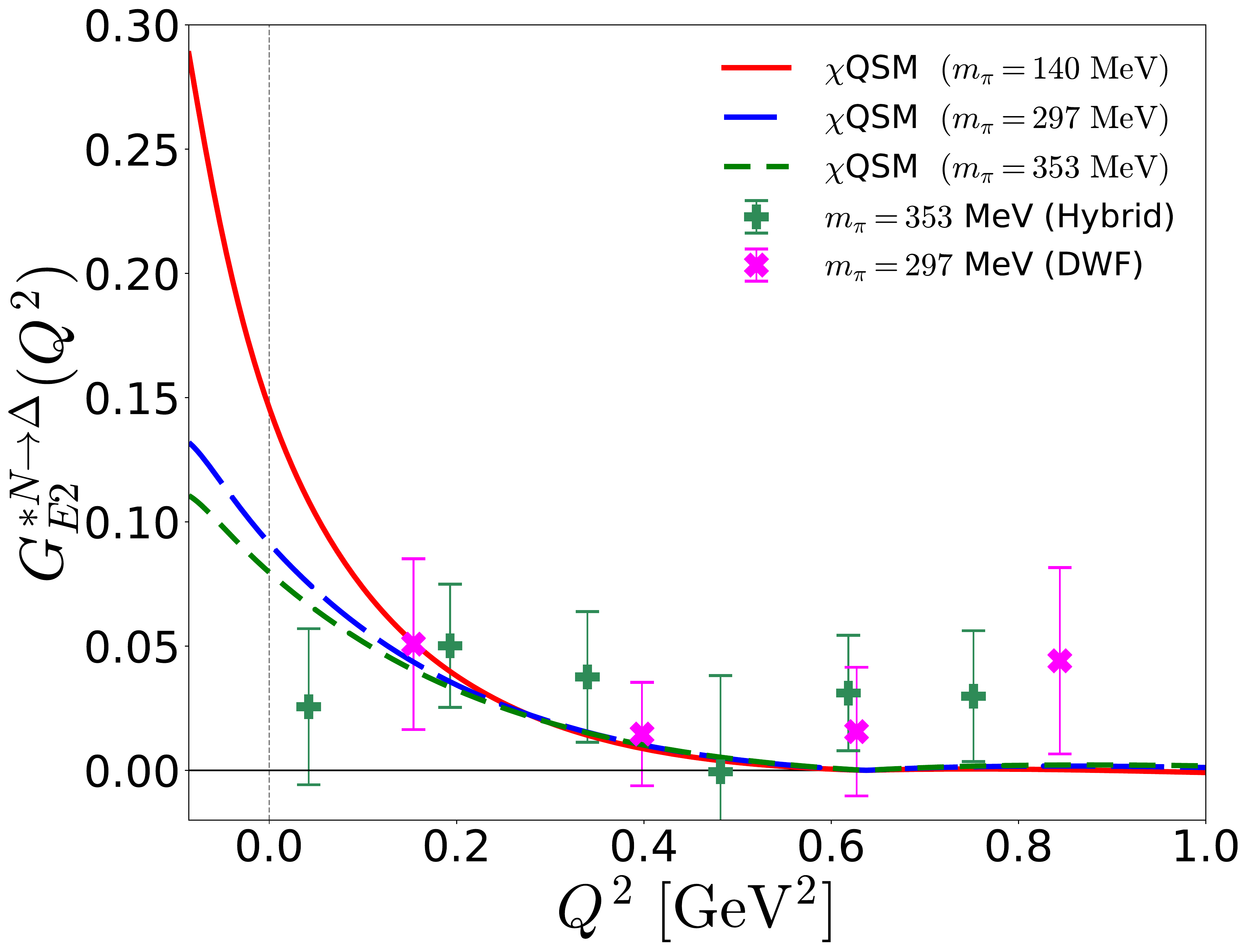}
\includegraphics[scale=0.235]{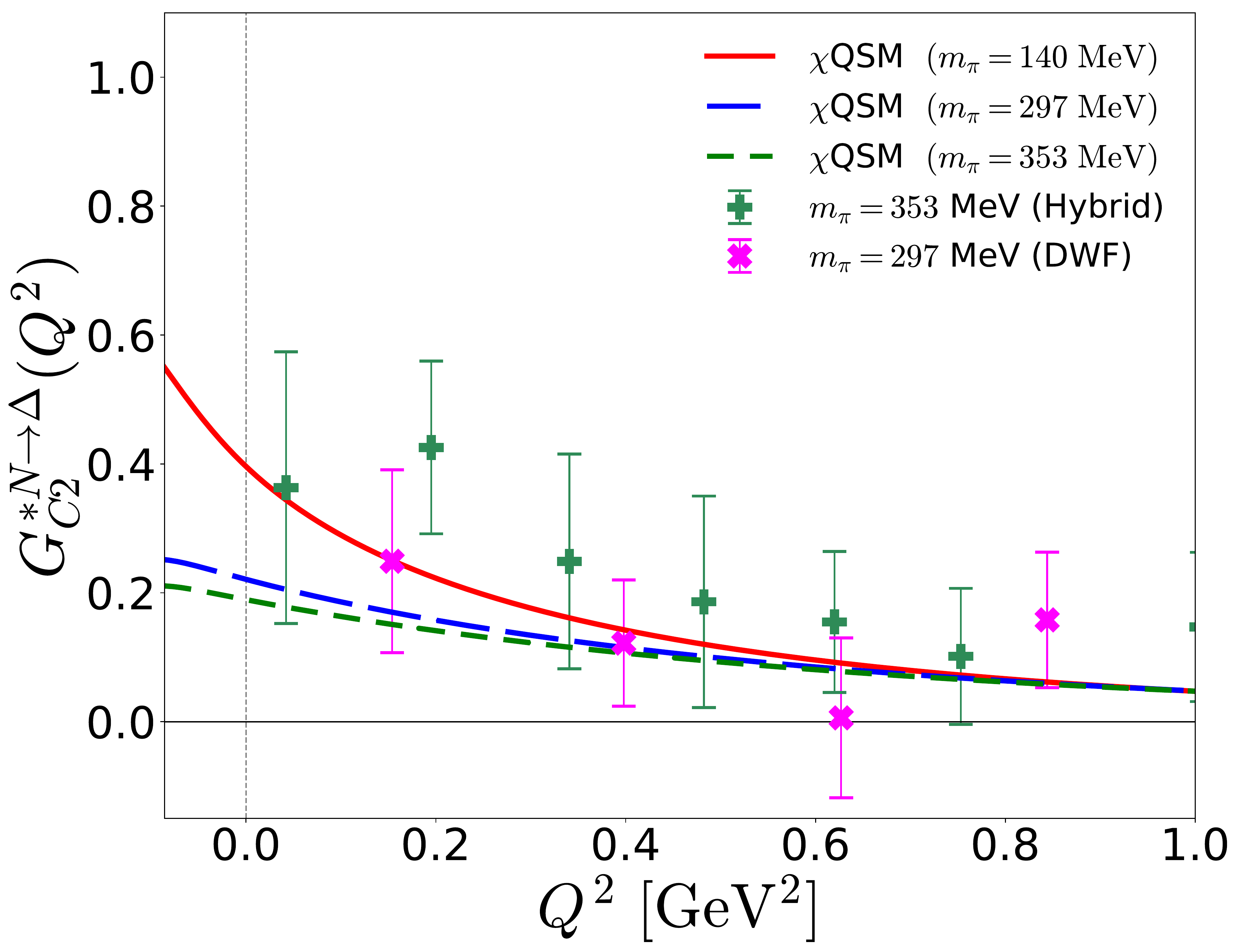}
\caption{The normalized results for the magnetic dipole  transition
  form factor for the $N\to \Delta$ EM transition as a function of
  $Q^2$ with the strange current quark mass   $m_{\mathrm{s}}$ taken
  to be 180 MeV. In the upper left and right panels, we draw the
  results of the magnetic dipole and electric quadrupole transition
  form factors respectively, whereas in the lower panel, we depict
  those of the Coulomb quadrupole form factor.  They are compared with
  the lattice data taken from  Ref.~\cite{Alexandrou:2010uk}. We
  normalize the numerical results by the lattice data with
  $m_{\pi}=297$ MeV, i.e. the $M1$ form factors by 1.51, the $E2$ ones
  by 6.93, and $C2$ ones by 1.57. The solid curve illustrates the  
  result of the EM transition form factors with the value of the
  physical pion mass, the long-dashed one shows those with $m_\pi=297$
  MeV, and the short-dashed one draws those with $m_\pi=353$ MeV. The 
  ``$\bm{+}$''  and ``$\bm{\times}$'' symbols denote respectively the
  lattice data with $m_\pi=353$ MeV and $m_\pi=297$ MeV. }  
\label{fig:3}
\end{figure}
In order to compare the present results with the lattice
data~\cite{Alexandrou:2010uk}, we have to compute the EM transition
form factors, employing the values of the unphysical pion mass, which
were used by Ref.~\cite{Alexandrou:2010uk}. To do that, we have to
derive the solutions of the pion mean fields with specific values of
the unphysical pion mass, i.e. $m_\pi=297$ MeV and 353 MeV. Then 
we can compute the EM transition form factors, using these solutions
with the values of $m_\pi$. In fact, Goeke et al.~~\cite{Goeke:2005fs}
showed that the stable mean-field soliton still exists in the wide
range of the pion mass $0\le m_\pi \le 1500$ MeV. It indicates that
the results from the $\chi$QSM can be directly compared with those
from lattice QCD, the unphysical pion mass being employed. They indeed 
described remarkably the mass of the nucleon in comparison with the
lattice data. The same method was extended to the description of the
energy-momentum form factors of the nucleon~\cite{Goeke:2007fq} and
the EM form factors of singly heavy baryons \cite{Kim:2019wbg}. Thus,
we compare the present results with the lattice data, using the values
of the unphysical pion mass employed by the lattice calculation.

The upper left panel of Fig.~\ref{fig:3} illustrates the results of
the $M1$ transition form factors as functions of $Q^2$. The solid
curve depicts the original values of the $M1$ form factors with the
physical pion mass, whereas the long-dashed and short-dashed ones draw
those obtained by using $m_\pi=297$ MeV and 353 MeV, respectively. The
sizes of the form factors get smaller as the value of $m_\pi$
increases. Moreover, the $M1$ form factors fall off more slowly as the
pion mass increases, which are well-known behaviors in the lattice
results of the EM form factors of the proton and
$\Delta$~\cite{Alexandrou:2007we, Alexandrou:2008bn,
  Alexandrou:2009hs, Capitani:2015sba}.  While the general $Q^2$
dependence of results of the $M1$ transition form factor is similar to
those of the lattice calculation, the present results decrease still faster
than the lattice ones as $Q^2$ increases. In the upper right panel of
Fig.~\ref{fig:3} we represent the results of the $E2$ transition form
factors with the pion mass varied as in the case of the $M1$ form
factor. The sizes of the $E2$ form factor are drastically diminished
by increasing the value of the pion mass. This tendency was already
seen in the $E2$ form factors of the $\Delta$ and $\Omega^-$ in
Ref.~\cite{Kim:2019gka}. As shown in the upper right panel of
Fig.~\ref{fig:3}, the present results with larger pion masses are in
better agreement with the lattice data in the lower $Q^2$ region
($Q^2\lesssim 0.5\,\mathrm{GeV}^2$).
In the lower panel of Fig.~\ref{fig:3}, we depict the results of the
$C2$ transition form factor. As the pion mass increases, the sizes of
the $C2$ one decreases as in the case of the $E2$ form
factor. However, the results get underestimated compared with the
lattice data in the lower $Q^2$ region.

\begin{figure}[htp]
\includegraphics[scale=0.235]{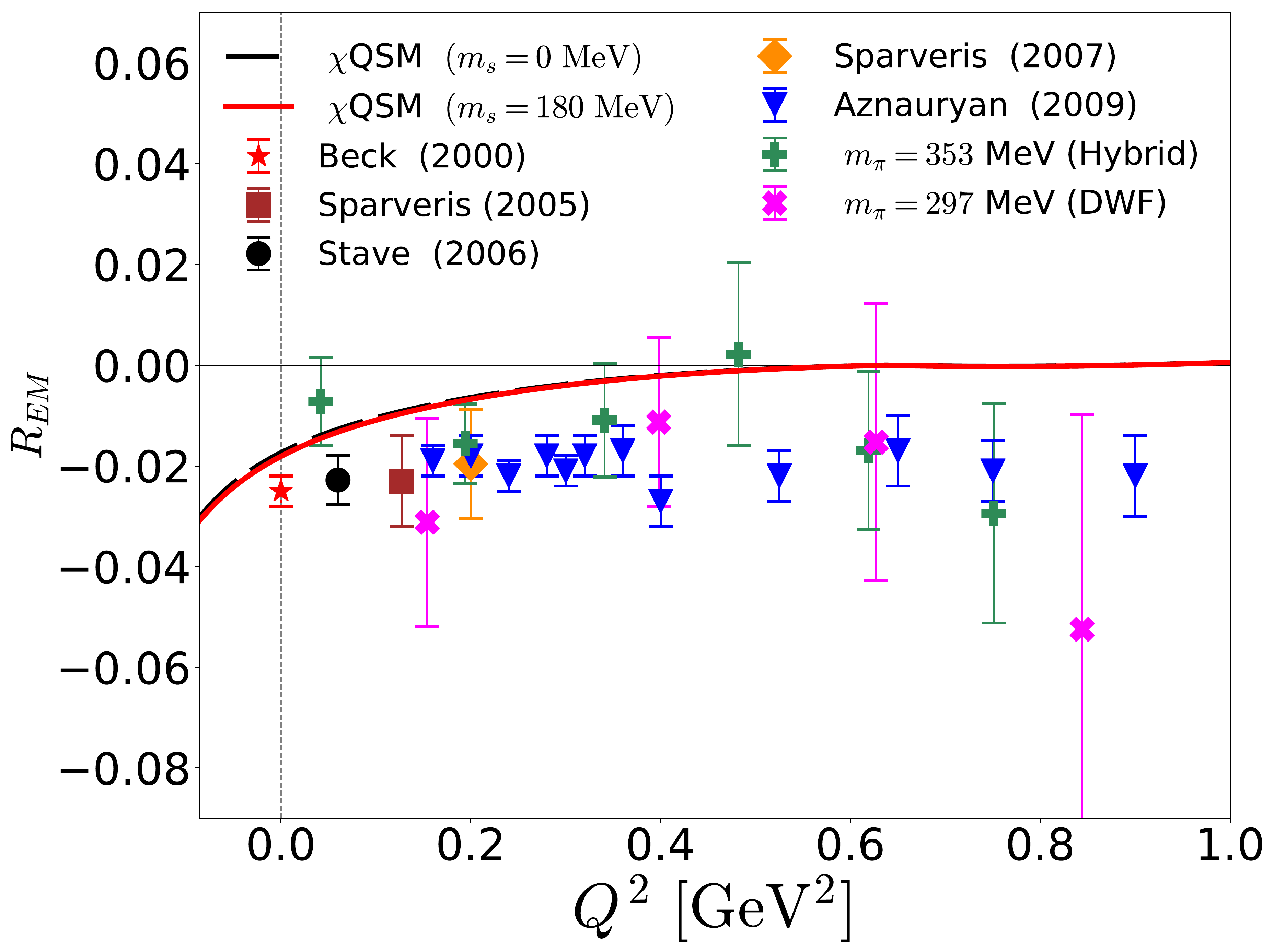}
\includegraphics[scale=0.235]{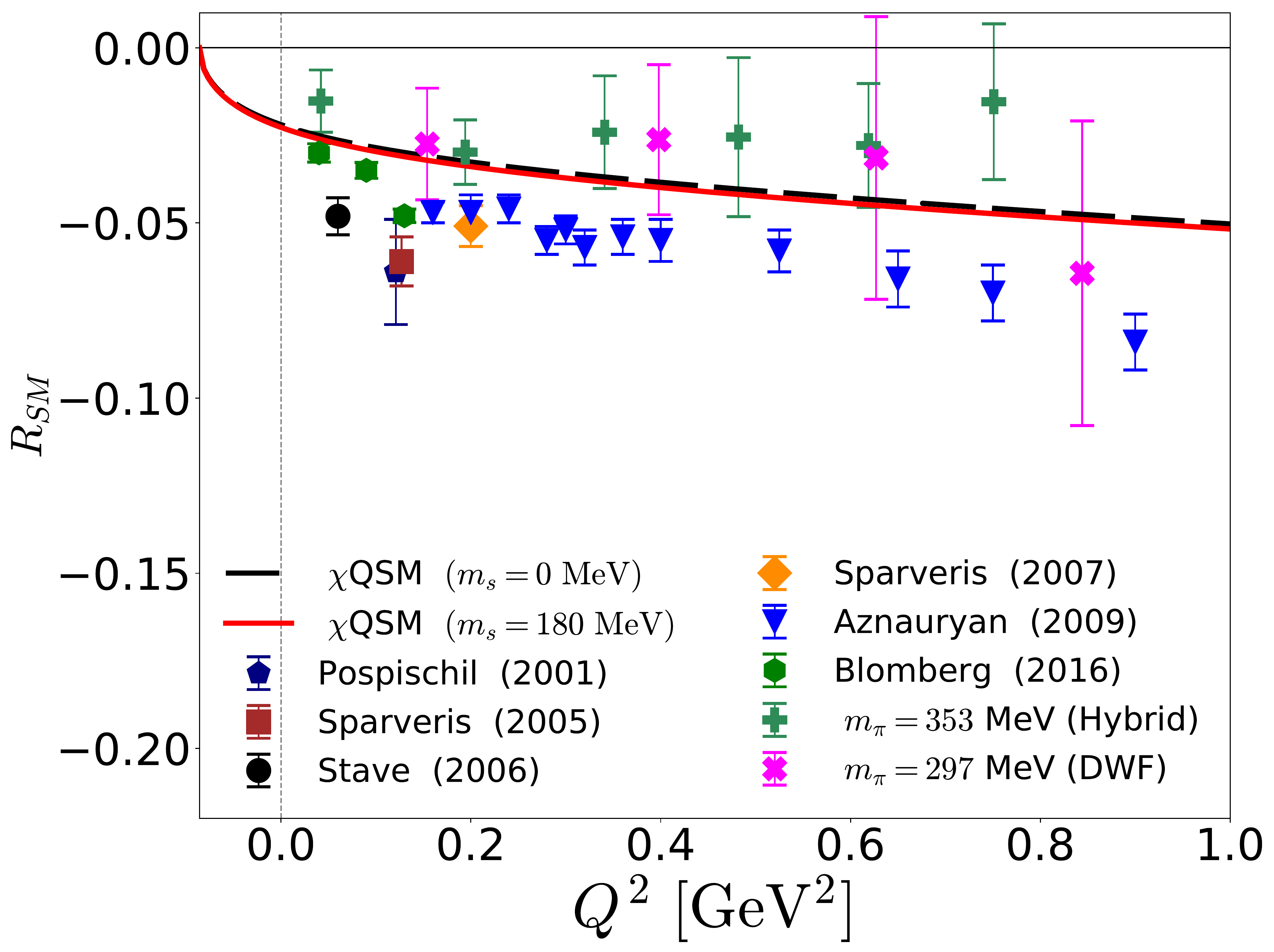}
\caption{The numerical results of the $E2/M1$ and $C2/M1$ ratios,
  i.e., $R_{EM}(Q^2)$ and $R_{SM}(Q^2)$ for the $N\gamma^*\to \Delta$
  excitation in the left and right panels, respectively, without any 
  normalization. The solid curves draws the results with the effects
  of flavor SU(3) symmetry breaking whereas the dashed ones depict
  those in exact SU(3) symmetry. The results   are compared with the
  experimental and empirical data taken from 
  Refs.~\cite{Beck:1999ge, Aznauryan:2009mx, Sparveris:2004jn,
    Sparveris:2006uk, Stave:2006ea, Pospischil:2000ad,
    Blomberg:2015zma} as well as the lattice
  data taken from~\cite{Alexandrou:2010uk}. }  
\label{fig:4}
\end{figure}
The left and right panels of Fig.~\ref{fig:4} show respectively the
results of the $E2/M1$ and $C2/M1$ ratios for the$N\gamma^* \to \Delta$
excitation as functions of $Q^2$, being compared with the experimental
and empirical data~\cite{Beck:1999ge, Aznauryan:2009mx,
  Sparveris:2004jn, Sparveris:2006uk, Stave:2006ea, Pospischil:2000ad,
  Blomberg:2015zma} as well as those of the lattice
calculation~\cite{Alexandrou:2010uk}. The results for the $E2/M1$ ratios
are in qualitatively good agreement with the experimental data near
$Q^2\approx 0$, the present ones fall off faster than the experimental
and empirical data. This arises from the $Q^2$ dependence of the $E2$
transition form factors, for which the results decrease much faster 
than those of the $M1$ form factors. On the other hand, the results for
the $C2/M1$ form factors are more or less in agreement with the
data.  Note that the lattice data on the $C2/M1$ are underestimated in
comparison with the experimental data. This may be due to the
unphysical values of the pion mass used in the lattice calculations. 

\begin{figure}[htp]
\includegraphics[scale=0.24]{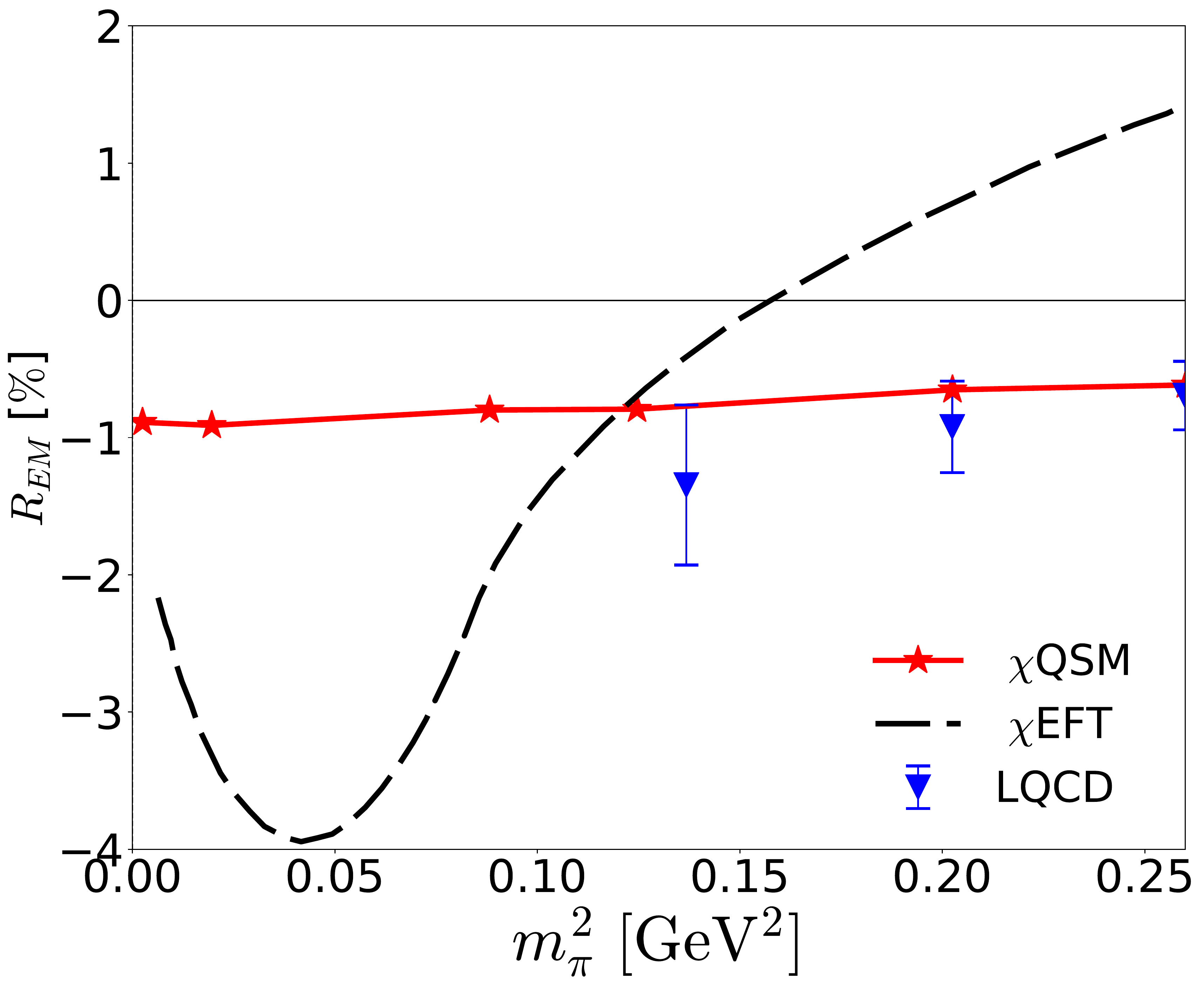}
\includegraphics[scale=0.24]{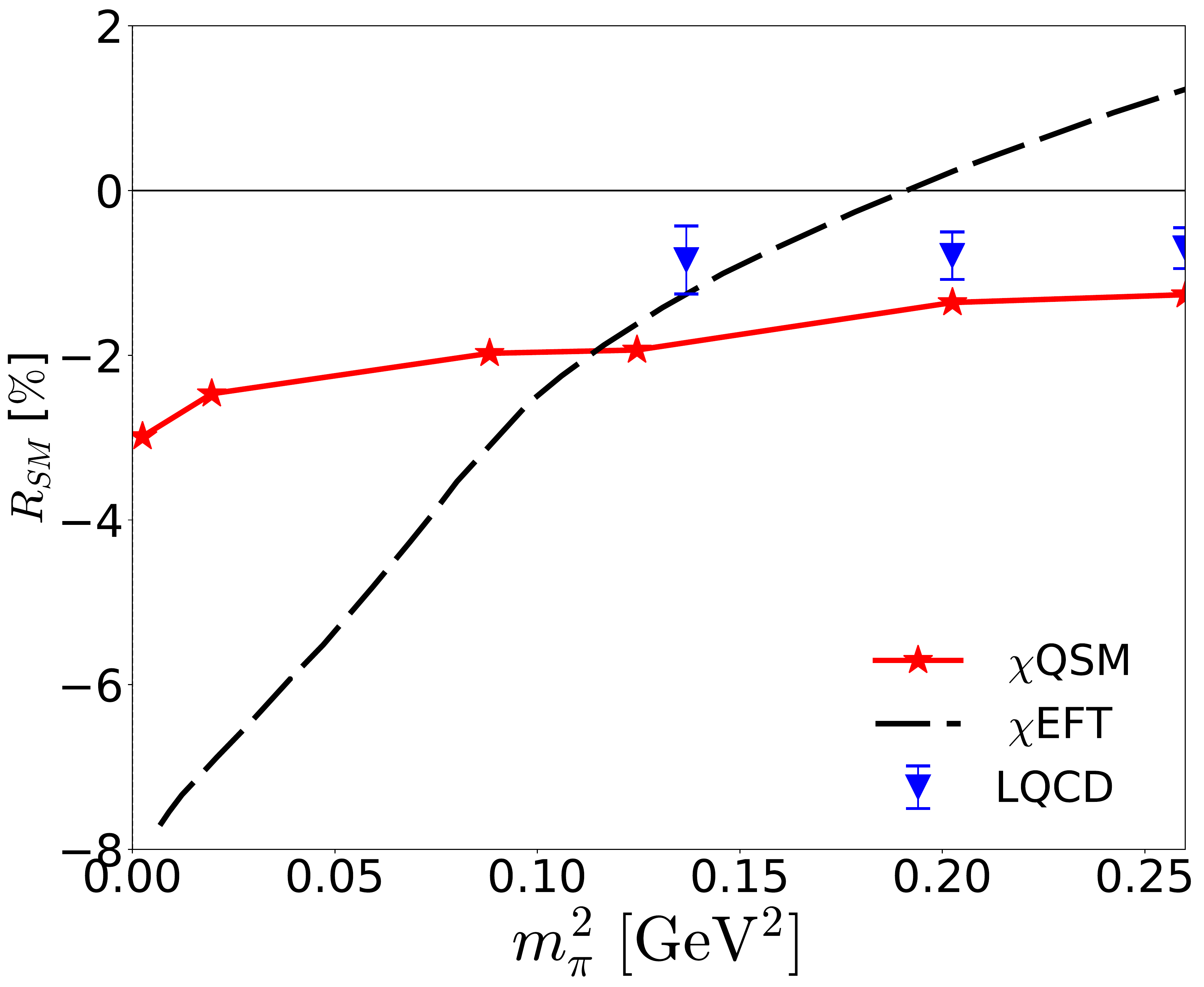}
\caption{The results for $R_{EM}$ and $R_{SM}$ given in percentage
  with the pion mass squared varied from 0 to
  $0.26\,\mathrm{GeV}^2$. The stars ({\color{red}$\star$}) denote the
  present results whereas the inverted triangles ({\color{blue}
    $\blacktriangledown$}) stand for the lattice data. The long-dashed
  curves illustrate the results from chiral effective theory
  ($\chi$EFT). The lattice data are taken from
  Ref.~\cite{Alexandrou:2004xn}, and the results from $\chi$EFT are 
  taken from Ref.~\cite{Pascalutsa:2005ts}.} 
\label{fig:5}
\end{figure} 
Since larger unphysical values of the pion mass are taken in lattice
calculations, it is of great importance to employ the same values of
the pion mass used in lattice QCD to make a quantitative comparison with
the lattice data. Moreover, it is also very interesting to examine the
dependence of the numerical results on the pion mass. In
Fig.~\ref{fig:5}, we show the present results for $R_{EM}$ and
$R_{SM}$ at $Q^2=0.1\,\mathrm{GeV}^2$ with the
pion mass squared varied from $0$ to $0.26\,\mathrm{GeV}^2$. As the
pion mass increases, the numerical results for $R_{EM}$ decrease
mildly. Those for $R_{SM}$ show a very interesting feature. As
$m_\pi^2$ decreases, the present results get closer to the lattice
data. The results from chiral effective field theory ($\chi$EFT)
behave very differently from both the present and lattice results,
which show very strong dependence on the pion mass.  

\begin{table}[htp]
  \setlength{\tabcolsep}{5pt}
\renewcommand{\arraystretch}{1.5}
\caption{The $R_{EM}$ and $R_{SM}$ on $B_{8}\gamma\to B_{10}$ within
  the chiral quark-soliton model with and without flavor SU(3)
  symmetry breaking in comparison those from Experimental data
  ~\cite{Beck:1999ge}, Skyrme 
  model~\cite{Haberichter:1996cp}, Linear sigma
  model(LSM)~\cite{Fiolhais:1996bp}, non-relativistic quark
  model(NQM)~\cite{Buchmann:1996bd}, QCD sum
  rule(QCDSR)~\cite{Wang:2009ru}, chiral constituent quark
  model($\chi$CQM)~\cite{Wagner:1998bu} and chiral perturbation
  theory($\chi$PT)~\cite{Li:2017vmq}.} 
  \scalebox{0.81}{
\begin{tabular}{c | c c c c c c c c c c c}
\hline
\hline
 & \multicolumn{2}{c}{$\chi$QSM($m_{s}=0$ MeV)}  &
\multicolumn{2}{c}{$\chi$QSM($m_{s}=180$ MeV)} &
PDG~\cite{Tanabashi:2018oca} & Skyrme~\cite{Haberichter:1996cp} &
LSM~\cite{Fiolhais:1996bp} & NQR~\cite{Buchmann:1996bd} &
QCDSR~\cite{Wang:2009ru} & $\chi$CQM~\cite{Wagner:1998bu}
  & $\chi$PT~\cite{Li:2017vmq} \\  
 $[\%]$&$R_{EM}$ & $R_{SM}$ &$R_{EM}$ & $R_{SM}$  &$R_{EM}$ &
$R_{EM}$&$R_{EM}$&$R_{EM}$&$R_{EM}$&$R_{EM}$&$R_{EM}$\\  
 \hline
 ${p\gamma\to\Delta^{+}}$            & -1.7 & -2.2 & -1.8 & -2.3 &
-2.5 & -2.1 & -1.8 & -3.5 & -1.7 &-3.7 & -2.5\\ 
 ${n\gamma\to\Delta^{0}}$            & -1.7 & -2.2 & -1.8 & -2.3 &
-2.5 & -2.1 & -1.8 & -3.5 & -1.7 &-3.7 & -2.5\\ 
  ${\Sigma^{+}\gamma\to\Sigma^{*+}}$ & -1.5 & -1.7 & -1.0 & -1.1 & --
                           & -1.2 & --   & --   & -2.9 &-2.9 & -1.1\\ 
 ${\Sigma^{0}\gamma\to\Sigma^{*0}}$  & -1.5 & -1.7 & -0.9 & -0.9 & --
                           & -1.0 & --   & --   & -2.3 &-2.3 & -0.9\\ 
 ${\Sigma^{-}\gamma\to\Sigma^{*-}}$  & -1.5 & -1.7 & -2.5 & -3.0 & --
                           & -1.9 & --   & --   & -8.0 &-5.5 & 3.7\\ 
 ${\Xi^{0}\gamma\to\Xi^{*0}}$        & -1.6 & -1.8 & -2.1 & -2.5 & --
                           & -1.2 & --   & --   & -1.6 &-1.3 & -0.9\\ 
 ${\Xi^{-}\gamma\to\Xi^{*-}}$        & -1.6 & -1.8 &  0.9 &  1.5 & --
                           & -2.0 & --   & --   & -12.4 &-2.8 & -3.9\\ 
 ${\Lambda^{0}\gamma\to\Sigma^{*0}}$ & -1.7 & -2.1 & -2.0 & -2.5 & --
                           & -1.8 & --   & --   & -4.6 &-2.0 & -1.8\\ 
 \hline
\end{tabular}
}
\label{tab:5}
\end{table}
The ratios $R_{EM}$ and $R_{SM}$ ratios for all the members of the
baryon decuplet have not been much investigated. The $R_{EM}$ is 
an only known ratio experimentally~\cite{Beck:1999ge,
  Tanabashi:2018oca}. Moreover, there are no experimental data and are 
few theoretical results on the $C2/M1$ ratios for the whole baryon
decuplet. In Table~\ref{tab:5}, we list the numerical results of the
$E2/M1$ ($R_{EM}$) and $C2/M1$ ($R_{SM}$) ratios at $Q^2=0$ in
comparison with those from other models. The second and fourth columns 
list the results of the $R_{EM}$ without and with the effects of
flavor SU(3) symmetry breaking. Comparing the results in these two
columns with each other, we find that the contributions of the
$m_{\mathrm{s}}$ corrections seem to be not at all small. However, one
has to keep in mind that the effects of flavor SU(3) symmetry breaking
to the $M1$ form factors are smaller than to the $E2$ and $C2$
form factors, as we will show later explicitly. Thus, the
effects of flavor SU(3) symmetry breaking apparently look
amplified. We want to mention that while the $M1$, $E2$, and $C2$ form
factors for the EM $\Sigma^- \to \Sigma^{*-}$ and $\Xi^- \to \Xi^{*-}$
transitions vanish in exact flavor SU(3) symmetry due to the $U$-spin
symmetry, the ratios $R_{EM}$ and $R_{SM}$ do not vanish. The reason
can be easily understood by examining Eqs.~\eqref{eq:M1leadingcon}
and~\eqref{eq:E2leading}, which are the SU(3) symmetric leading
contributions to the $M1$ and $E2$ form factors, respectively. The
matrix elements of the collective operators for both the $M1$ and $E2$
form factors have basically the same structures, so that the ratios of
these form factors are proportional to the ratios of the densities
given in terms of $\mathcal{Q}_0$ and so on. Therefore, even though
form factors for the $\Sigma^- \to \Sigma^{*-}$ and $\Xi^- \to \Xi^{*-}$
photo-transitions vanish in exact flavor SU(3) symmetry, the ratios
$R_{EM}$ and $R_{SM}$ turn out finite.

The present value of $R_{EM}$ for the $N\gamma \to \Delta$ is
underestimated by about 20~\% in comparison with the experimental 
data. This discrepancy may be overcome by going beyond the pion
mean-field approximation, as was hinted by the results of
$\chi$PT. The results of $R_{EM}$ for all decuplet hyperons are
comparable with those of chiral perturbation theory
($\chi$PT)~\cite{Li:2017vmq} except for those of  
$R_{EM}$ for the $\Sigma^- \gamma \to \Sigma^{*-}$ and $\Xi^-\gamma
\to \Xi^{*-}$ excitations. Interestingly, both channels are forbidden
by the $U$-spin symmetry. On the other hand, the results from the
chiral constituent quark model~\cite{Wagner:1998bu} are overall larger
than the present ones.  The results for the $\Sigma^- \gamma \to
\Sigma^{*-}$ and $\Xi^-\gamma \to \Xi^{*-}$ transitions from the QCD sum 
rules~\cite{Wang:2009ru} are very large, compared with those of the
present work.

The EM transition form factors should comply with the $U$-spin
symmetry in the exact flavor-SU(3) symmetric case. The $U$-spin
symmetry is inherited in Eqs.~\eqref{eq:M1leadingcon}
and~\eqref{eq:E2leading} as it should be.  The $U$-spin relations for
the magnetic transition moments were given in Refs.~\cite{Beg:1964nm,
  Kim:2005gz}. In particular, the magnetic transition form factors for
the negatively charged decuplet baryons should vanish in exact flavor
SU(3) symmetry, which one can easily see from
Eqs.~\eqref{eq:M1leadingcon}. Some years ago, the SELEX Collaboration
measured the upper limit of the partial width for the radiative decay
of $\Sigma^{*-}$, which is given as
$\Gamma (\Sigma^{*-}\to \Sigma^- \gamma) < 9.5\,\mathrm{keV}$.  
It indicates that the corresponding magnetic transition moment should
satisfy the upper limit
$|\mu_{\Sigma^{*-}\Sigma^-} | < 0.82\,\mu_N$~\cite{Kim:2005gz}. Thus,
the experimental data can provide a clue as to how much the $U$-spin
symmetry is broken in the case of the EM transitions for the baryon
decuplet.  

\begin{figure}[htp]
\centering
\includegraphics[scale=0.234]{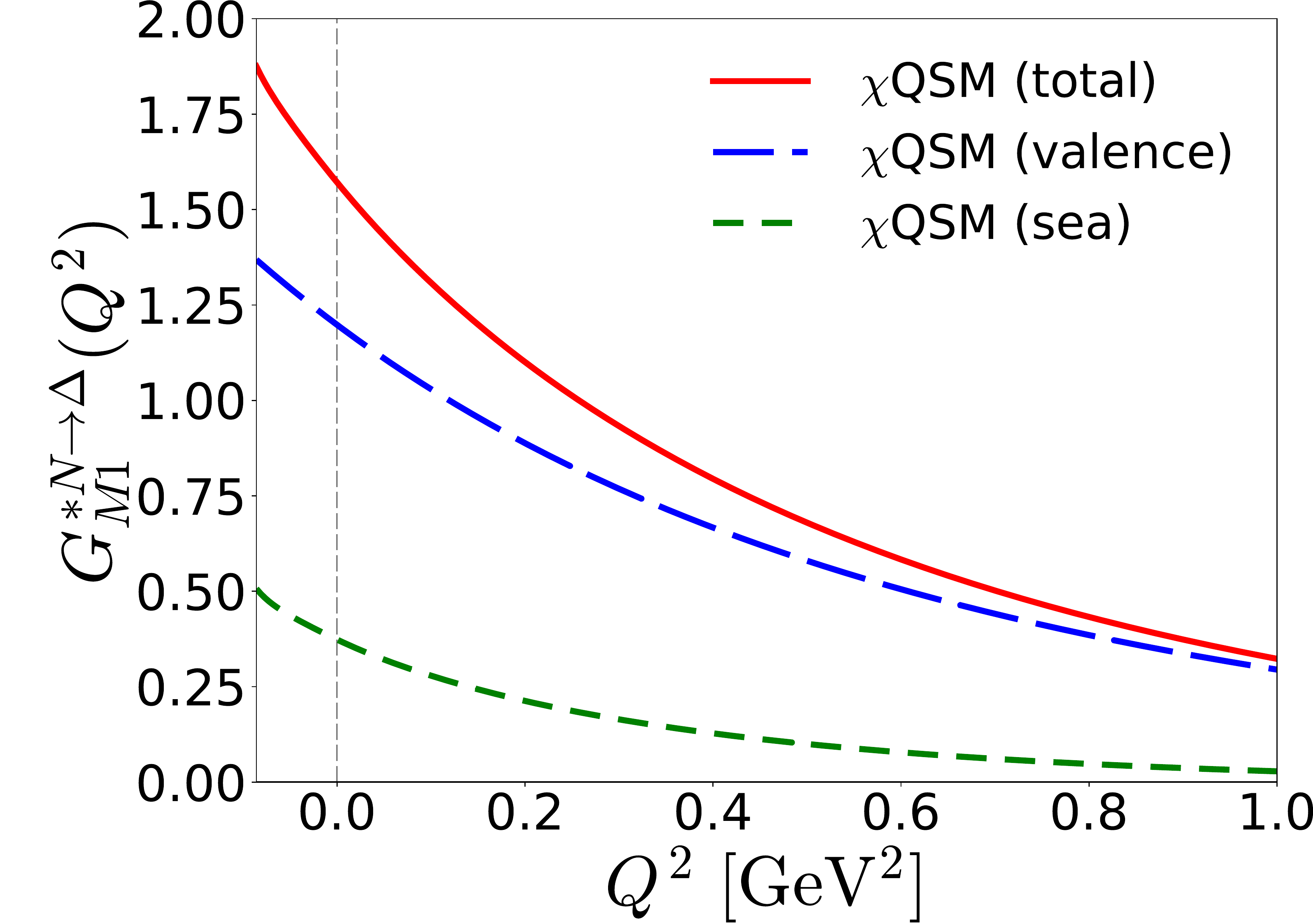}
\includegraphics[scale=0.234]{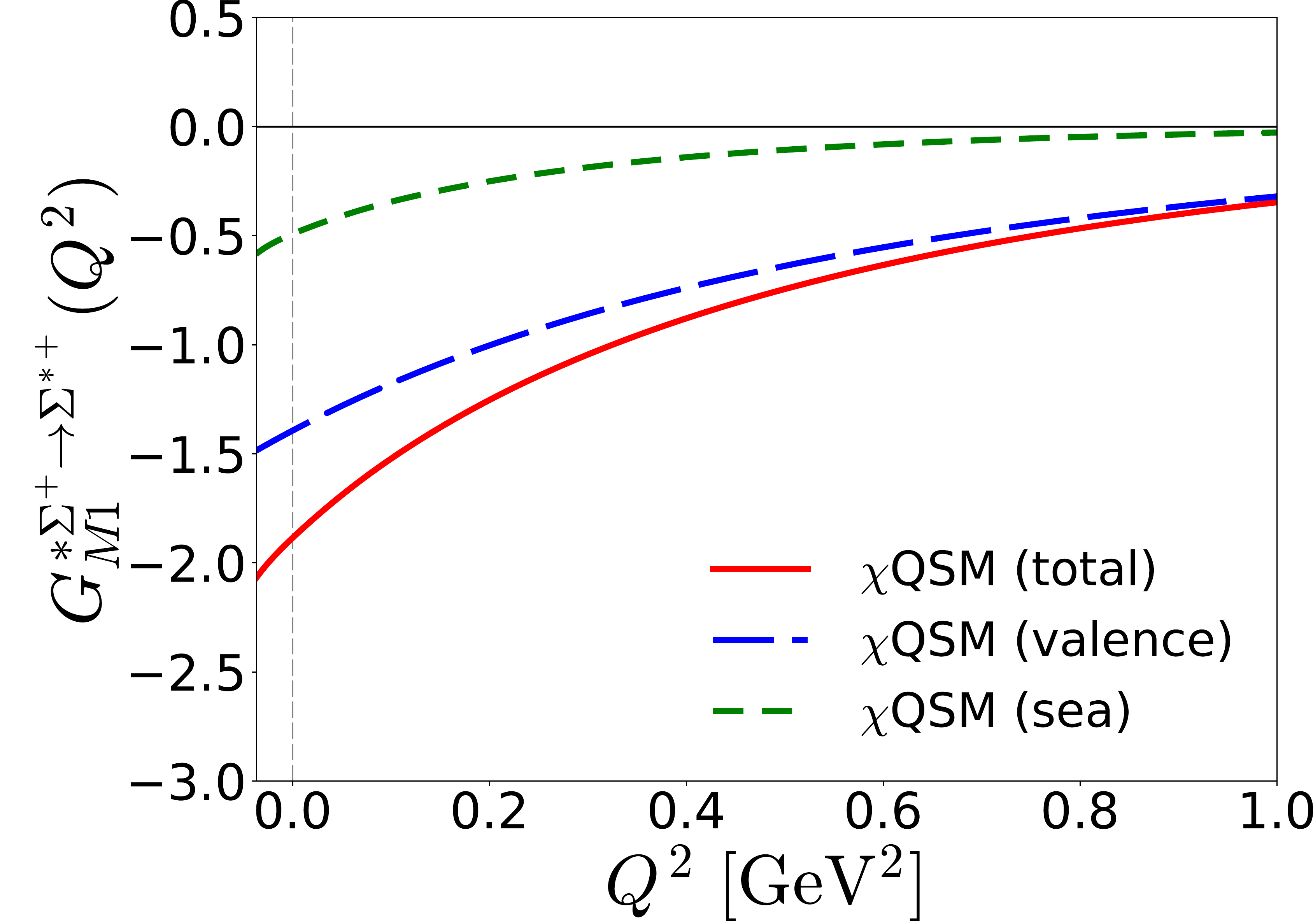}
\includegraphics[scale=0.234]{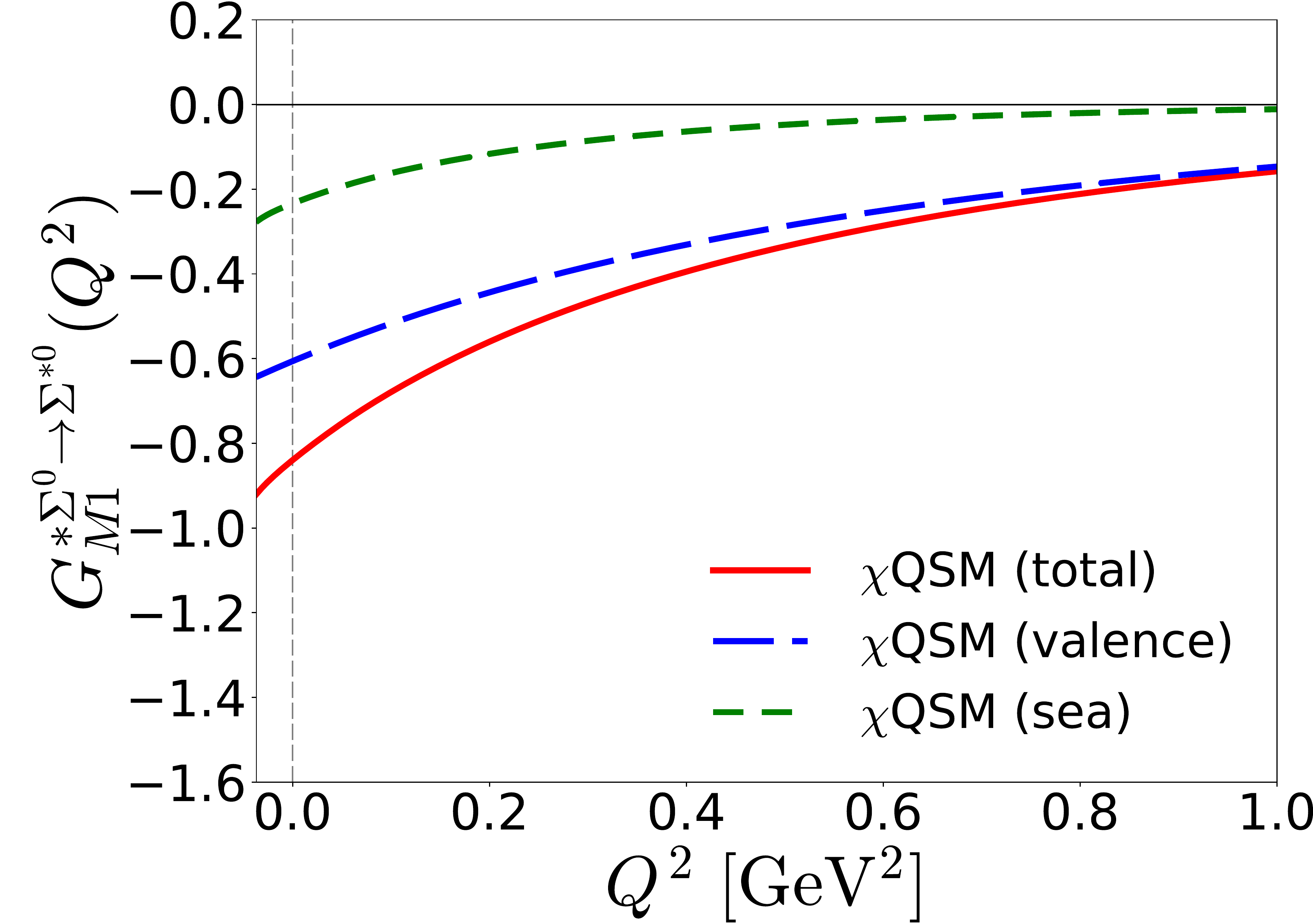}
\includegraphics[scale=0.234]{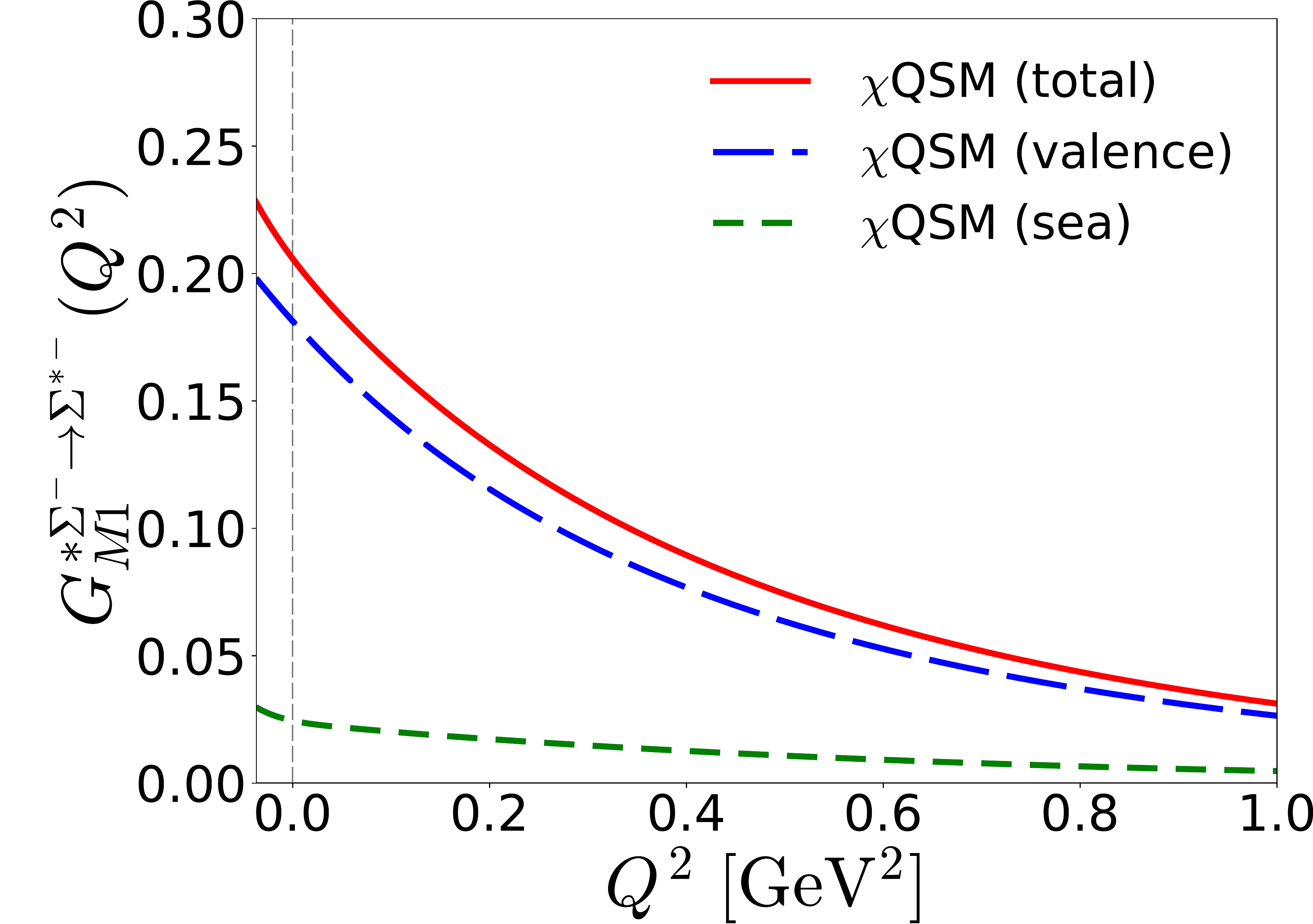}
\includegraphics[scale=0.234]{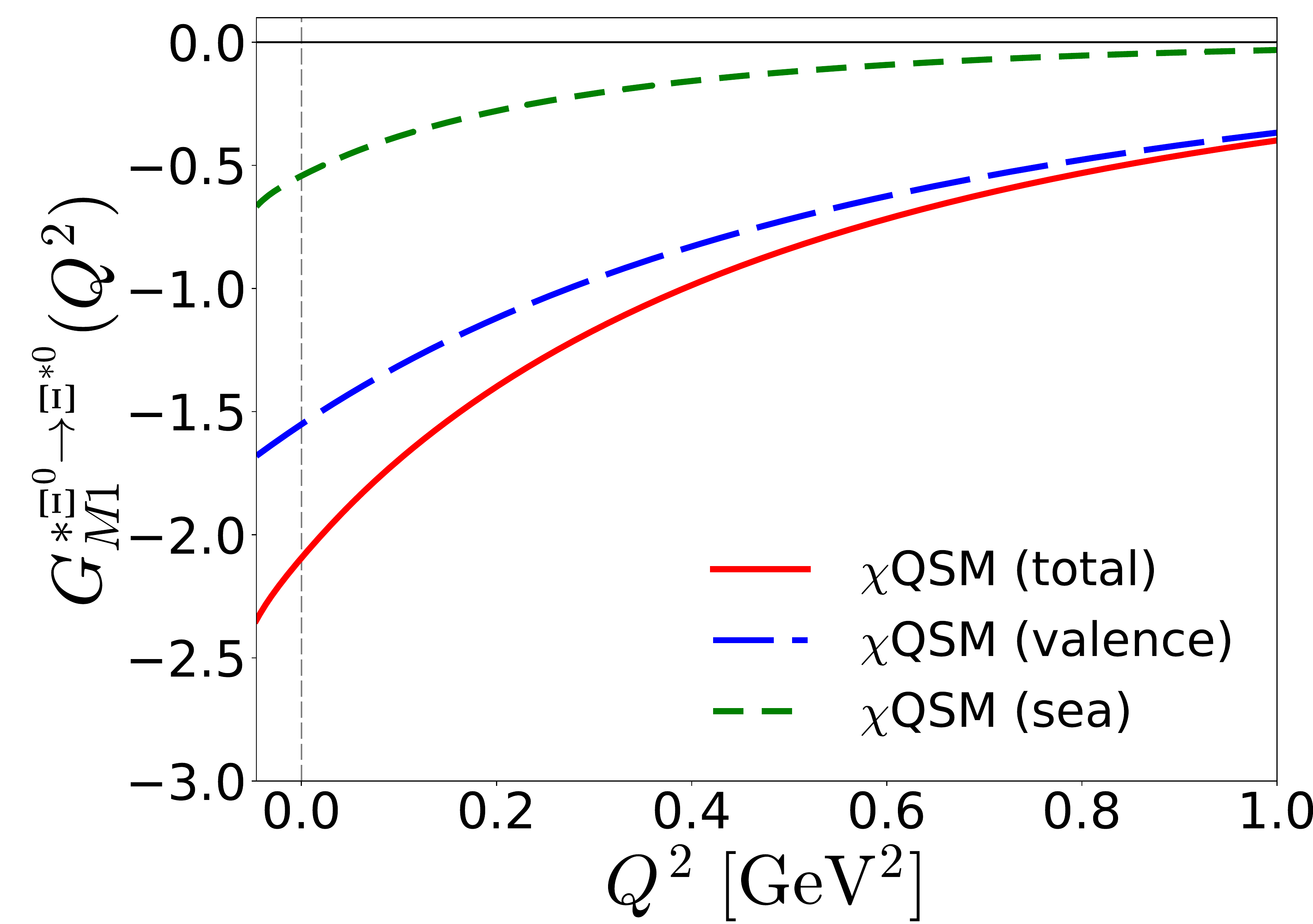}
\includegraphics[scale=0.234]{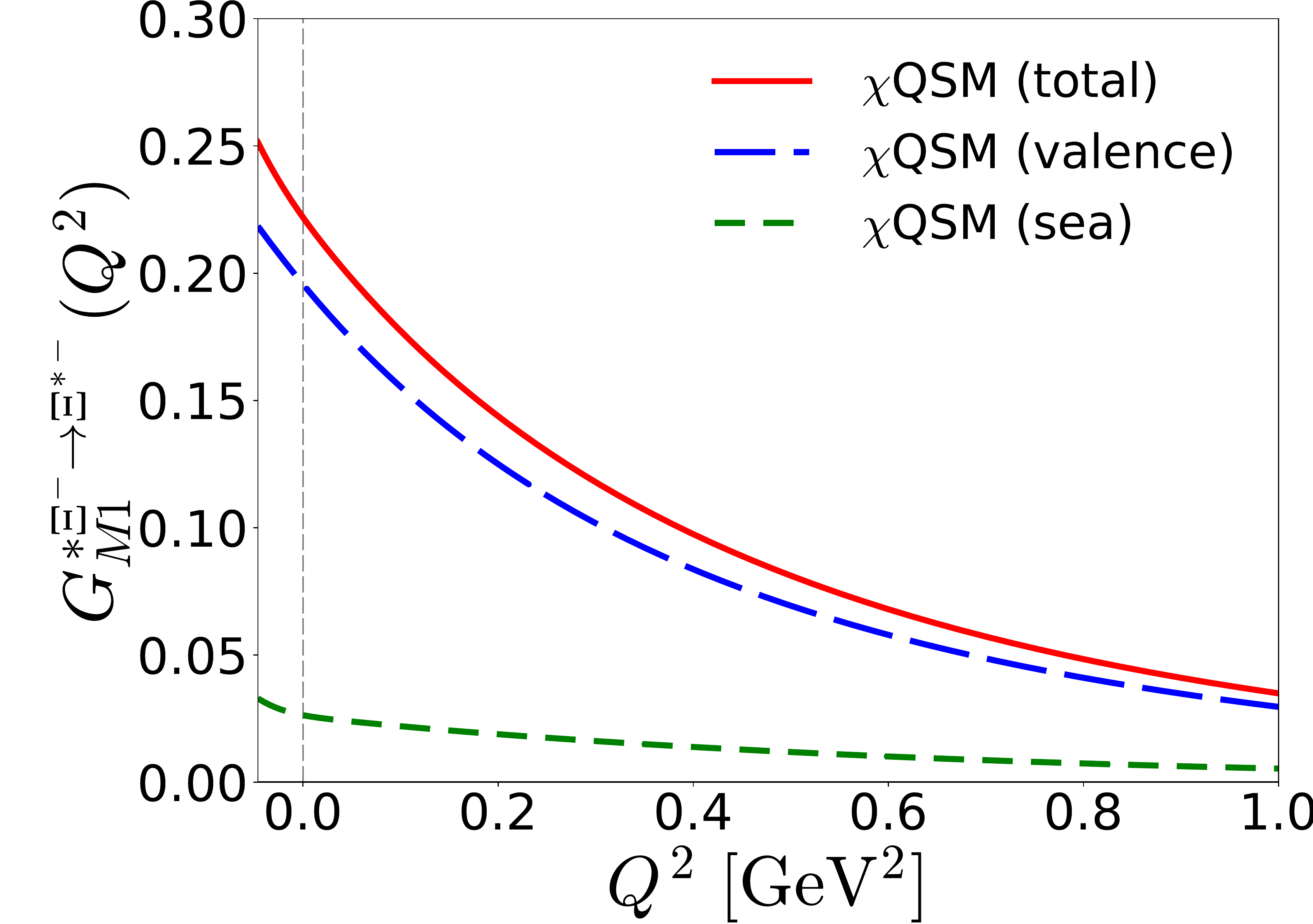}
\includegraphics[scale=0.234]{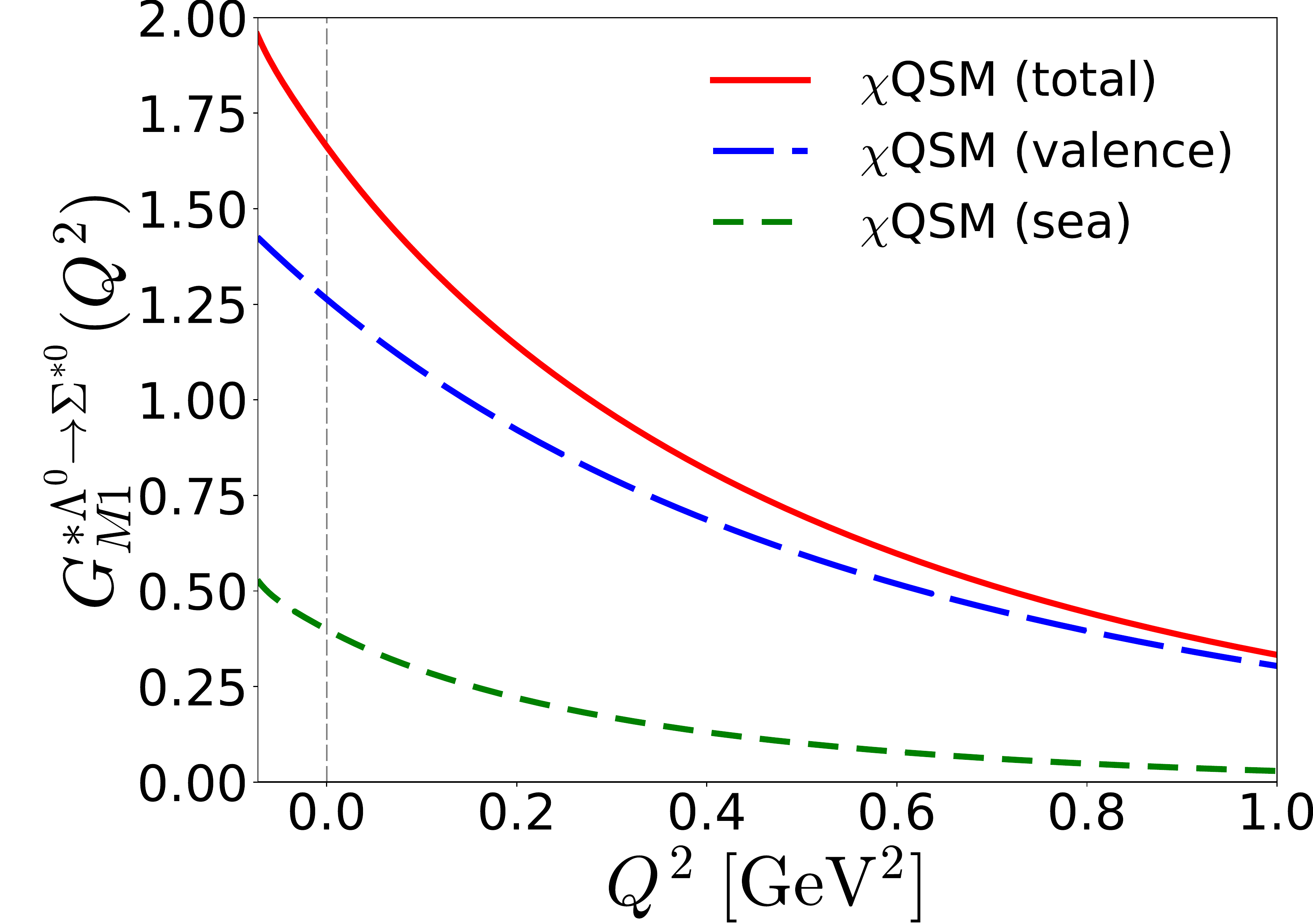}
\caption{Results for the magnetic dipole transition form factors of all
  the other members of the baryon decuplet with the valence and
  sea-quark contributions separated. The long-dashed curves draw the
  valence-quark (level-quark) contributions whereas the short-dashed ones depict
  the sea-quark (continuum) contributions. The solid curves show the
  total results.} 
\label{fig:6}
\end{figure}
\begin{figure}[htp]
\includegraphics[scale=0.234]{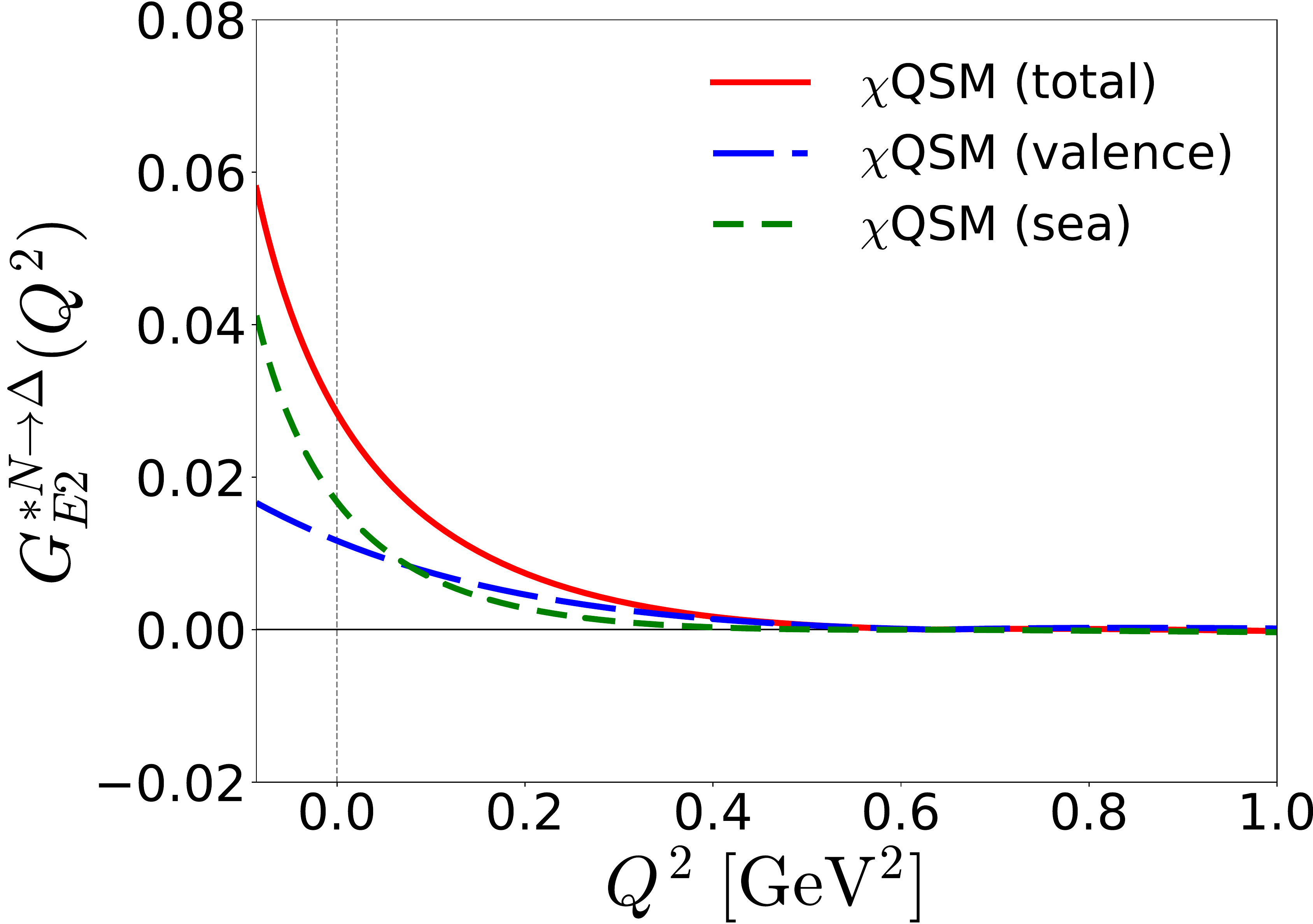}
\includegraphics[scale=0.234]{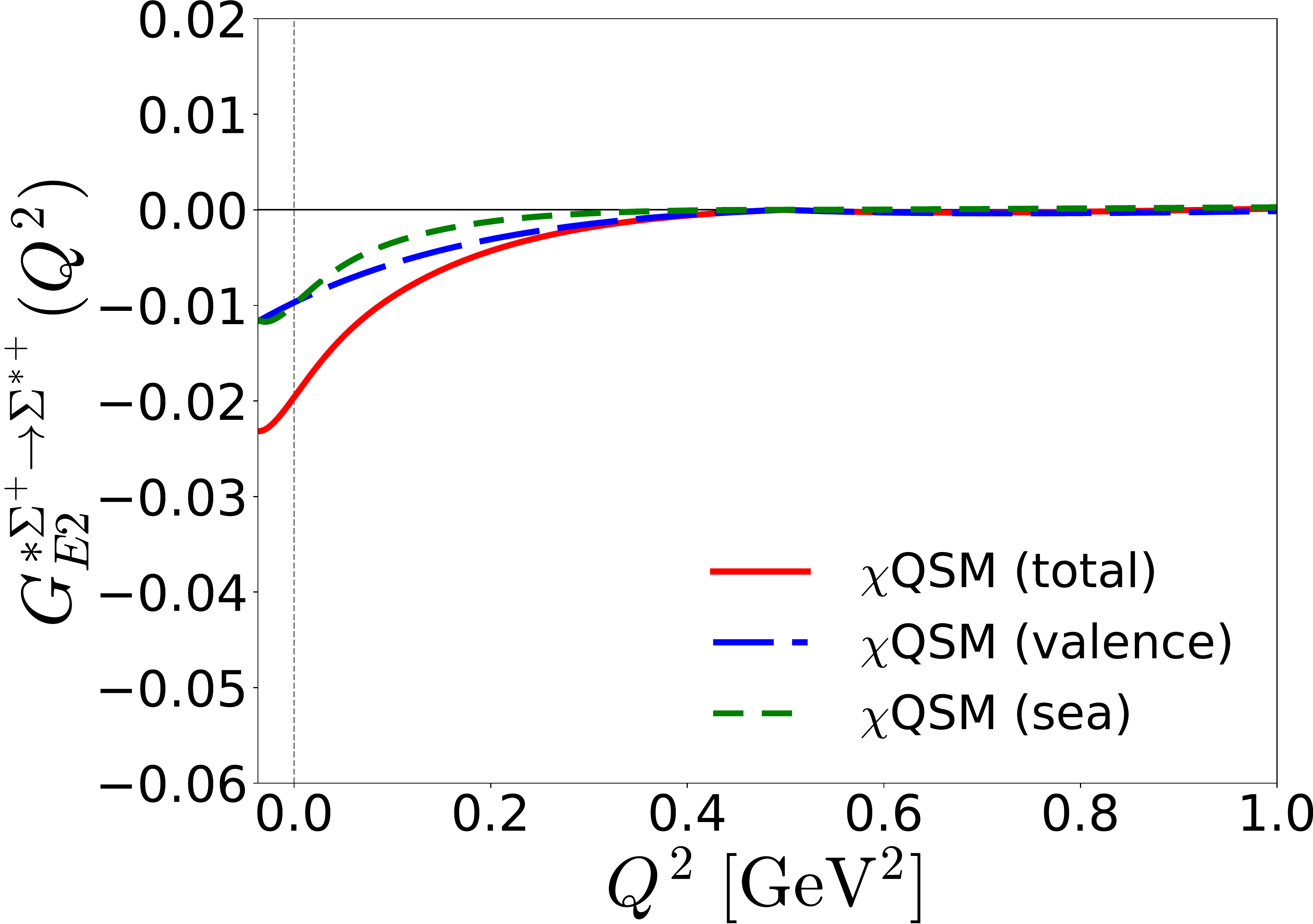}
\includegraphics[scale=0.234]{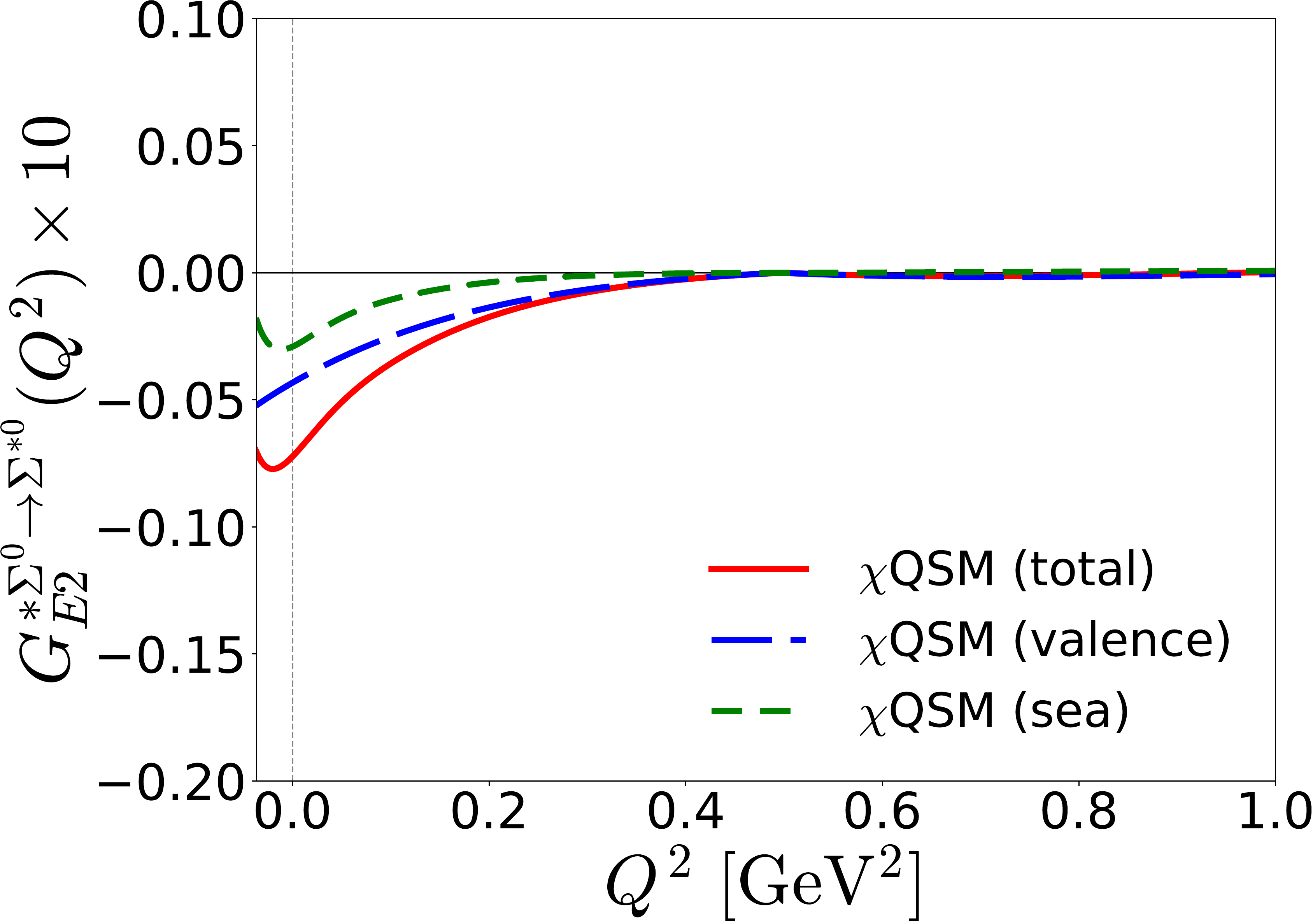}
\includegraphics[scale=0.234]{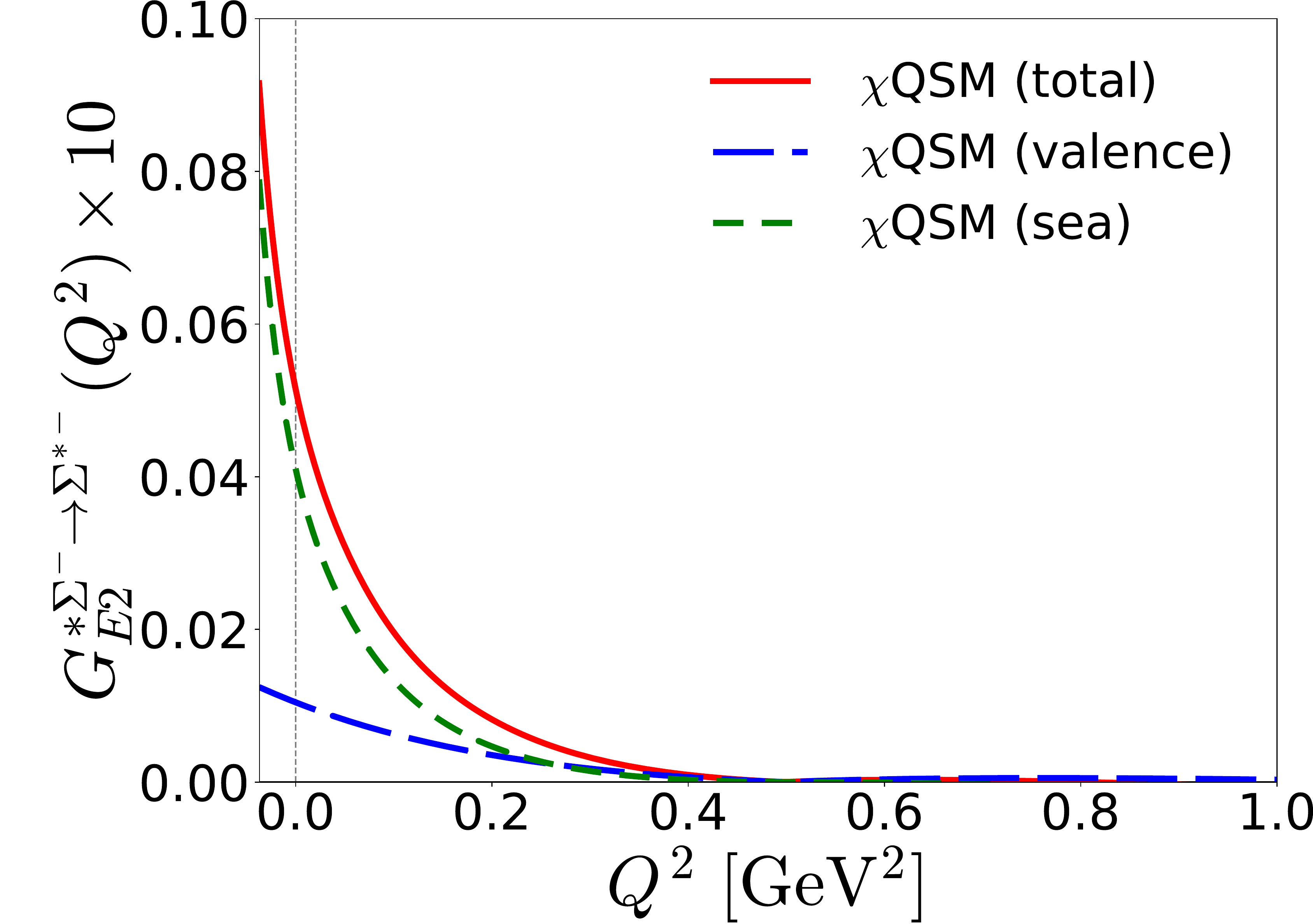}
\includegraphics[scale=0.234]{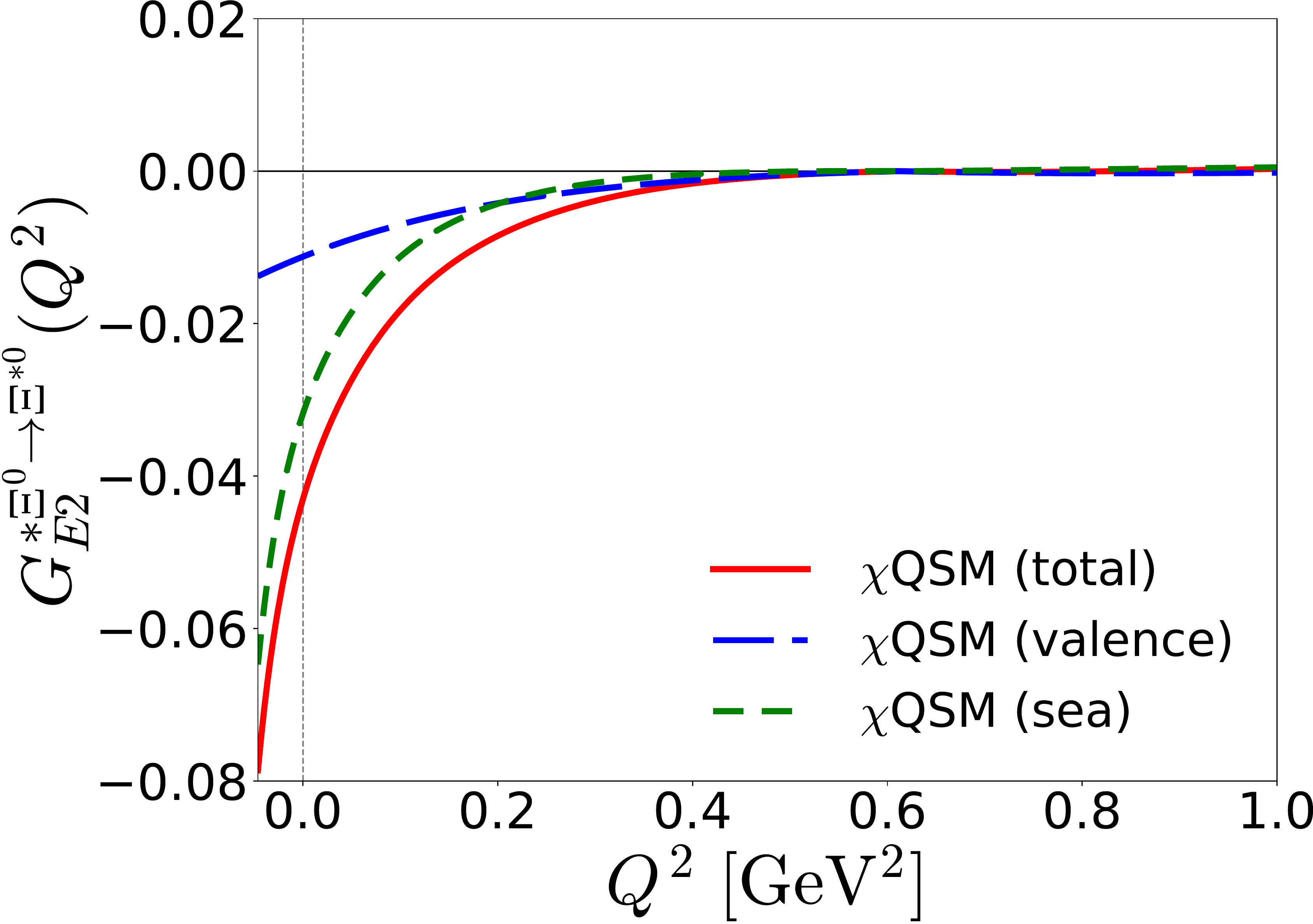}
\includegraphics[scale=0.234]{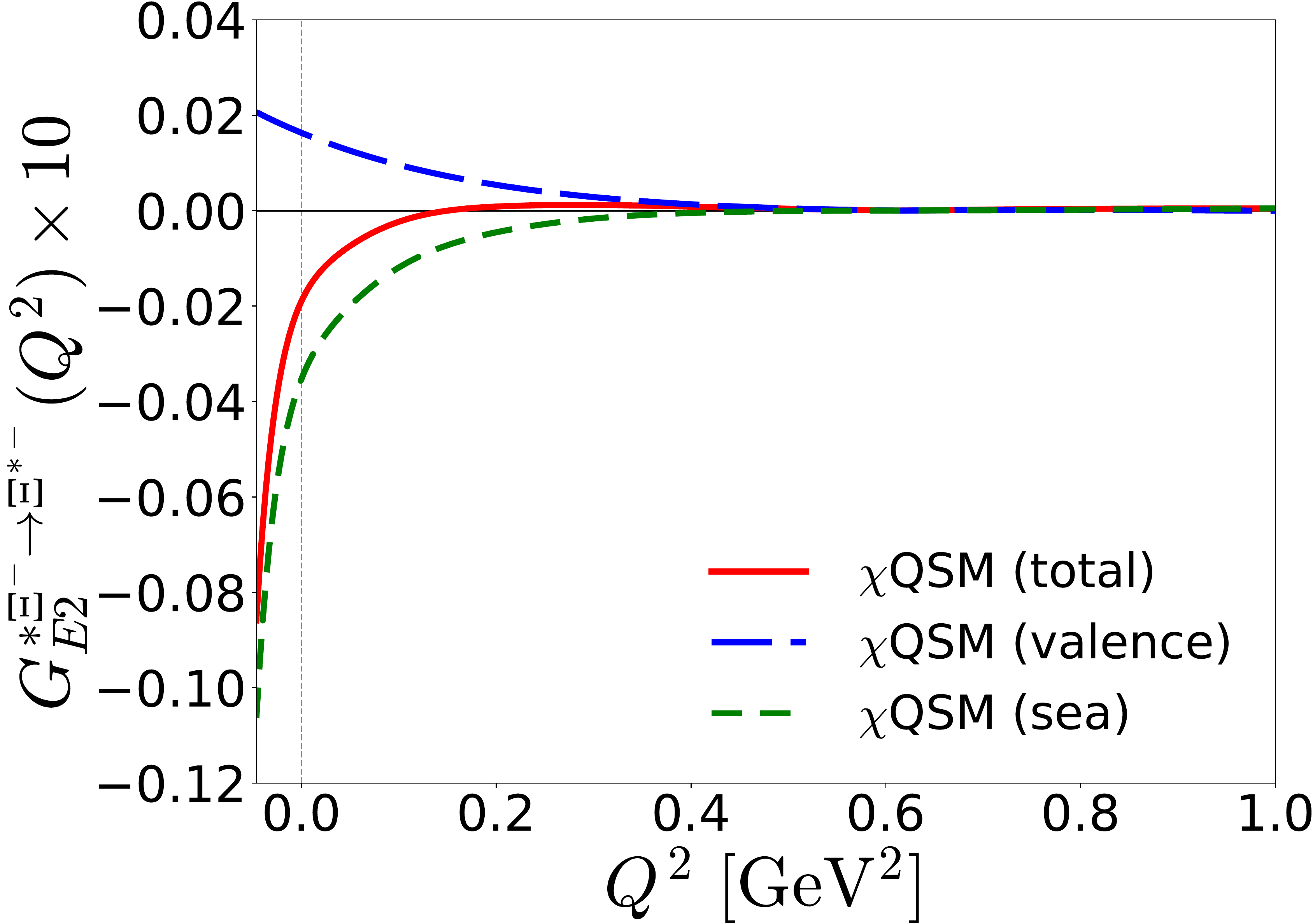}
\includegraphics[scale=0.234]{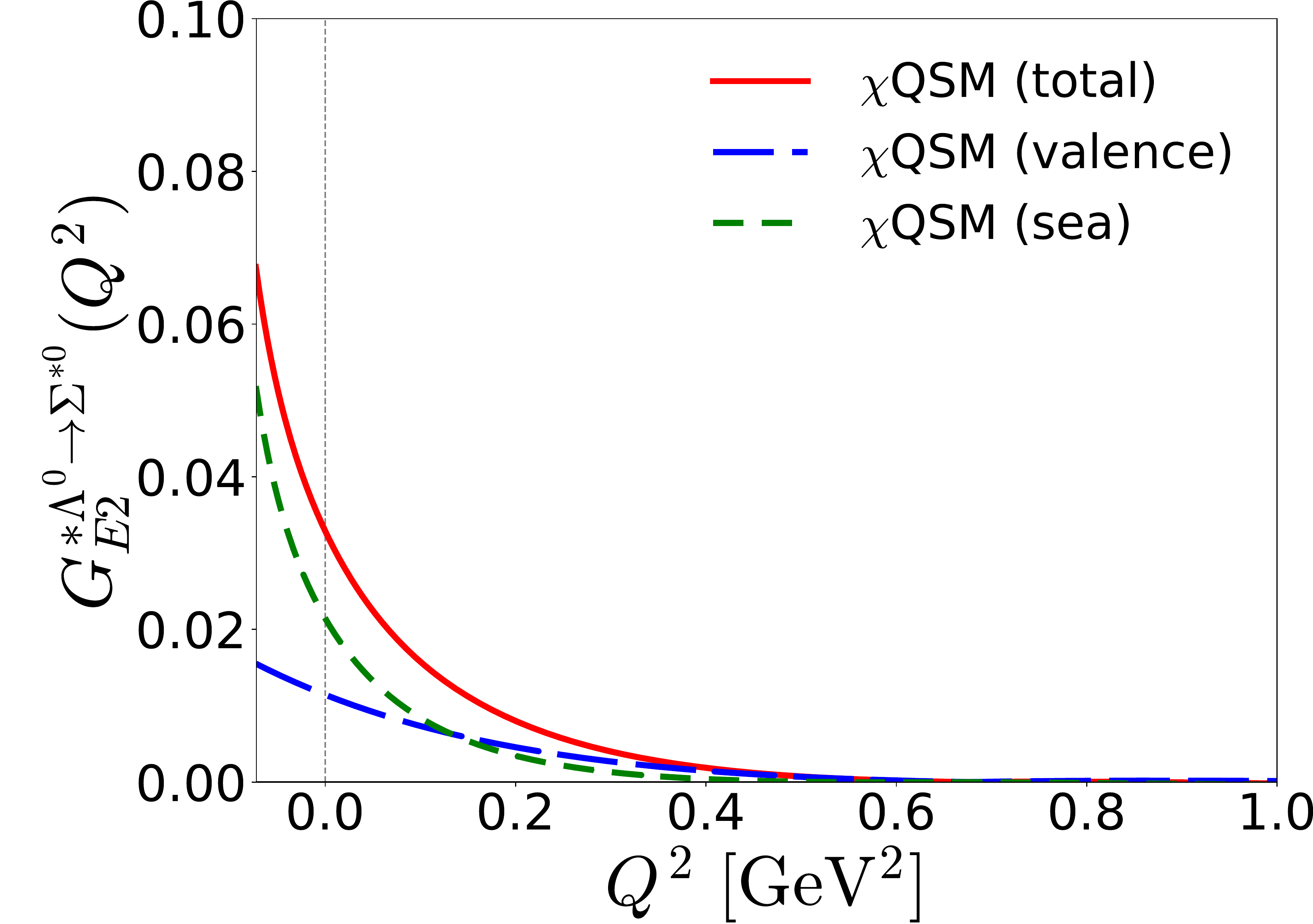}
\caption{Results for the electric quadrupole transition form factors
  of all the other members of the baryon decuplet with the valence and 
  sea-quark contributions separated. The long-dashed curves draw the
  valence-quark (level-quark) contributions whereas the short-dashed
  ones depict the sea-quark (continuum) contributions. The solid
  curves show the total results.}
\label{fig:7}
\end{figure}
\begin{figure}[htp]
\includegraphics[scale=0.234]{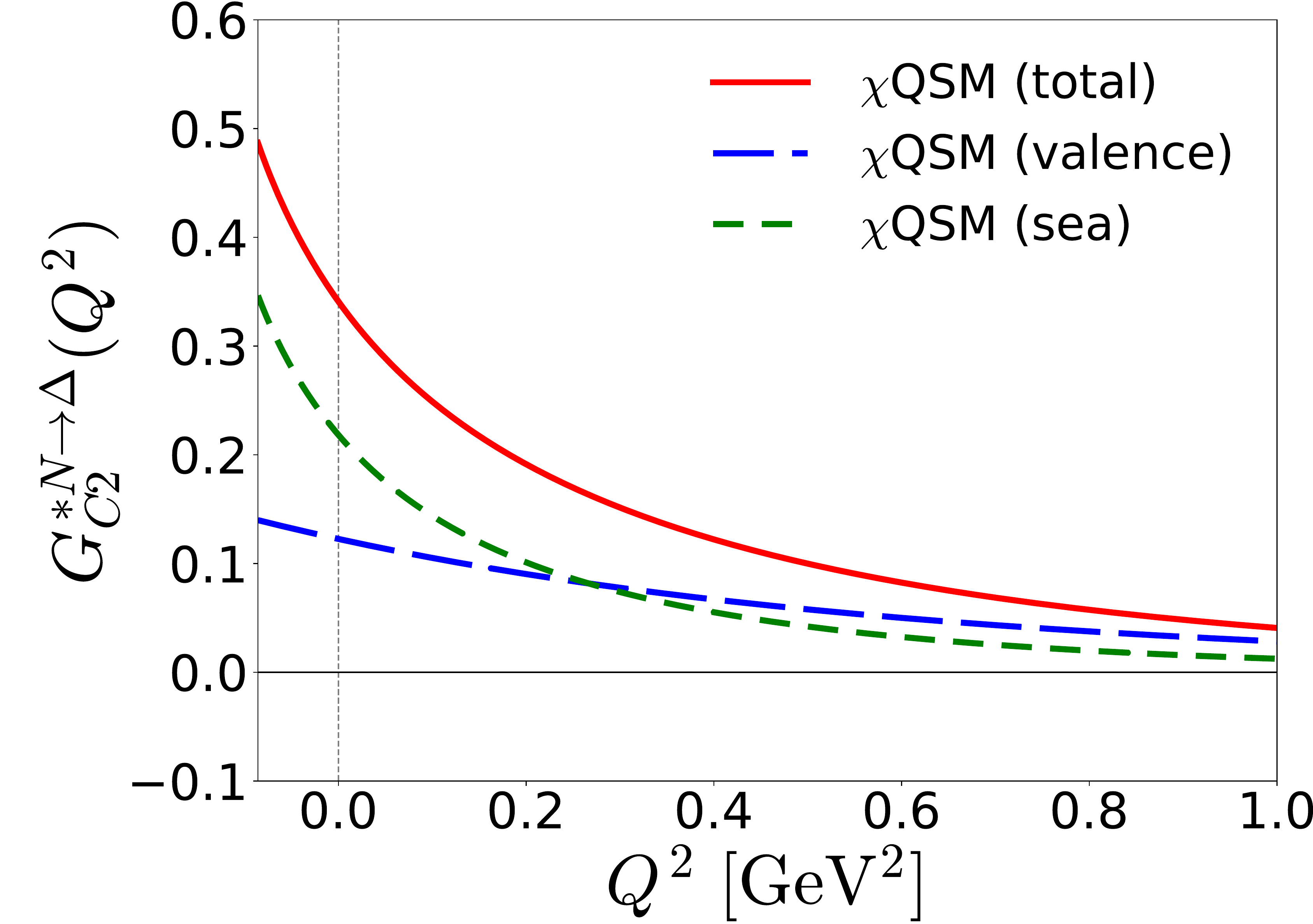}
\includegraphics[scale=0.234]{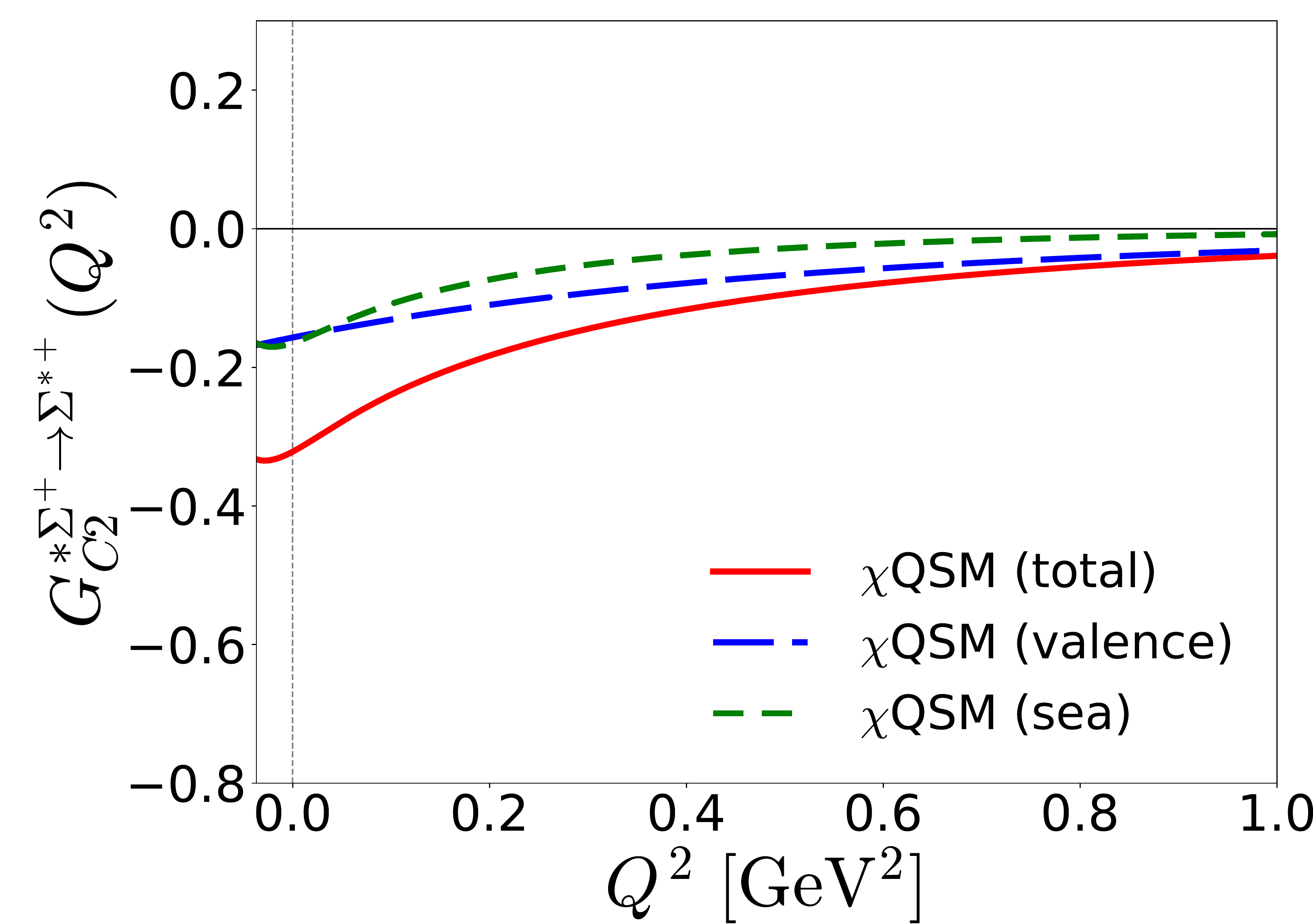}
\includegraphics[scale=0.234]{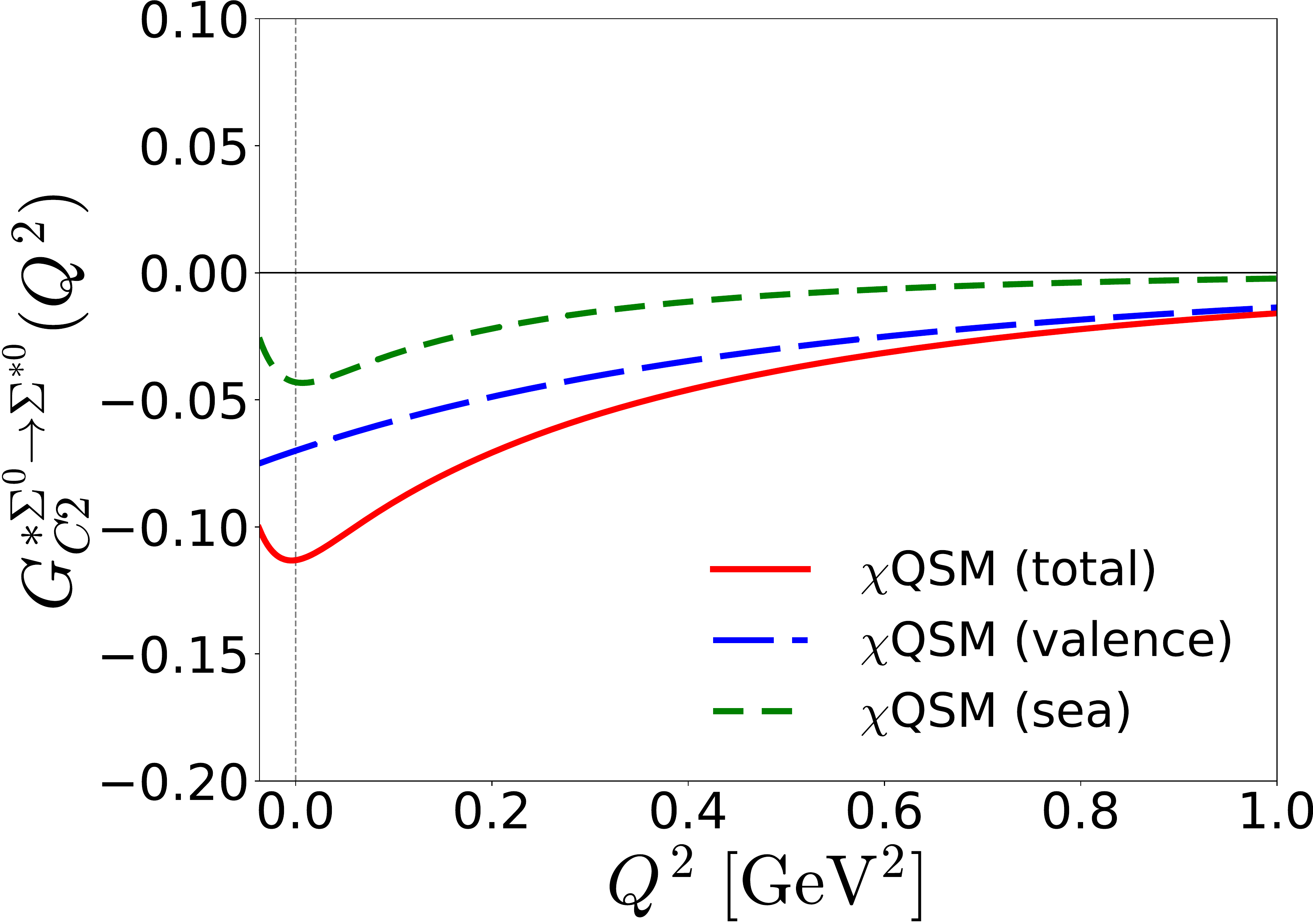}
\includegraphics[scale=0.234]{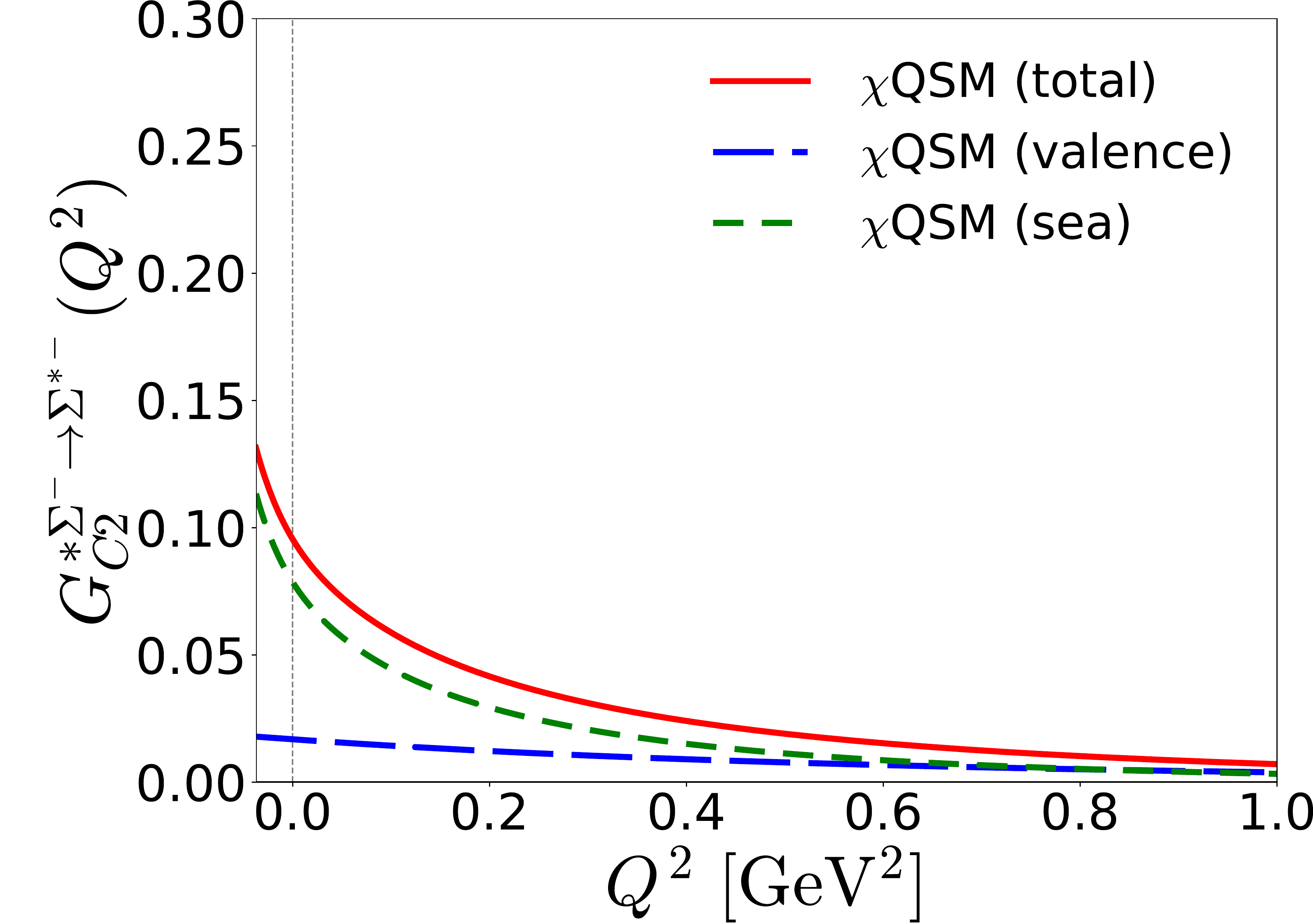}
\includegraphics[scale=0.234]{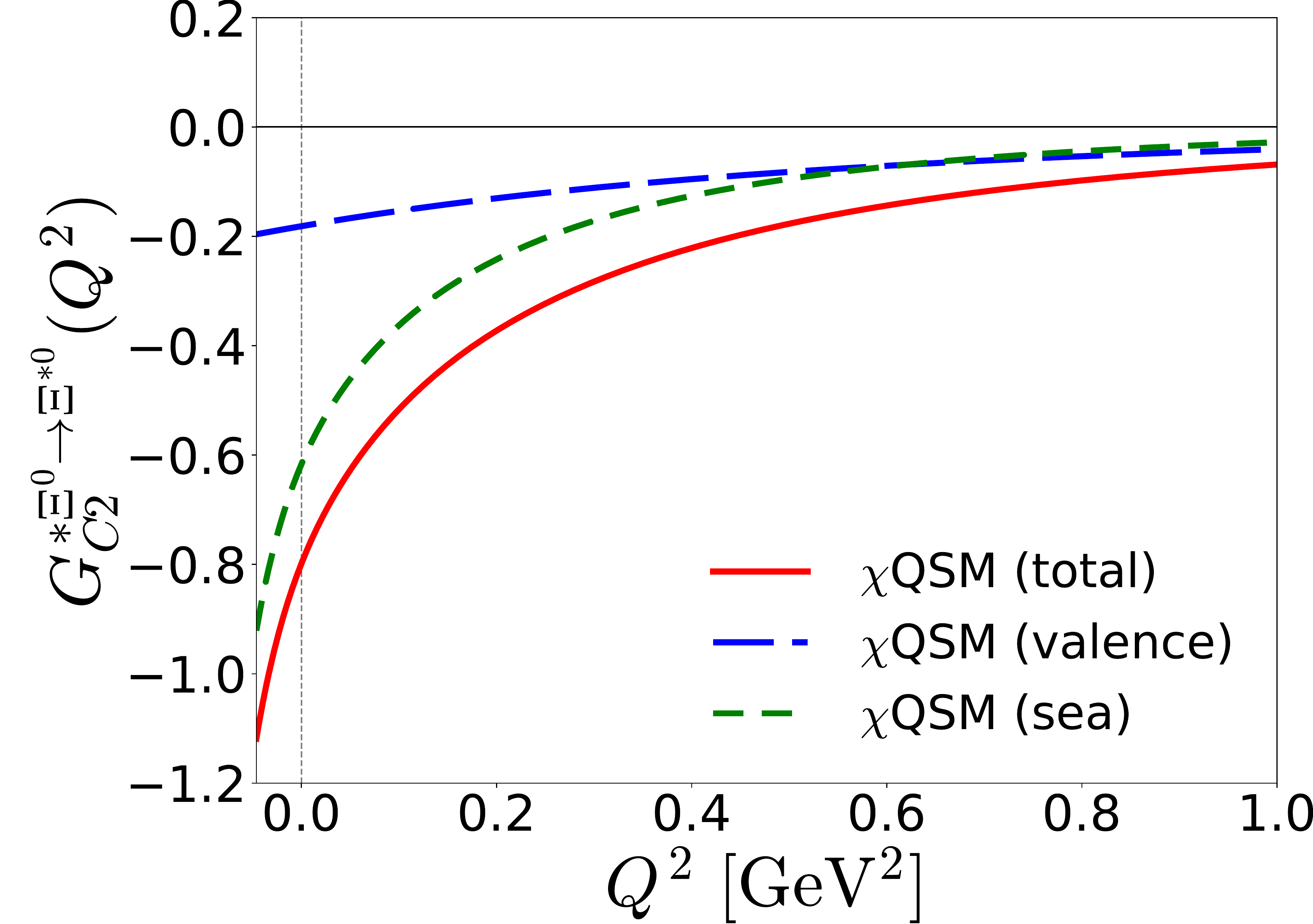}
\includegraphics[scale=0.234]{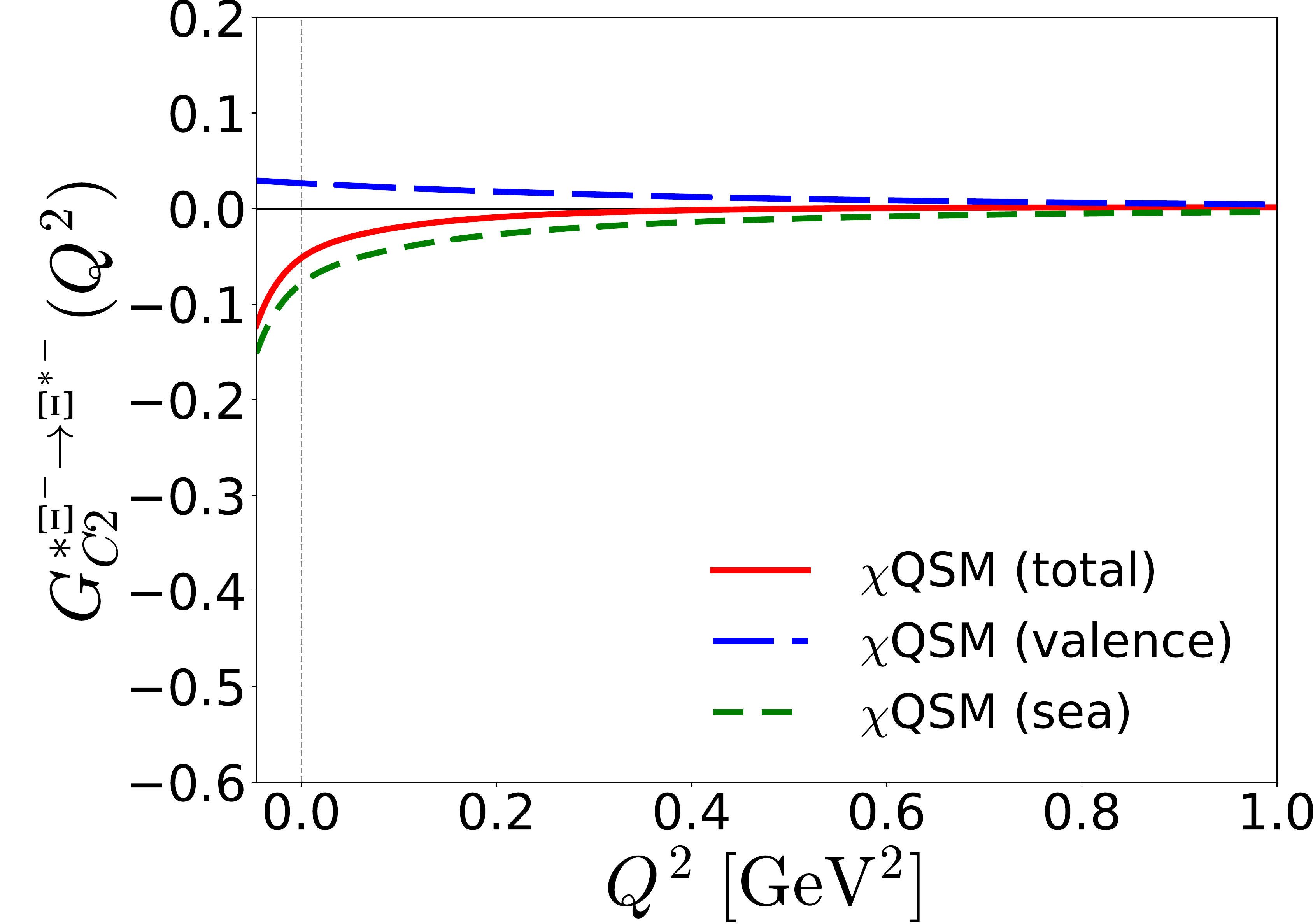}
\includegraphics[scale=0.234]{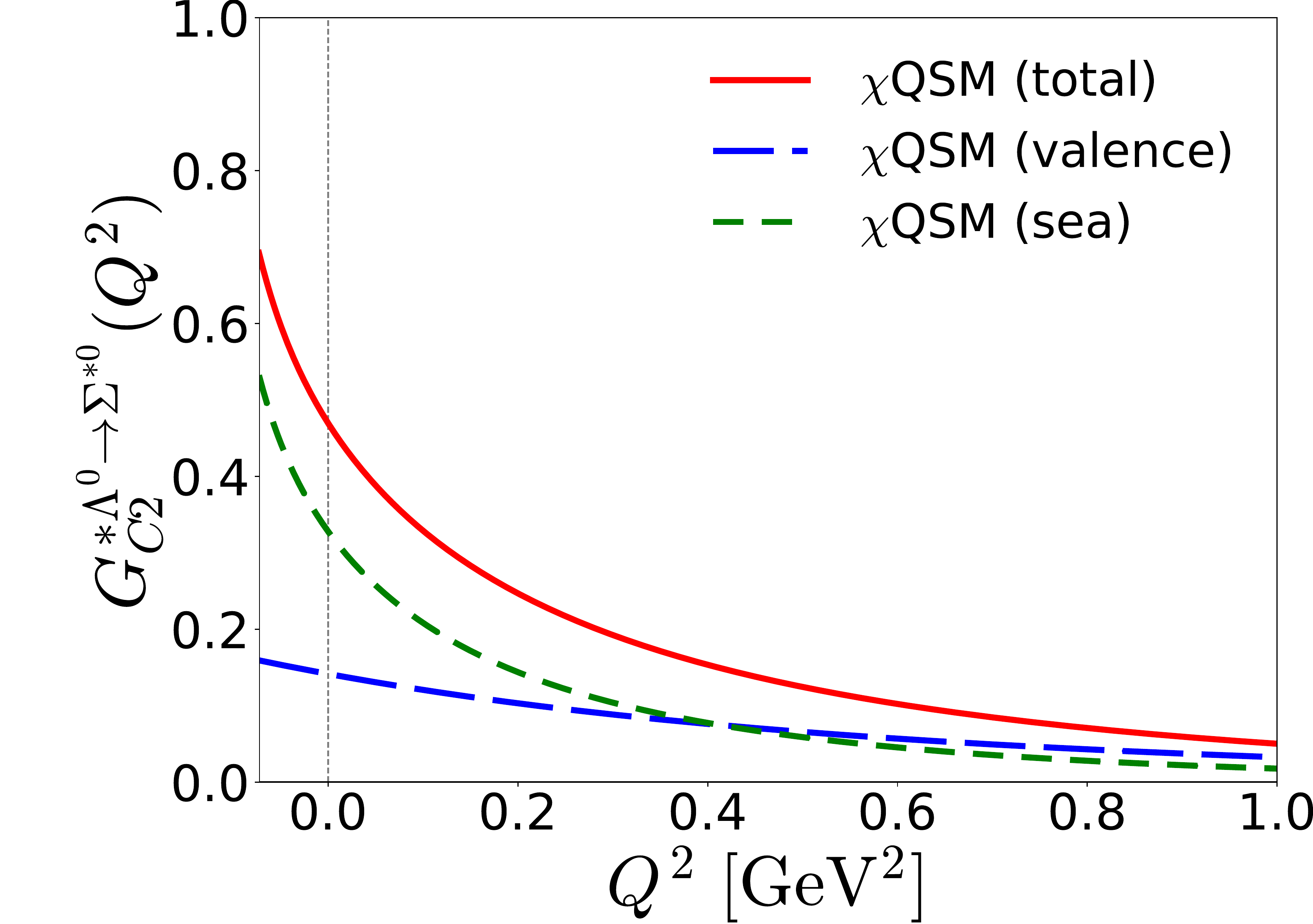}
\caption{Results for the Coulomb quadrupole transition form factors
  of all the other members of the baryon decuplet with the valence and 
  sea-quark contributions separated. The long-dashed curves draw the
  valence-quark (level-quark) contributions whereas the short-dashed
  ones depict the sea-quark (continuum) contributions. The solid
  curves show the total results.}
\label{fig:8}
\end{figure}
As mentioned previously, it has been known that the effects of the
pion clouds play an essential role in describing the structure of the
$\Delta$ isobar~\cite{Pascalutsa:2005ts}. In particular, they
contribute noticeably to the $E2$ form factor of $\Delta$. Thus, it is
of great importance to consider the pion clouds in explaining a
deformed shape of the $\Delta$. It was already shown in the $\chi$QSM
that the $E2$ form factor of the $\Delta$ is mainly governed by the
contributions to from pion clouds or the polarized sea quarks in lower
$Q^2$ regions. This implies that the effects of the polarized sea
quarks will be also dominant on the $E2$ transition form factors of
the baryon decuplet. In Figs.~\ref{fig:6}, \ref{fig:7}, and
\ref{fig:7}, we draw the results for the $M1$, $E2$, and $C2$
transition form factors of the baryon decuplet with the valence- and
sea-quark contributions separated. As shown in Fig.~\ref{fig:6}, the
sea-quark effects  contribute to the $M1$ transition form factors by
about $(20-30)\,\%$. However, when it comes to the $E2$ and $C2$
transition form factors, the sea-quark effects 
are dominant over those of the valence quarks in lower $Q^2$ regions
for all possible radiative transitions. The sea-quark contributions
fall off in general drastically as $Q^2$ increases, compared with
those of the valence quarks. So, the valence-quark contributions take
over those of the sea quarks in higher $Q^2$ regions. 
Thus, we conclude that the sea-quark contributions or the effects of
the pion clouds play indeed a major role in explaining how the
decuplet baryons are deformed.  

\begin{figure}[htp]
\centering
\includegraphics[scale=0.234]{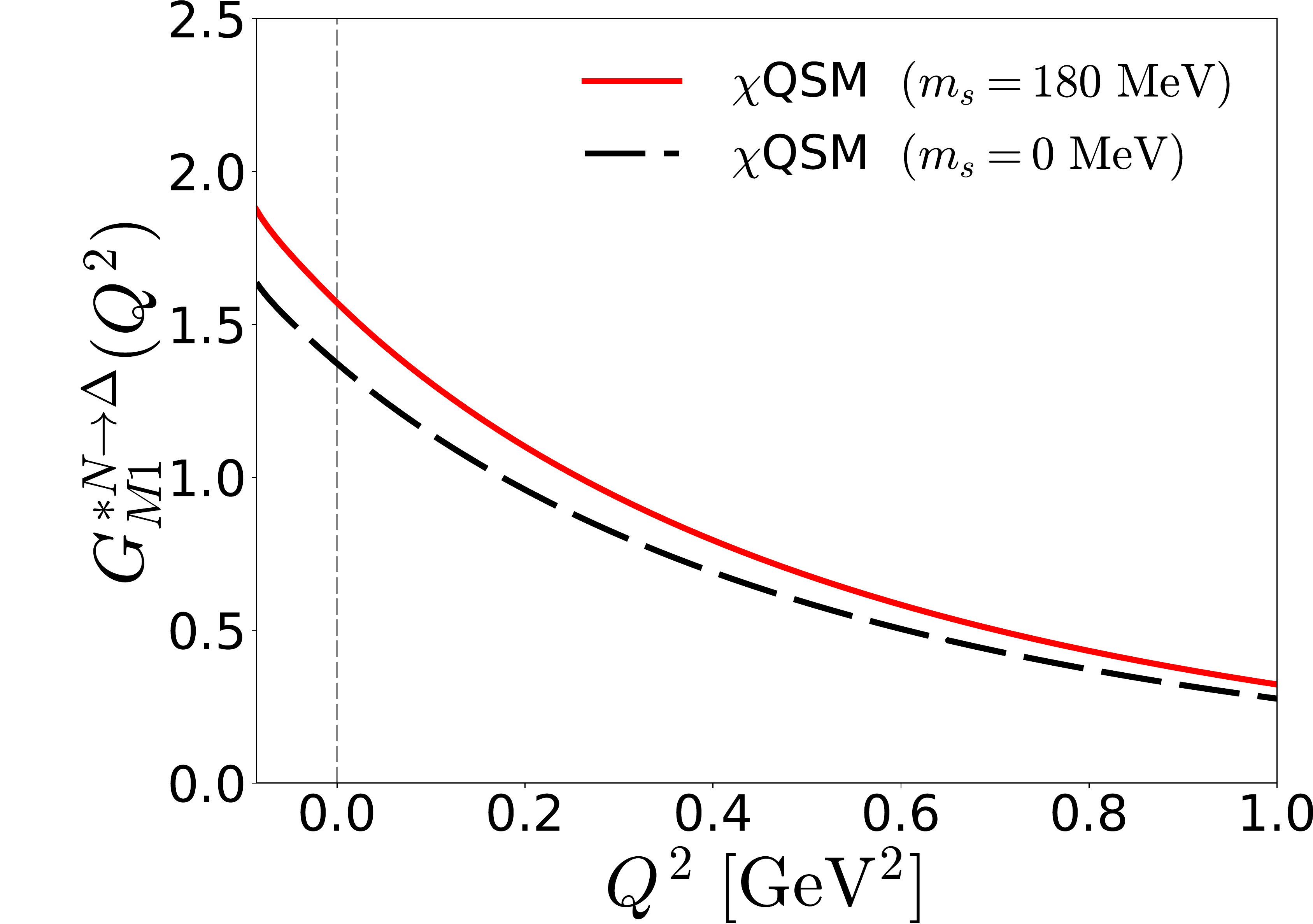}
\includegraphics[scale=0.234]{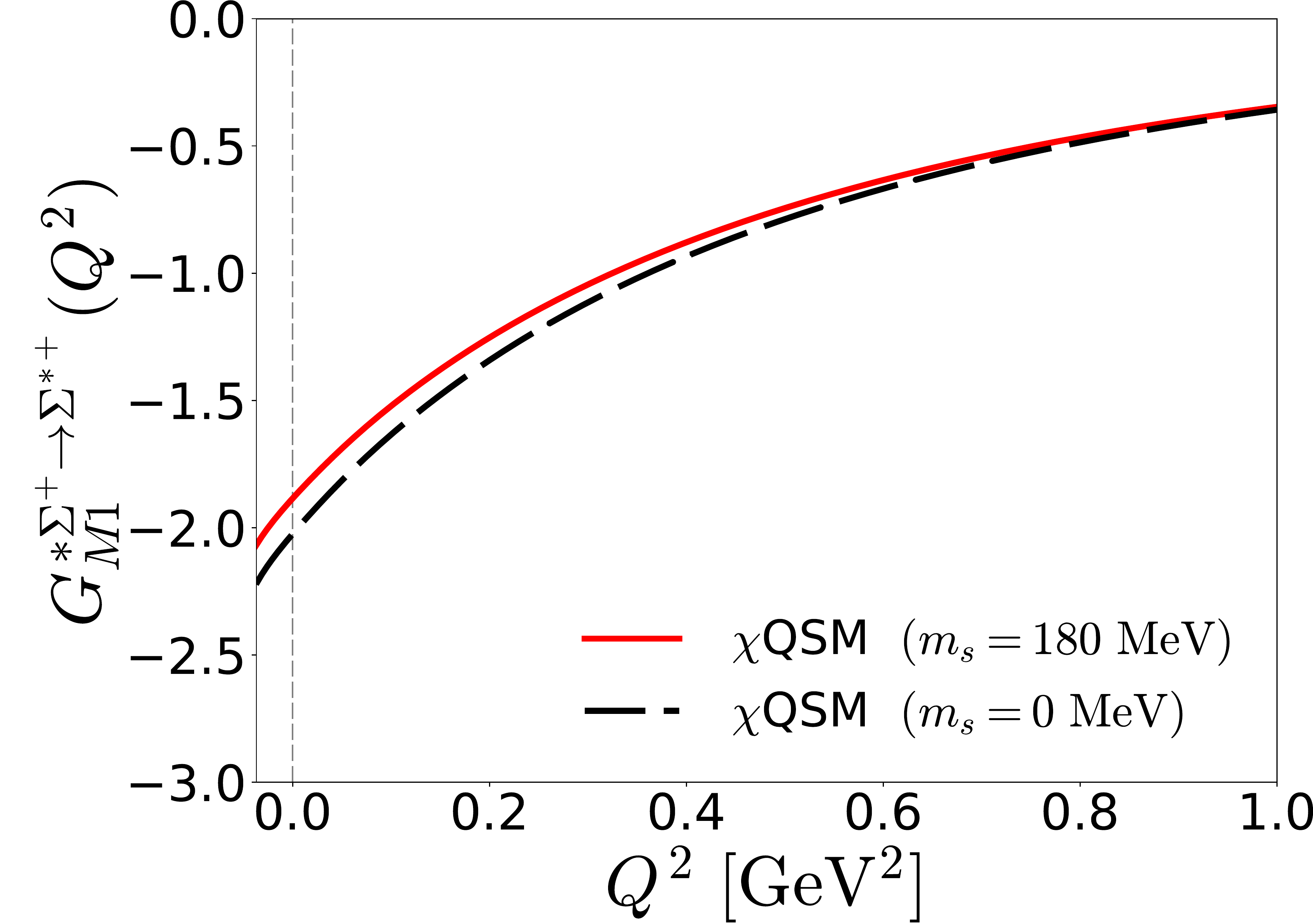}
\includegraphics[scale=0.234]{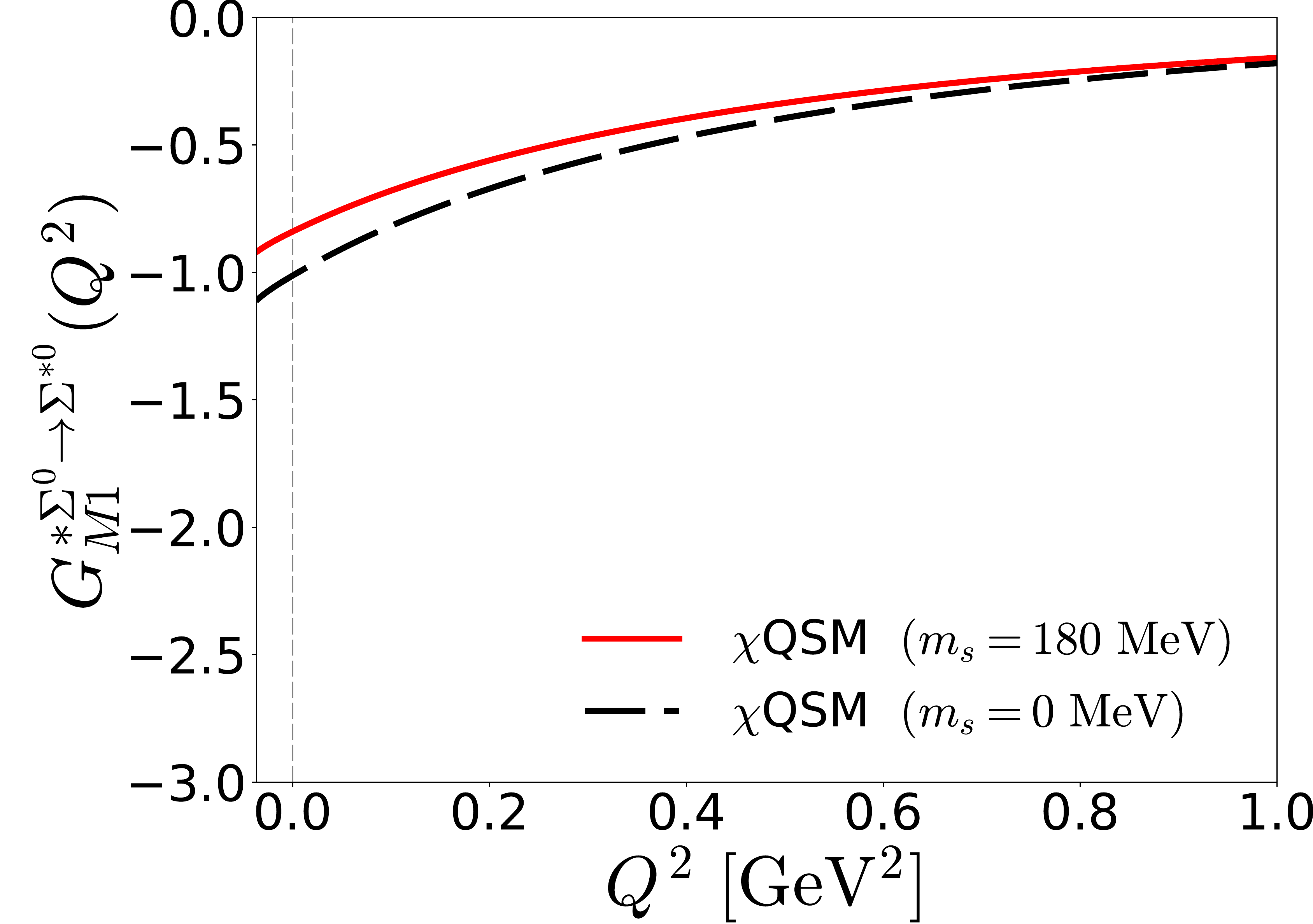}
\includegraphics[scale=0.234]{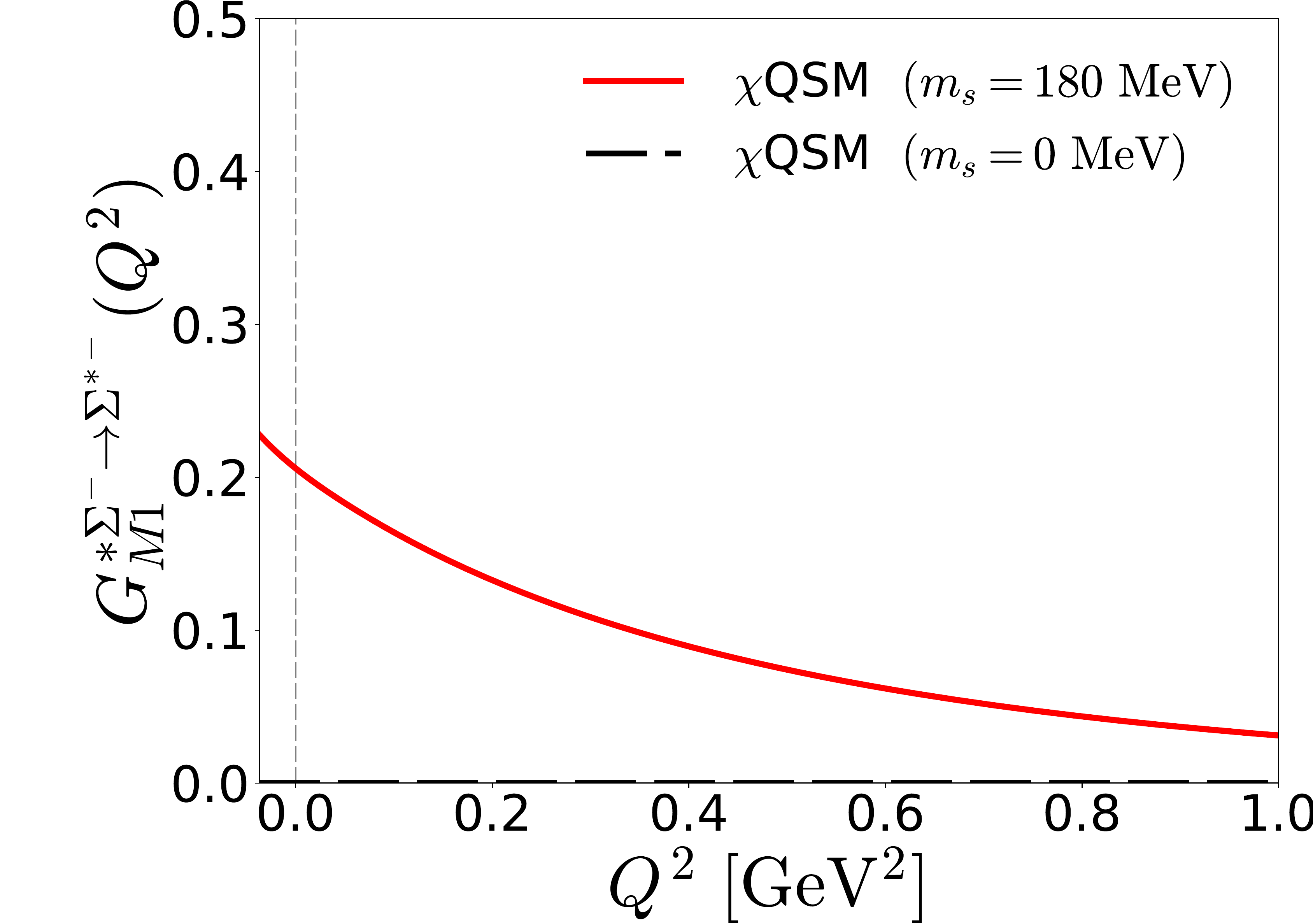}
\includegraphics[scale=0.234]{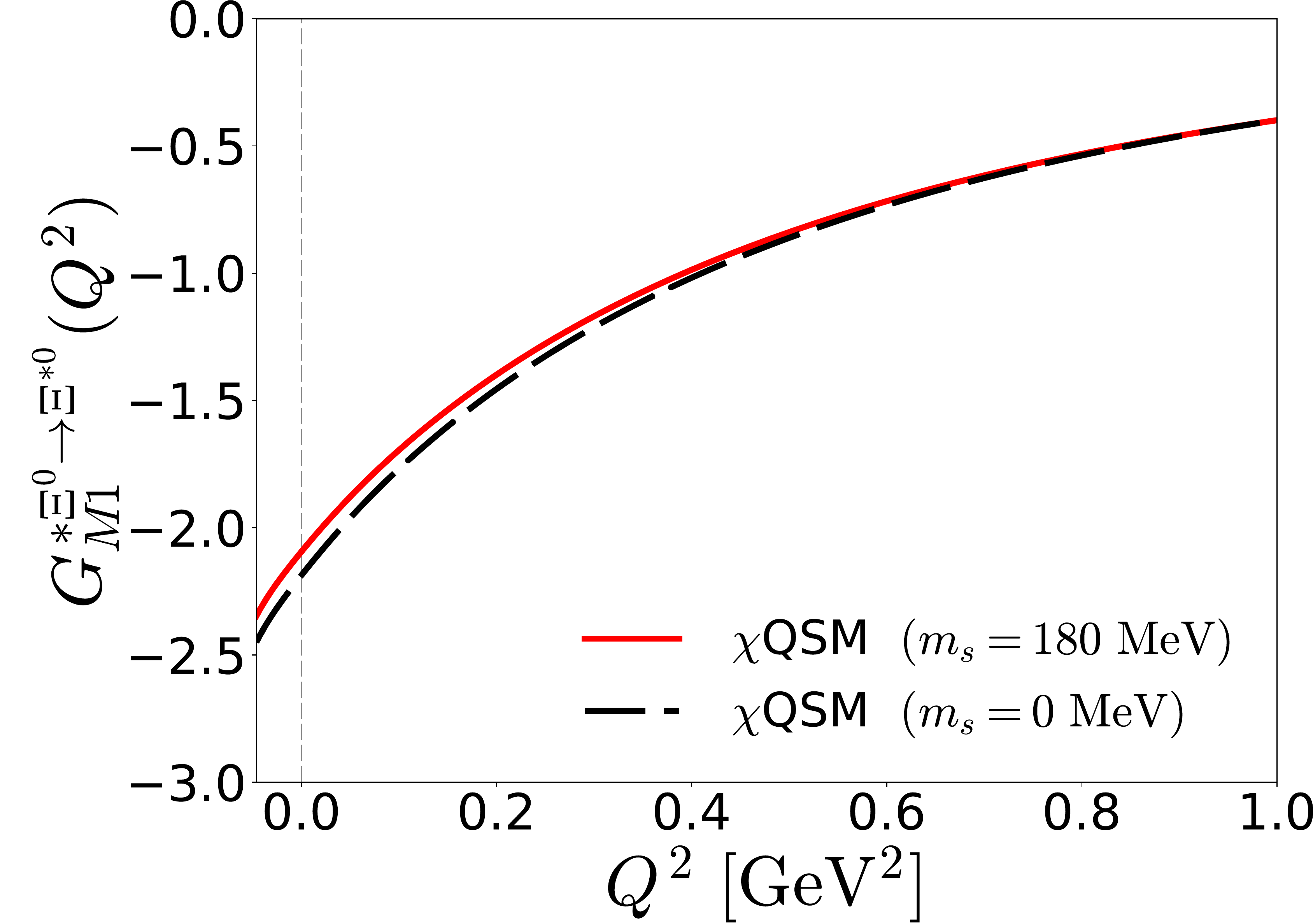}
\includegraphics[scale=0.234]{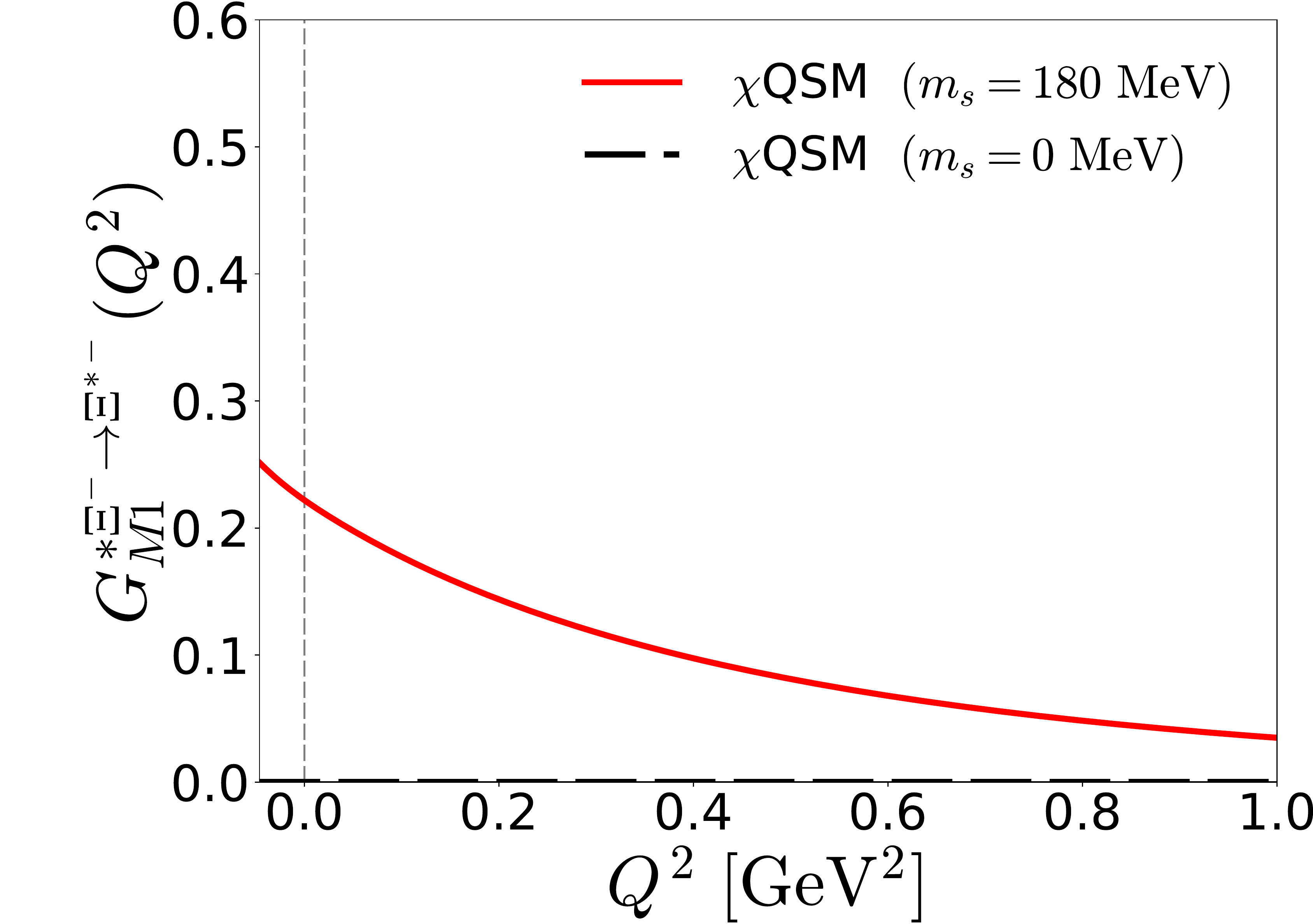}
\includegraphics[scale=0.234]{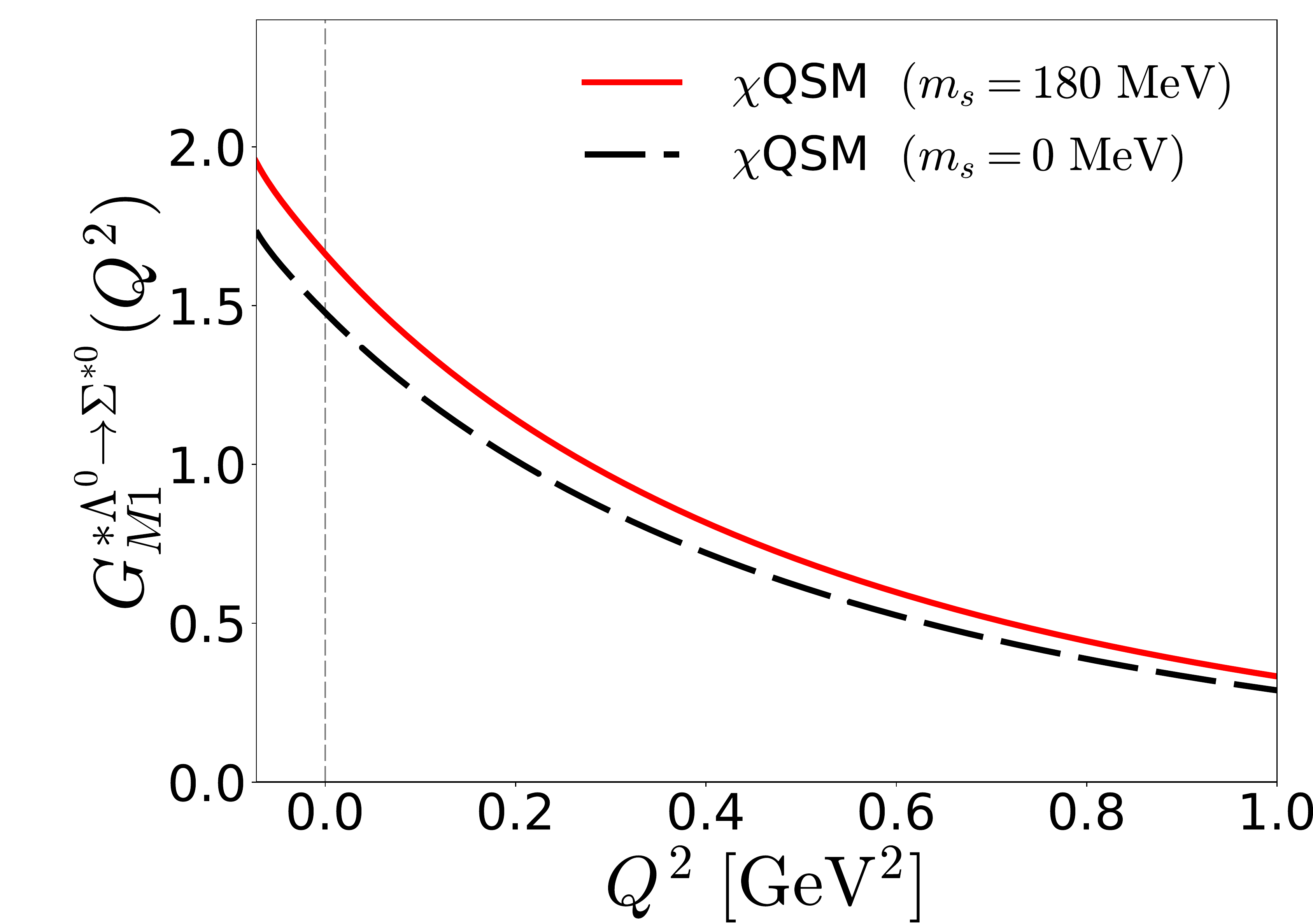}
\caption{Results for the $M1$ transition form factors of all
  the other members of the baryon decuplet with
  and without the effects of flavor SU(3) symmetry breaking. The solid
curves depict the results with $m_{\mathrm{s}}=180$ MeV, whereas the
dashed ones draw those in exact flavor SU(3) symmetry.} 
\label{fig:9}
\end{figure}
\begin{figure}[htp]
\includegraphics[scale=0.234]{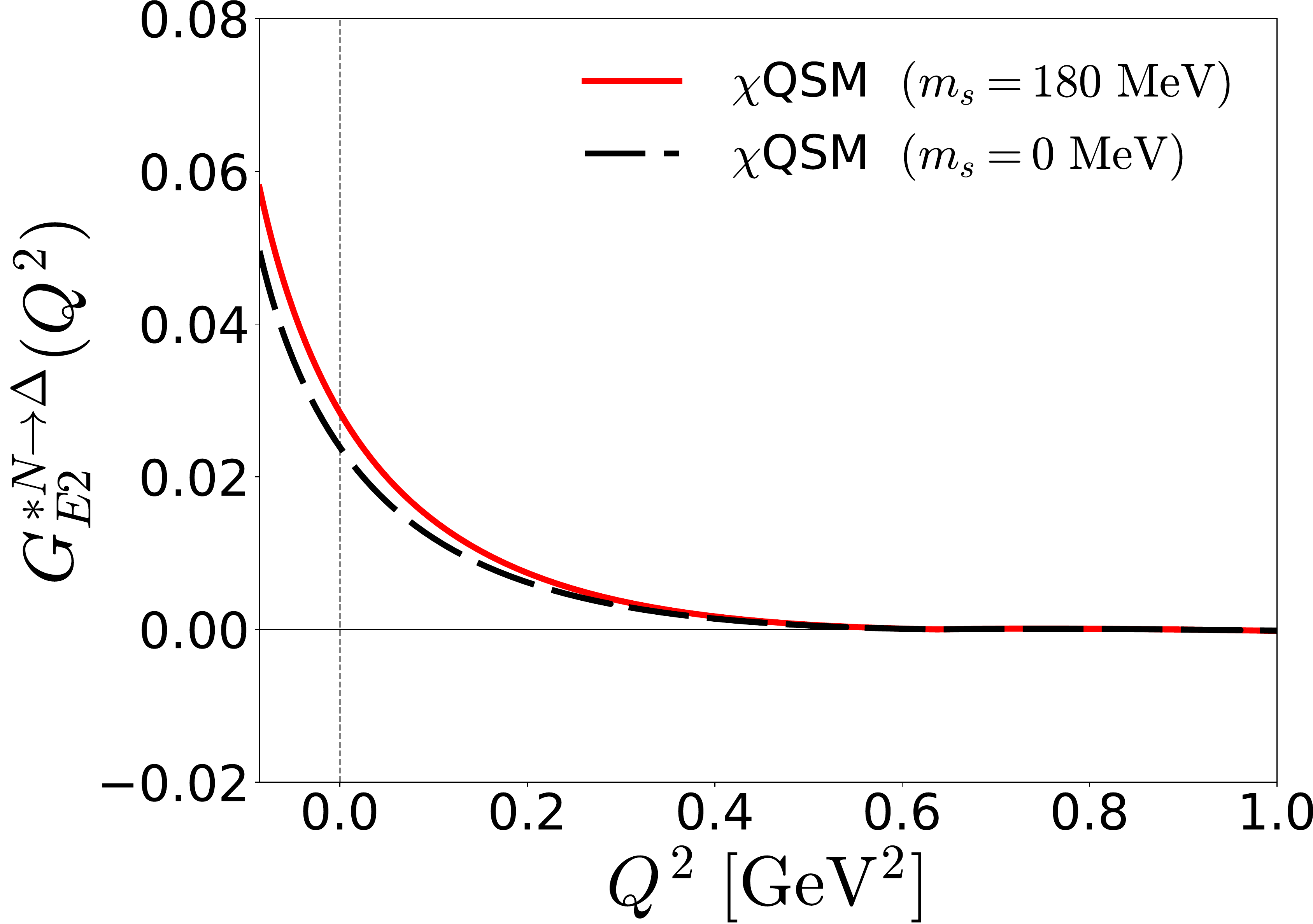}
\includegraphics[scale=0.234]{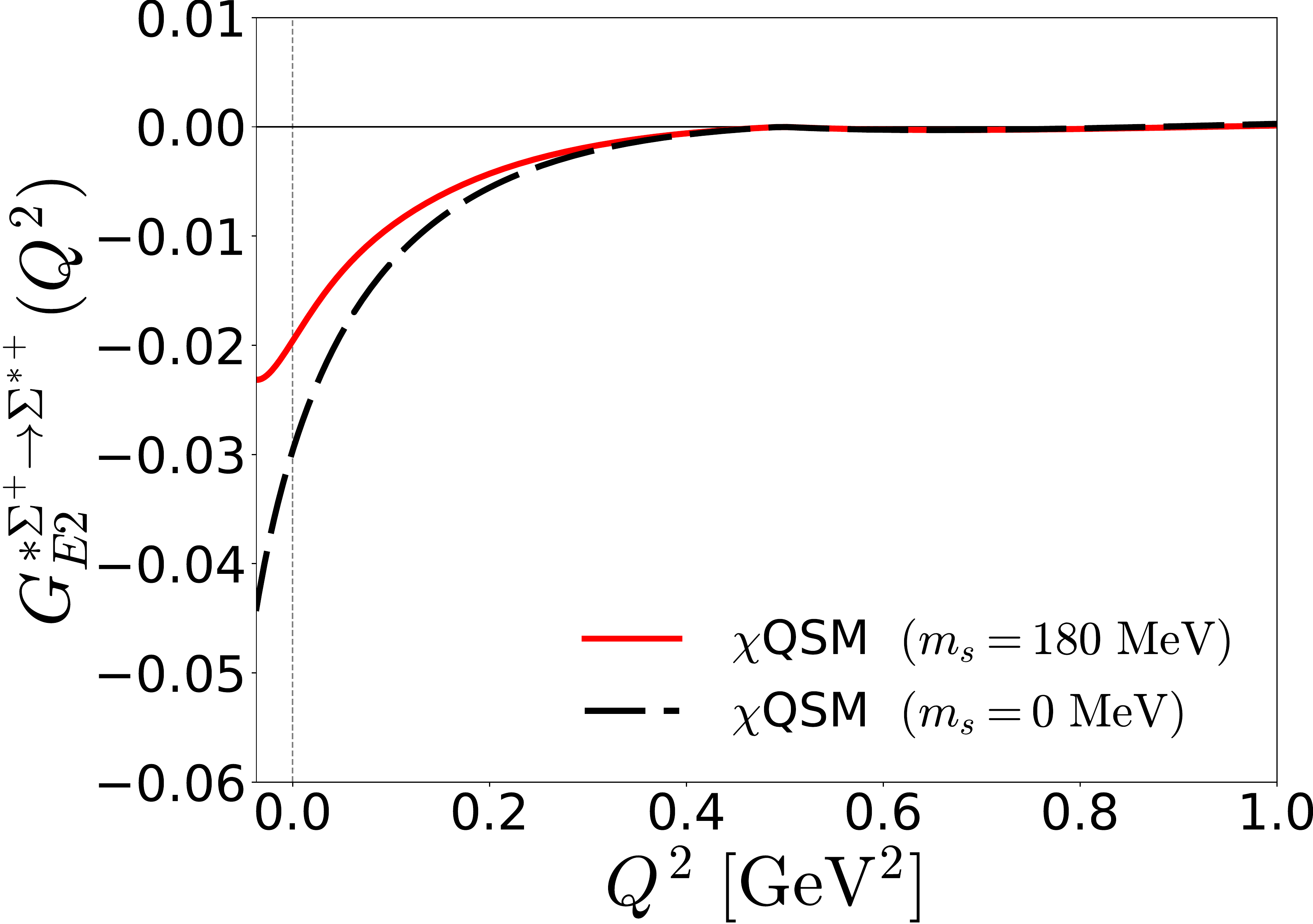}
\includegraphics[scale=0.234]{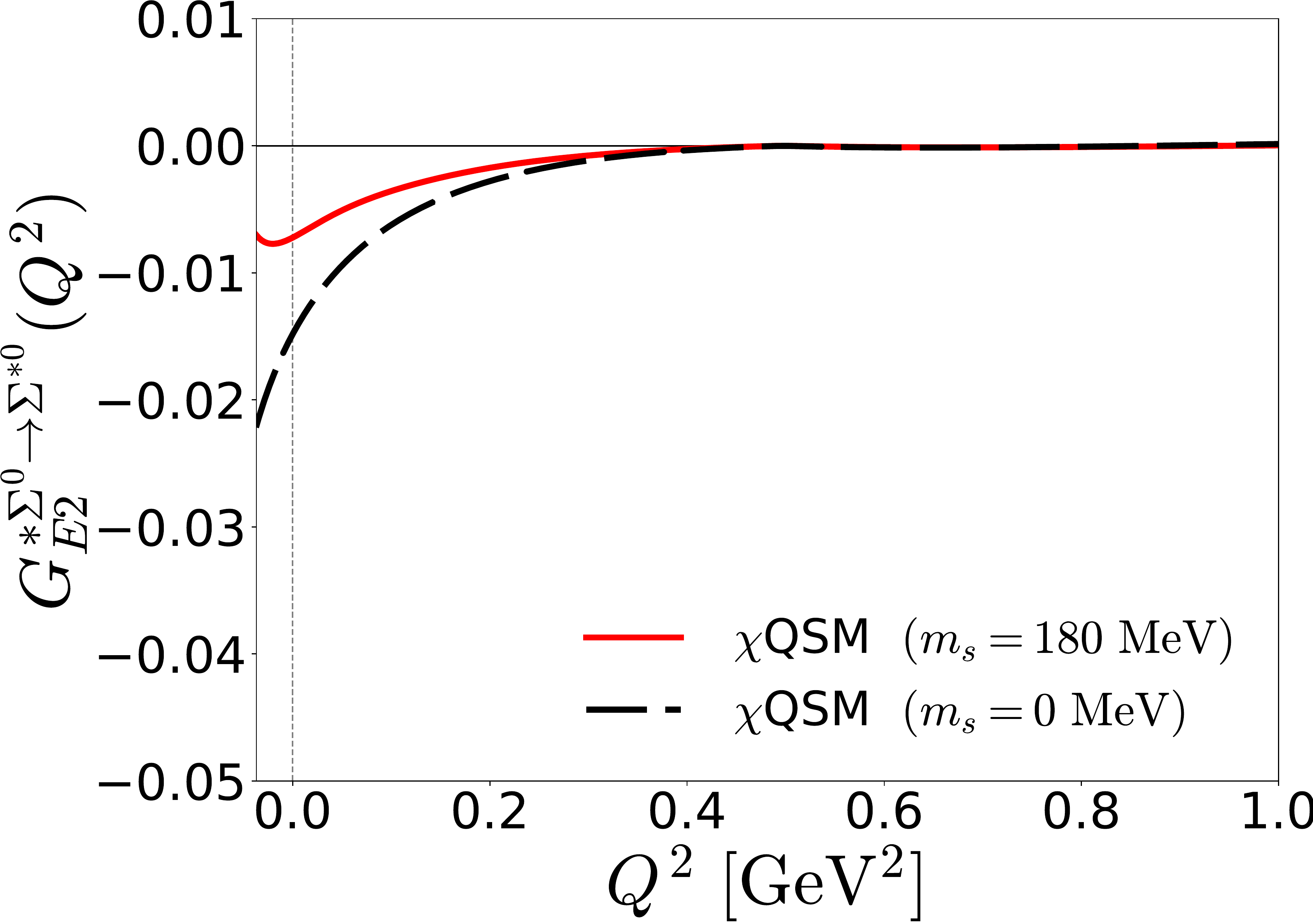}
\includegraphics[scale=0.234]{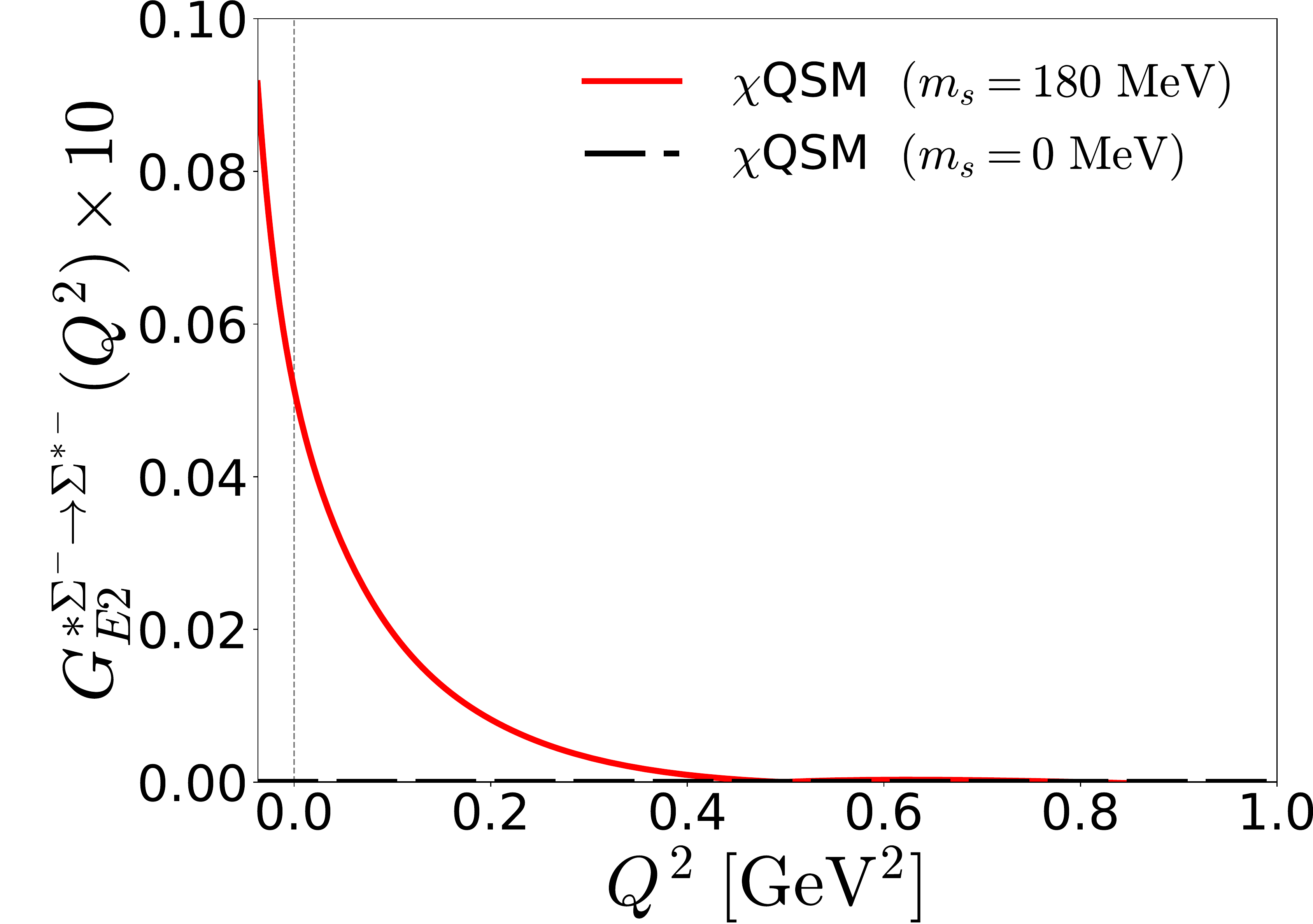}
\includegraphics[scale=0.234]{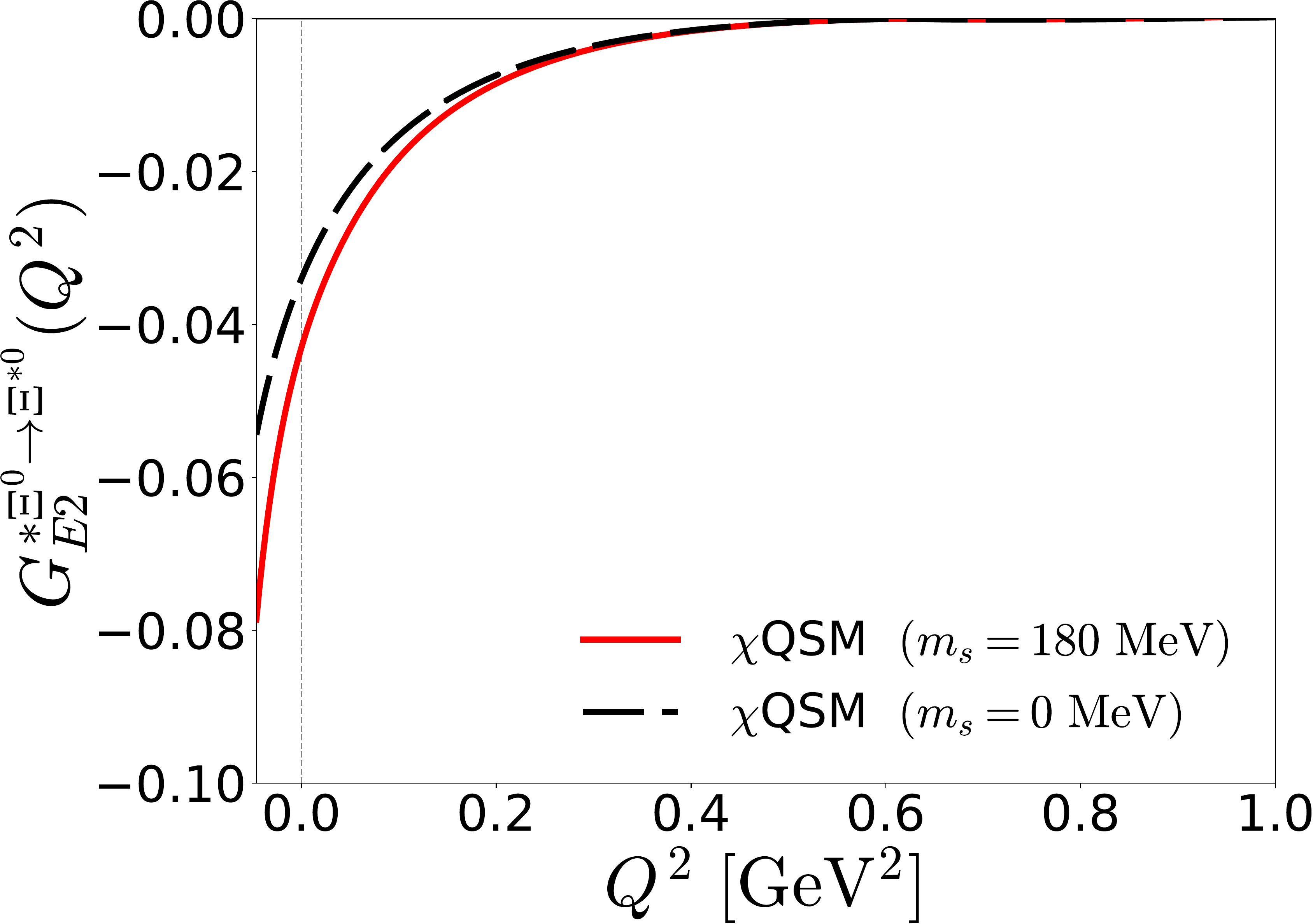}
\includegraphics[scale=0.234]{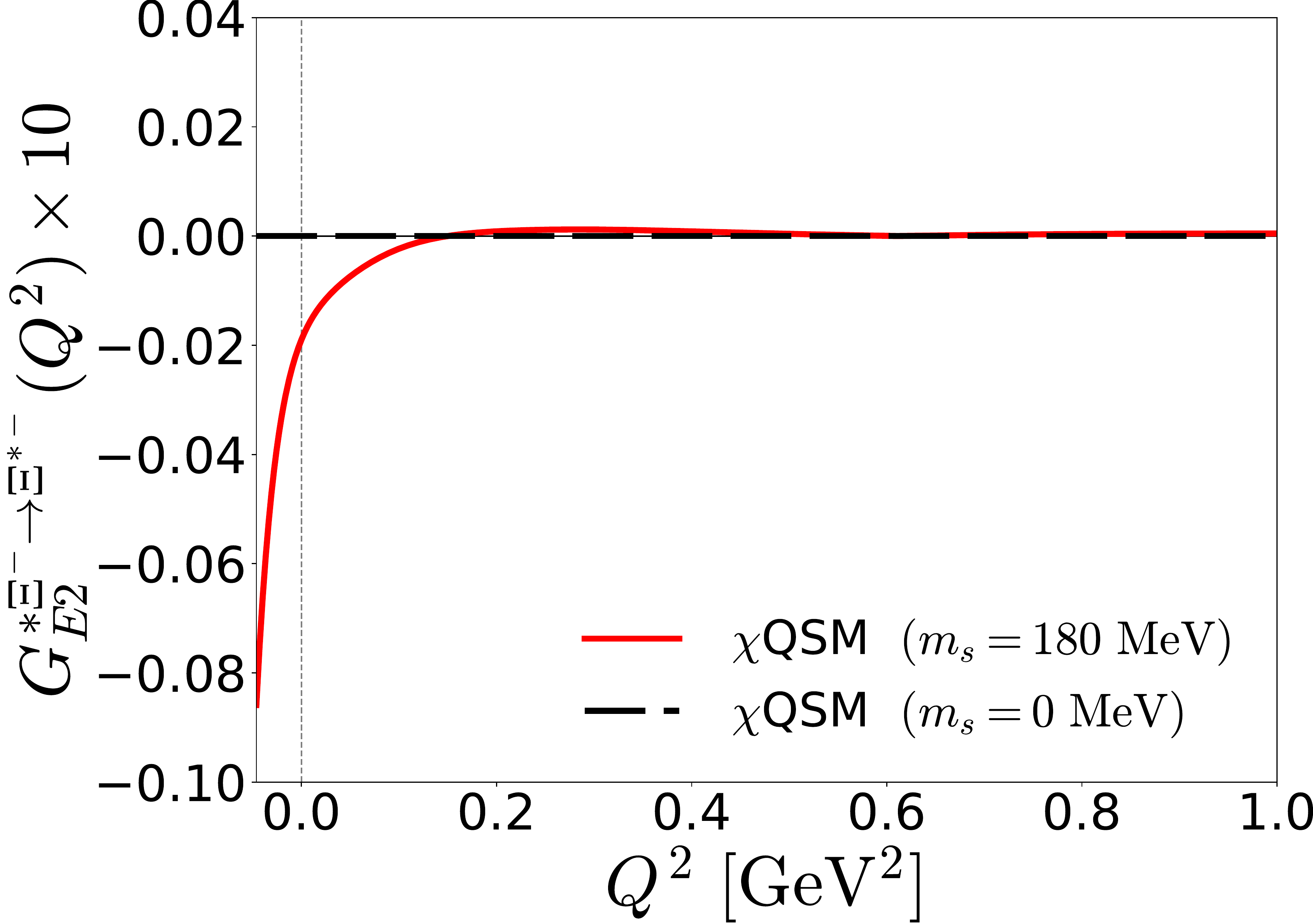}
\includegraphics[scale=0.234]{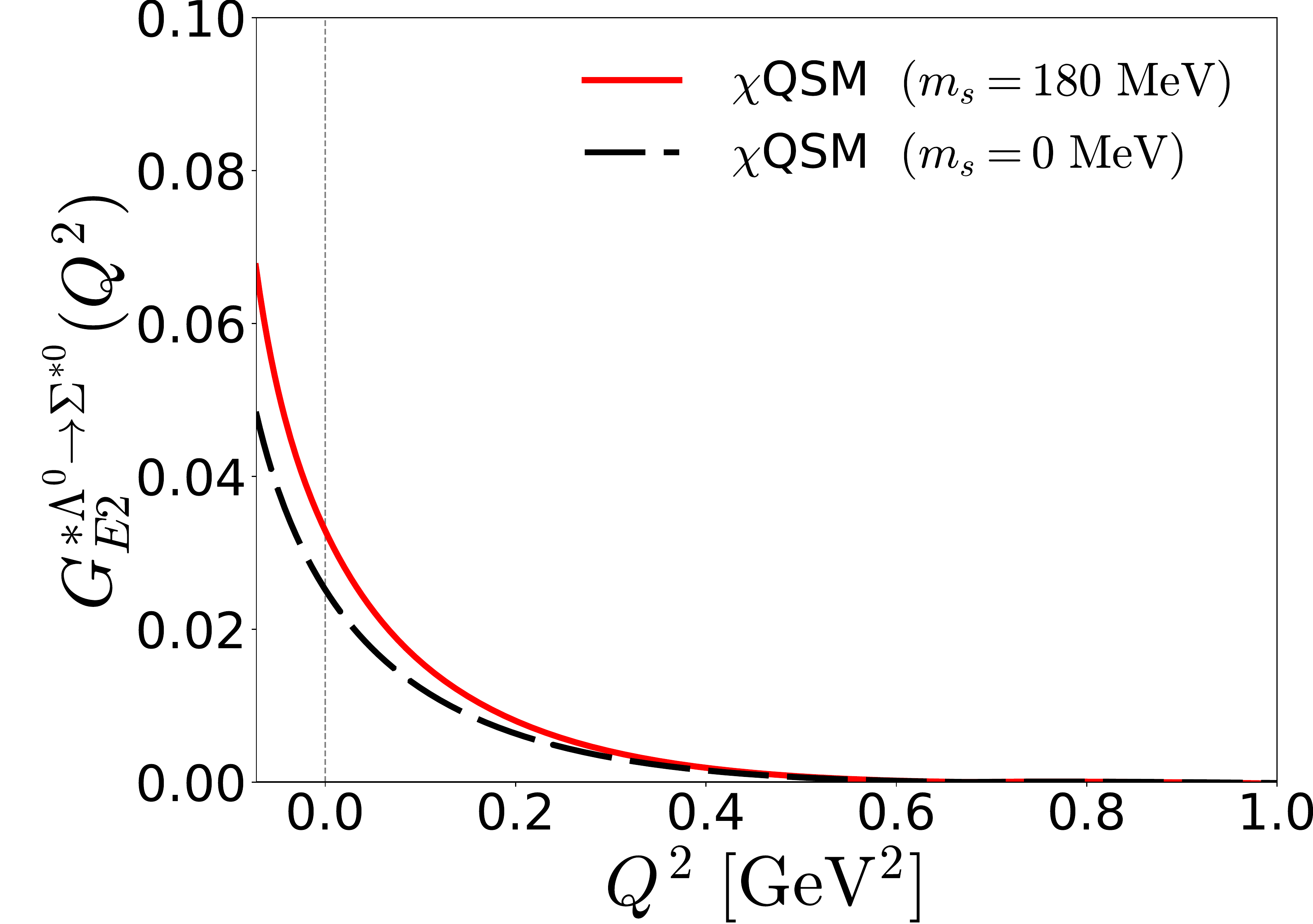}
\caption{Results for the $E2$ transition form factors of all
  the other members of the baryon decuplet with
  and without the effects of flavor SU(3) symmetry breaking. The solid
curves depict the results with $m_{\mathrm{s}}=180$ MeV, whereas the
dashed ones draw those in exact flavor SU(3) symmetry. }
\label{fig:10}
\end{figure}
\begin{figure}[htp]
\centering
\includegraphics[scale=0.234]{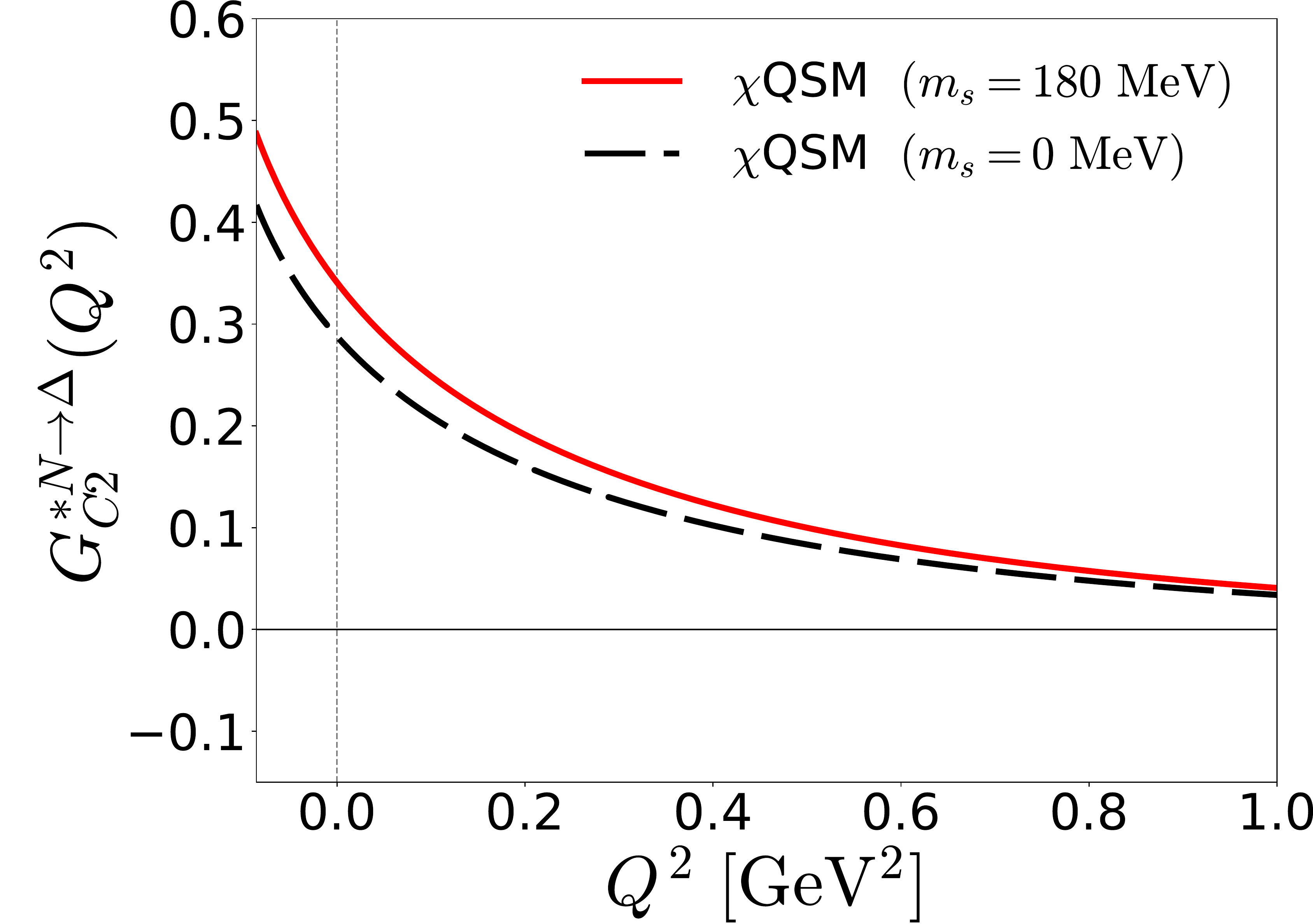}
\includegraphics[scale=0.234]{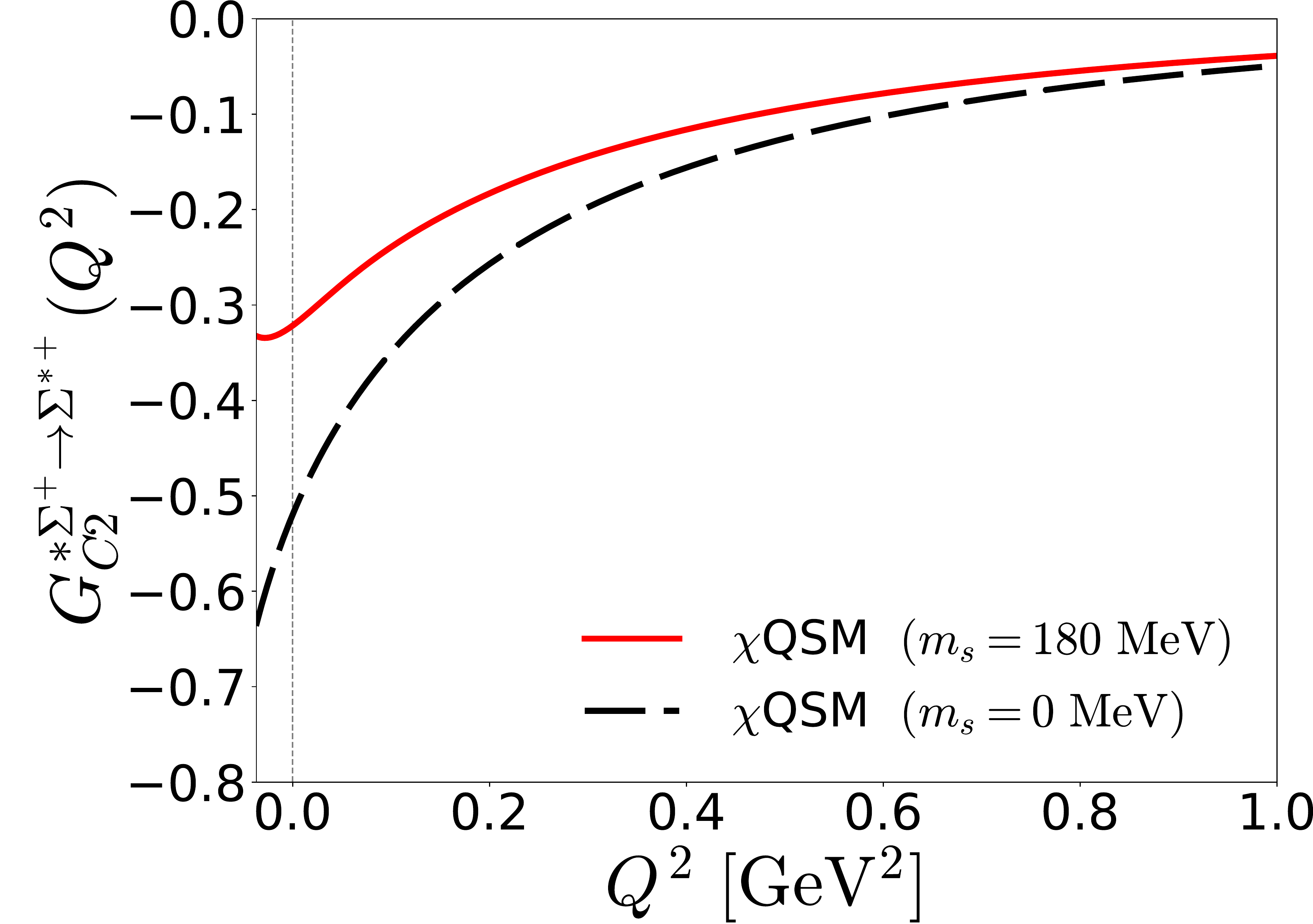}
\includegraphics[scale=0.234]{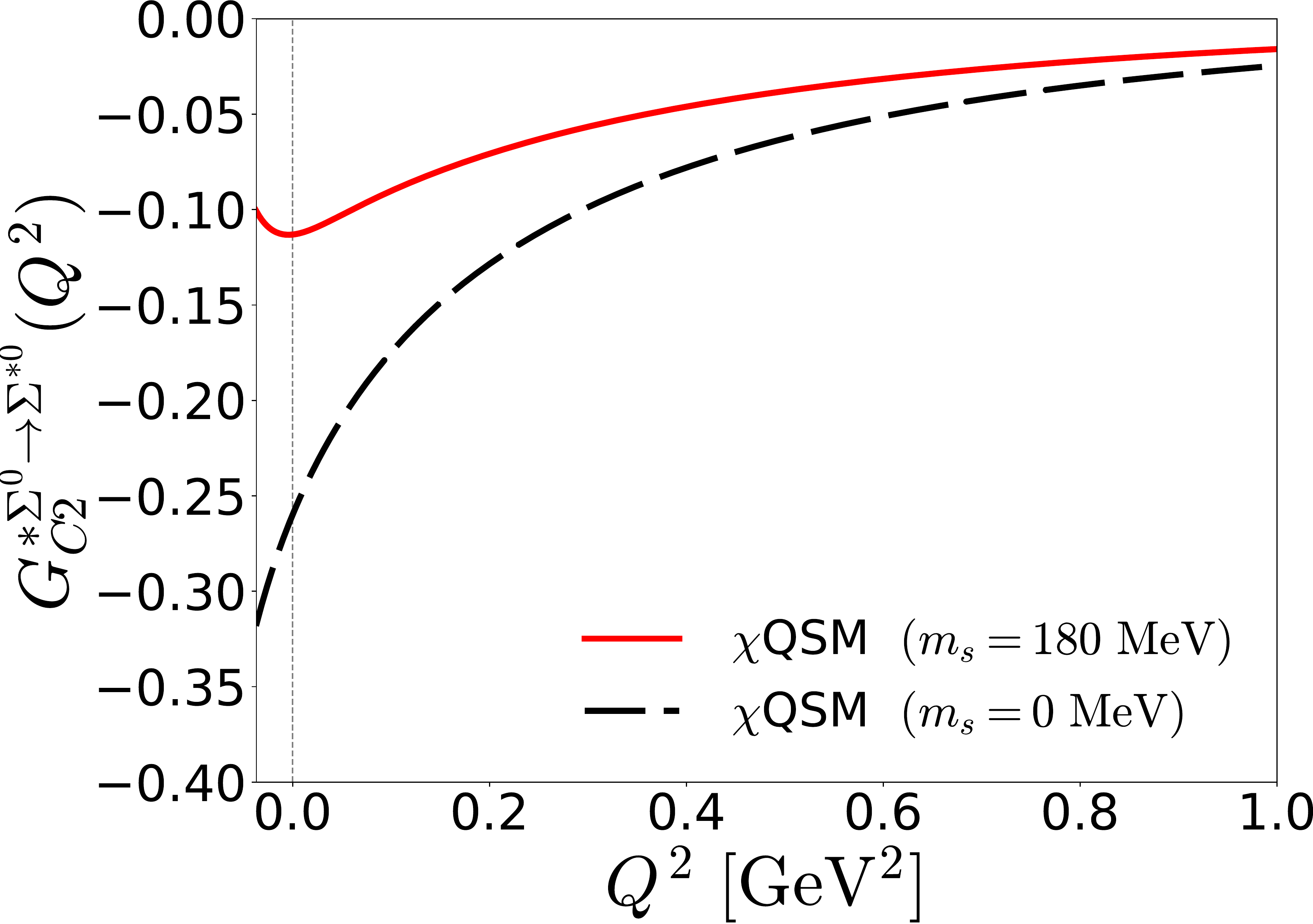}
\includegraphics[scale=0.234]{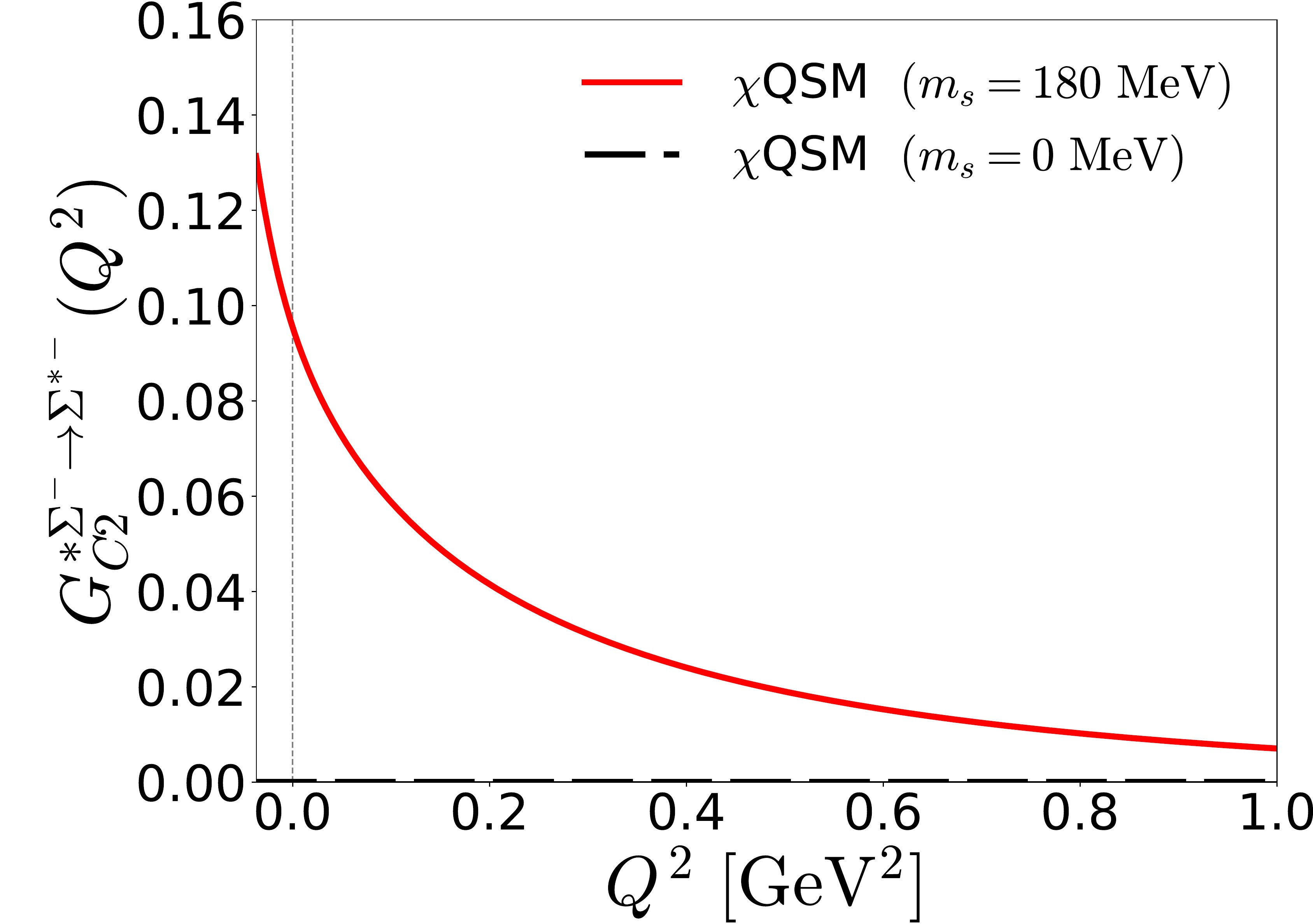}
\includegraphics[scale=0.234]{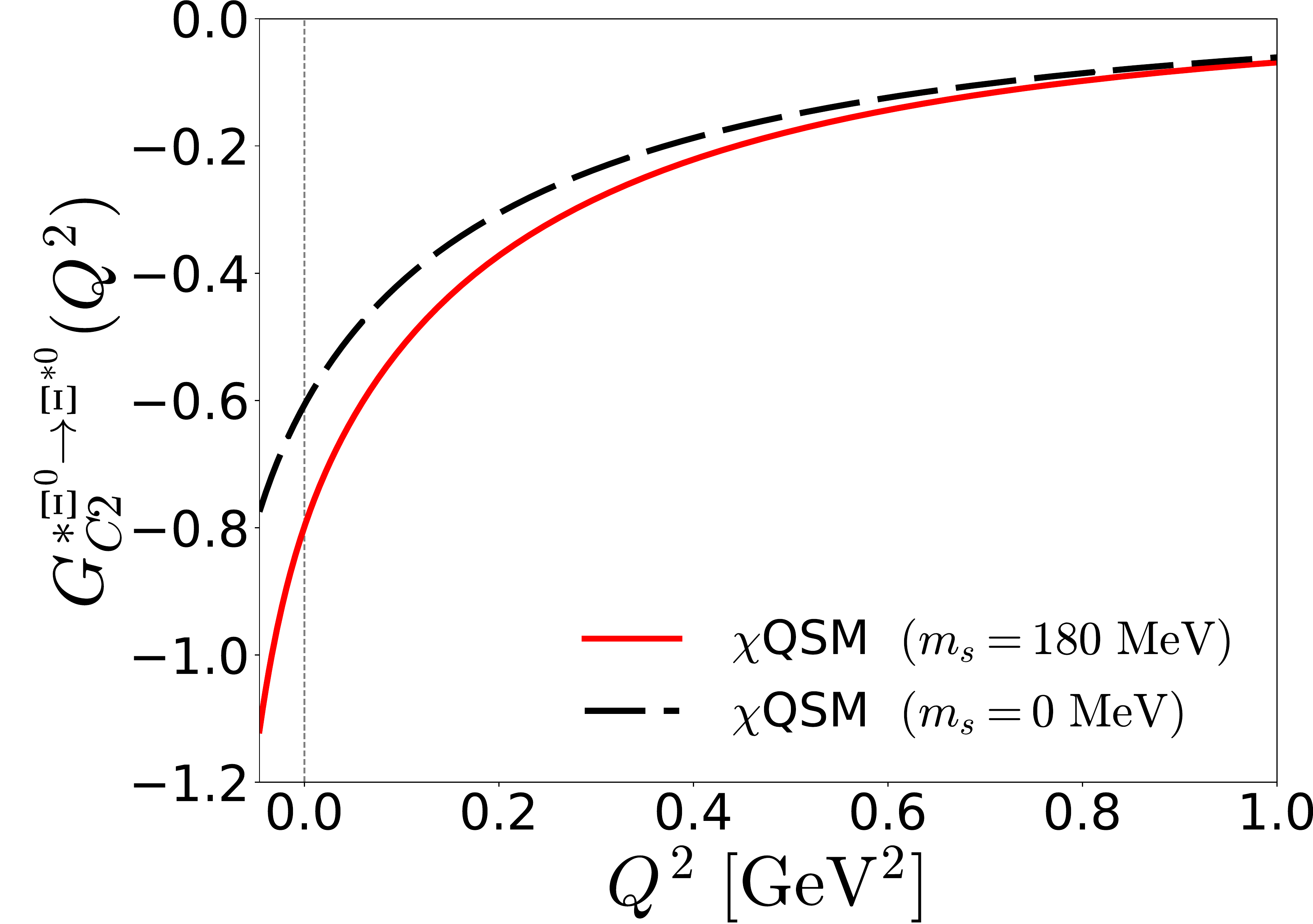}
\includegraphics[scale=0.234]{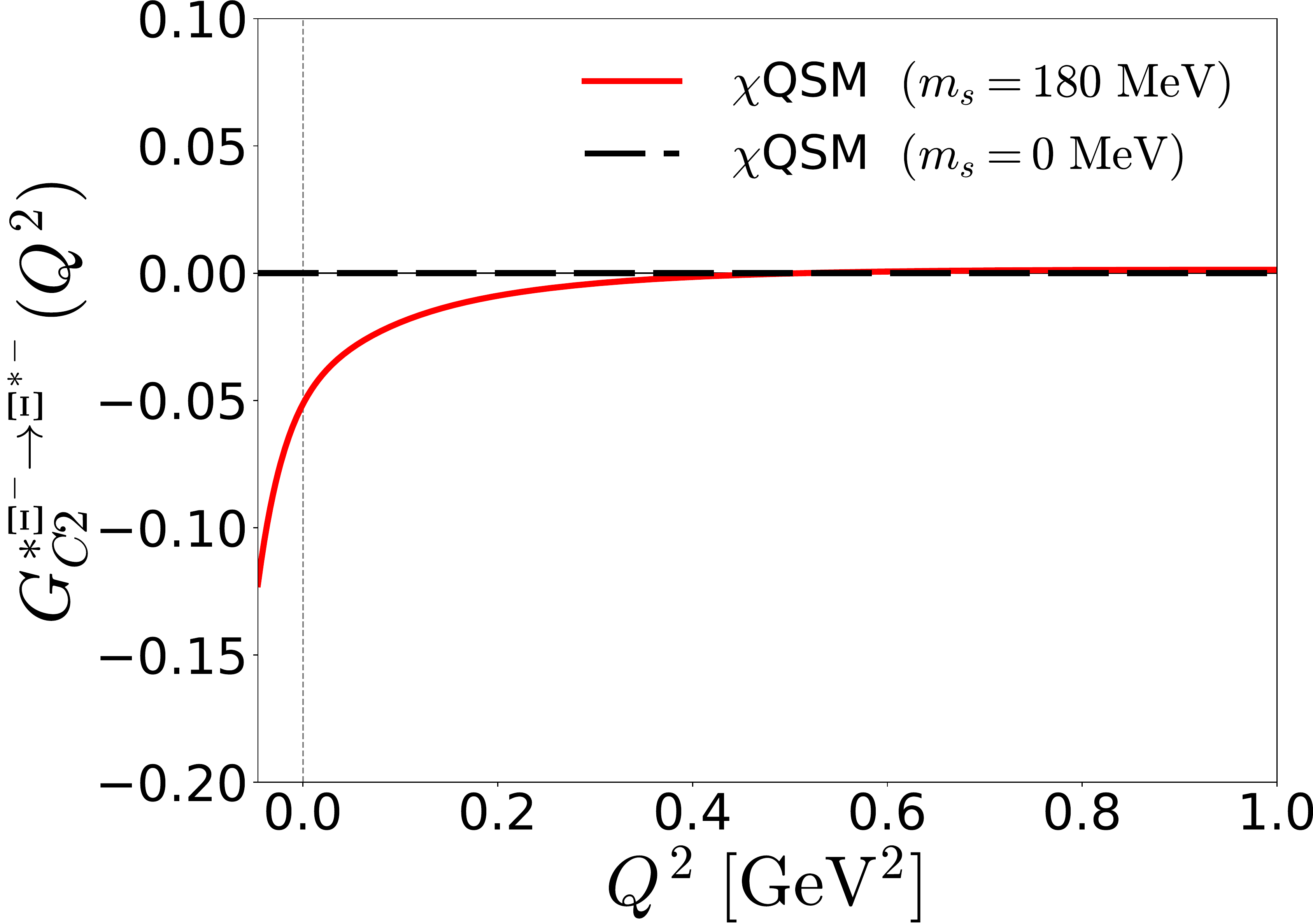}
\includegraphics[scale=0.234]{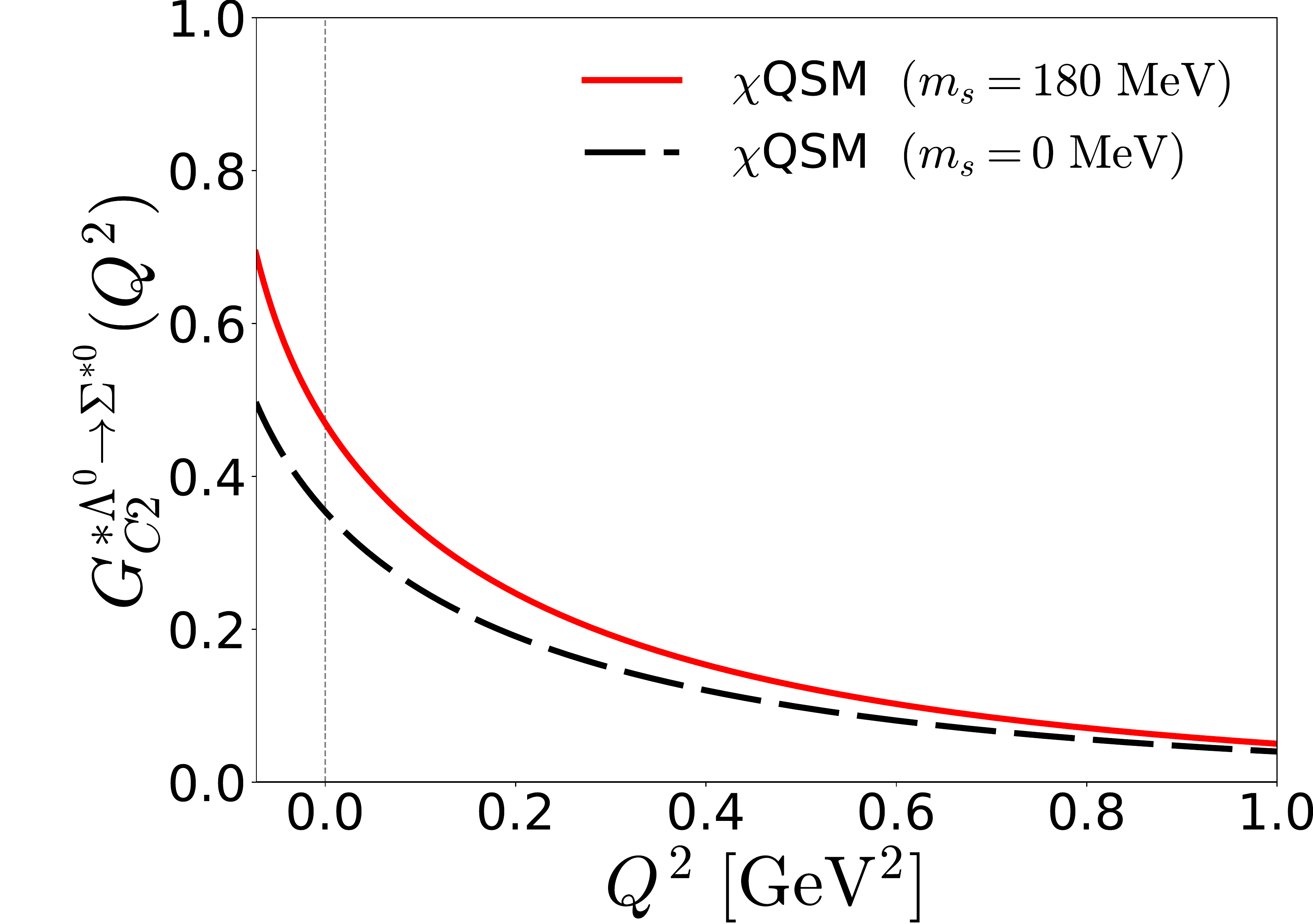}
\caption{Results for the $C2$ transition form factors of all
  the other members of the baryon decuplet with
  and without the effects of flavor SU(3) symmetry breaking. The solid
curves depict the results with $m_{\mathrm{s}}=180$ MeV, whereas the
dashed ones draw those in exact flavor SU(3) symmetry. } 
\label{fig:11}
\end{figure}
Figure~\ref{fig:9} draws the results of the magnetic dipole
transition form factors without and with the effects of flavor SU(3)
symmetric breaking. The solid curves depict those with the
linear $m_{\mathrm{s}}$ corrections whereas the dashed ones exhibit
those in the exact SU(3) symmetric case. One can see that the effects
of flavor SU(3) symmetry breaking contribute to the $M1$ form factors
in general below $10~\%$, which is in agreement with the quark-model 
prediction~\cite{Lipkin:1973rw}.  Note that they have almost a
negligible contribution to the $\Xi\gamma^*\to \Xi^*$
transition. However, when it comes to the EM transitions for the
negatively charged decuplet hyperons, the linear $m_{\mathrm{s}}$
terms take a leading role, since the flavor-SU(3) 
symmetric contributions vanish because of the $U$-spin symmetry. The
magnitudes of these forbidden transition form factors lie below the
upper limit imposed by the SELEX experiment.
In Figs.~\ref{fig:10} and~\ref{fig:11}, we draw respectively the $E2$
and $C2$ transition form factors of all the hyperons of the baryon
decuplet. Since  the densities for the $E2$ and $C2$ form factors are
in fact the same each other, the general behaviors of these form
factors are very similar, as shown in Figs.~\ref{fig:10}
and~\ref{fig:11}. Except for the $N\gamma^*\to \Delta$ transition, the
effects of flavor SU(3) symmetry breaking are noticeable. In
particular, when it comes to the $\Sigma^0\gamma \to \Sigma^{*0}$
transition, the contributions of  flavor SU(3) symmetry breaking are
of almost the same order as the SU(3)-symmetric term. However, as we
mentioned already in the previous Section, the $E2$ and $C2$
transition form factors do not have any leading-order
contributions. It means that the rotational $1/N_c$ correction plays a
role of the leading-order contribution. While the linear
$m_{\mathrm{s}}$ corrections should be usually smaller than the
leading-order contributions as in the case of the $M1$ transition form
factor, they become rather important when the leading-order
contributions vanish. A typical example can be found in the
calculation of the singlet axial-vector charge~\cite{Blotz:1993am} for
which the leading-order contribution disappears too. Thus, the linear
$m_{\mathrm{s}}$ corrections come into significant play, when the $E2$
and $C2$ transition form factors are discussed. 

As already mentioned previously, the magnitudes of the
radiative decay rates for the baryon decuplet based on the $\chi$QSM
are quite underestimated. However, their ratios are still
interesting. For example, the results of the ratios for some
decay widths are given as
\begin{align}
\frac{\Gamma^{\Delta \to N\gamma}}{\Gamma^{\Sigma^* \to \Sigma \gamma}}
  = 3.27\,(\mbox{Exp: }2.64),\;\;\;
\frac{\Gamma^{\Delta \to N\gamma}}{\Gamma^{\Sigma^* \to \Lambda^0 \gamma}}
  = 1.50 \,(\mbox{Exp: }1.40),\;\;\;
  \frac{\Gamma^{\Sigma^* \to \Lambda^0 \gamma}}{\Gamma^{\Sigma^* \to
  \Sigma \gamma}} 
  = 2.18\,(\mbox{Exp: }1.88).
\end{align}
Thus, the present results for these ratios are in qualitative
agreement with the data.

\section{Summary and conclusions}
In the present work, we aimed at investigating the electromagnetic
transition form factors of the baryon decuplet within the framework of
the self-consistent SU(3) chiral quark-soliton model, taking into
account the effects of flavor SU(3) symmetry breaking. We emphasized
how the electromagnetic transition form factors depend on the momentum
transfer squared, comparing the corresponding results with both the
experimental and lattice data. The present results of the $Q^2$
dependence of the magnetic dipole and Coulomb quadrupole form factors
are more or less well reproduced in comparison with the experimental
and empirical data. On the other hand, the results of the electric
quadrupole form factors fall off faster than the data.  In order to
compare the present results of the form factors with the lattice data,
we employed the values of the unphysical pion mass, i.e. $m_\pi=297$
MeV and 353 MeV, so that we are directly able to compare the results
with the lattice data. In addition, we normalized the present results
with the lattice data at $Q^2=0.06\,\mathrm{GeV}^2$ such that we can
see how the $Q^2$ dependences are different from those of the lattice
calculation. The form factors fall off more slowly as the pion mass
increases. Moreover, the magnitudes of the $E2$ and $C2$ form factors
are much reduced by using the unphysical pion masses. We then computed
the $E2/M1$ and $C2/M1$ ratios as functions of $Q^2$. The
$E2/M1$ ratios fall off faster than the experimental data. On the
other hand, the results of the $Q^2$ dependence of the $C2/M1$
ratios are in good agreement with the experimental data.

We have examined the valence- and sea-quark contributions separately. 
While the sea-quark effects, which can be regarded as those of the
pion clouds, contribute to the $M1$ transition form 
factors by about $(20-30)\,\%$, they are dominant over those of the
valence quarks in lower $Q^2$ regions for all possible radiative $E$2
and $C2$ transition form factors. The sea-quark contributions fall off
fast as $Q^2$ increases, compared with those of the valence
quarks. This means that in higher $Q^2$ regions the valence-quark
contributions take over those of the sea quarks or the pion
clouds. Thus, the sea-quark contributions play an essential role in
describing how the decuplet baryons are deformed.  

We then presented the results of the $E2/M1$ and $C2/M1$ ratios at
$Q^2=0$.  There exists an experimental data only on the $E2/M1$ ratio
for the EM $N\to \Delta$ transition. The comparison of the present
result with the data shows around 20~\% deviation from it. We then
examined the effects of flavor SU(3) symmetry breaking on the
electromagnetic transition form factors of the decuplet
hyperons. While they are rather marginal on the magnetic dipole
transition form factors, they play an important role in describing the
$E2$ and $C2$ transition form factors. The reason is that the
leading-order contributions vanish for the $E2$ and $C2$ transition
form factors. Thus, the rotational $1/N_c$ and linear $m_{\mathrm{s}}$
corrections are equally important. 

While the present results of the EM transition form factors for the
baryon decuplet describe well the $Q^2$ dependence, the magnitudes of 
the $N\gamma^*\to \Delta$ form factors are still underestimated,
compared with the experimental data. This is already a well-known
feature of the $\chi$QSM. There is in fact a way of improving
the present work. One can combine the $Q^2$ behavior of the form
factors obtained by the present work with the magnetic transition
moments evaluated in a model-independent
approach~\cite{Kim:2005gz}. In principle, the quadrupole transition
moments can be determined in a similar way. Actually, it is of great
importance to describe the electromagnetic transition form factors of
the baryon decuplet, since all the determined dynamical parameters can
be employed when we compute the strangeness-changing
transitions. While there is no experimental information on 
semileptonic decays of the baryon decuplet except for the $\Omega^-$,
they are still very important in determining the strong vector and tensor
coupling constants for the baryon decuplet and the octet to the vector
meson vertices through the Goldberger-Treiman relations. The
corresponding work is under way.

\begin{acknowledgments}
The authors want to express M. V. Polyakov for valuable
discussions. Part of the work was carried out during HChK's visit to 
Institute for Theoretical Physics II at Ruhr-Universit{\"a}t Bochum.
The present work was supported by Basic Science Research Program
through the National Research Foundation of Korea funded by the
Ministry of Education, Science and Technology
(Grant-No. 2018R1A2B2001752 and 2018R1A5A1025563). JYK is supported 
by DAAD doctoral scholarship. 
\end{acknowledgments}

\appendix

\section{Densities for the EM transition form factors
  \label{app:a}} 
The densities of the magnetic dipole transition form factor are
expressed explicitly as follows:
\begin{align}
 \frac{1}{N_{c}}{\cal{Q}}_{0} (\bm{r})  =&    \langle \mathrm{val}| \bm{r}
 \rangle  \gamma^5\{\hat{\bm{r}} \times \bm{\sigma} \}\cdot\bm{\tau} \langle
  \bm{r} | \mathrm{val}  \rangle 
  +N_{c}\sum_{n}   \mathcal{R}_{1}(E_{n}) \langle n| \bm{r} \rangle
  \gamma^5  \{ \hat{\bm{r}} \times \bm{\sigma} \}\cdot \bm{\tau} \langle
 \bm{r} |  n \rangle,  \cr 
 \frac{1}{N_{c}}{\cal{Q}}_{1} (\bm{r})  =& i \frac{1}{2} \sum_{n \ne
 \mathrm{val}} \frac{\mathrm{sign}(E_{n})}{E_{n}-E_{\mathrm{val}}}  
 \langle n| \bm{r} \rangle\gamma^5[  \{\hat{\bm{r}} \times \bm{\sigma}
 \}\times\bm{\tau} ] \langle  \bm{r} | \mathrm{val} \rangle \cdot
 \langle  {\mathrm{val}} | \bm{\tau}| n \rangle
 \cr 
& +i \frac{1}{4}\sum_{n,m}  {\mathcal R}_{4}(E_{n},E_{m})  \langle
  m | \bm{r} \rangle\gamma^5 [\{\hat{\bm{r}} \times
  \bm{\sigma}\}\times\bm{\tau}] \langle \bm{r} | n \rangle \cdot
  \langle m |\bm{\tau}| n \rangle ,  \cr  
 \frac{1}{N_{c}}{\mathcal{X}}_{1} (\bm{r}) =&  \sum_{n \ne
  \mathrm{val}}\frac{1}{E_{n}-E_{\mathrm{val}}}
   \langle \mathrm{val}| \bm{r} \rangle\gamma^5
  \{\hat{\bm{r}}\times\bm{\sigma}\} \langle \bm{r} | \mathrm{val}
 \rangle\cdot \langle n|  \bm{\tau} | \mathrm{val}  \rangle \cr  
&+\frac{1}{2}\sum_{n, m} \mathcal{R}_{5}(E_{n},E_{m})\langle n |  
  \bm{r} \rangle \gamma^5  \{\hat{\bm{r}}\times\bm{\sigma}\}  \langle \bm{r}
  | m  \rangle \cdot  \langle m | \bm{\tau} | n \rangle , \cr 
\frac{1}{N_{c}}{\cal{X}}_{2} (\bm{r})   =&     \sum_{n^{0}}
  \frac{1}{E_{n^{0}}-E_{\mathrm{val}}}
  \langle \mathrm{val}| \bm{r} \rangle\gamma^5 \{\hat{\bm{r}} \times
  \bm{\sigma} \}\cdot\bm{\tau}  \langle \bm{r} | n^0 \rangle \langle
 n^0   |  \mathrm{val} \rangle  \cr 
& +\sum_{n^0,m}\mathcal{R}_{5}(E_{m},E_{n^{0}}) \langle m |
  \bm{r} \rangle \gamma^5  \{\hat{\bm{r}} \times
  \bm{\sigma} \}\cdot\bm{\tau} \langle  \bm{r}| n^0 \rangle \langle
  n^0 | m \rangle ,  \cr 
 \frac{1}{N_{c}}{\cal{M}}_{0} (\bm{r})   =&   \sum_{n \ne
\mathrm{val}} \frac{1}{E_{n}-E_{\mathrm{val}}}  \langle
\mathrm{val}| \bm{r} \rangle \gamma^5  \{\hat{\bm{r}} \times
\bm{\sigma} \}\cdot\bm{\tau} \langle \bm{r} |  n \rangle \langle n|
\gamma^{0} |   \mathrm{val} \rangle  \cr 
& - \frac{1}{2}\sum_{n,m} \mathcal{R}_{2}(E_{n},E_{m}) \langle m |
  \bm{r} \rangle \gamma^5  \{\hat{\bm{r}} \times
\bm{\sigma} \}\cdot\bm{\tau} \langle \bm{r} |  n \rangle \langle n|
\gamma^{0} | m \rangle  ,  \cr 
\frac{1}{N_{c}}{\cal{M}}_{1} (\bm{r})=& {}  \sum_{n \ne
 \mathrm{val}}\frac{1}{E_{n}-E_{\mathrm{val}}}
 \langle \mathrm{val}| \bm{r} \rangle  \gamma^5  \{\hat{\bm{r}} \times
  \bm{\sigma} \} \langle \bm{r} | n \rangle\cdot \langle n| \gamma^{0}
 \bm{\tau}  | \mathrm{val}  \rangle    \cr 
& -\frac{1}{2}\sum_{n,m} {\cal R}_{2}(E_{n},E_{m})
  \langle m | \bm{r} \rangle \gamma^5  \{\hat{\bm{r}} \times
  \bm{\sigma} \} \langle \bm{r} | n \rangle\cdot \langle n| \gamma^{0} 
 \bm{\tau} | m \rangle,  \cr 
\frac{1}{N_{c}}{\cal{M}}_{2} (\bm{r}) =& {} \sum_{n^{0}}
  \frac{1}{E_{n^{0}}-E_{\mathrm{val}}}
  \langle \mathrm{val}| \bm{r} \rangle  \gamma^5  \{\hat{\bm{r}} \times
\bm{\sigma} \}\cdot\bm{\tau} \langle \bm{r} | n^{0} \rangle \langle
 n^{0}|  \gamma^{0} | \mathrm{val} \rangle  \cr 
& - \sum_{n^{0},m} \mathcal{R}_{2}(E_{n^0},E_{m})
  \langle m | \bm{r} \rangle   \gamma^5  \{\hat{\bm{r}} \times
\bm{\sigma} \}\cdot\bm{\tau}
\langle  \bm{r}| n^0 \rangle\langle n^0 | \gamma^{0} | m \rangle.
\label{eq:M1den}                                                   
\end{align}
The densities of the electric quadrupole transition form factors are
given as 
\begin{align}
(-\sqrt{10})\frac{2}{N_{c}} \mathcal{I}_{1E2}(\bm{r})&= \sum_{n \ne
\mathrm{val} }\frac{1}{E_{n}-E_{\mathrm{val}}}{\langle\mathrm{val}
| \bm{\tau} | n\rangle} \cdot{\langle n |\bm{r} \rangle
\{ \sqrt{4\pi}Y_{2}  \otimes\tau_{1}  \}_{1}\langle \bm{r} |
\mathrm{val}\rangle} \cr 
&  + \frac{1}{2} \sum_{n,m} \mathcal{R}_{3}(E_n,E_m)   {\langle n |
  \bm{\tau} | m \rangle} \cdot{\langle m | \bm{r} \rangle \{
  \sqrt{4\pi} Y_{2}  \otimes \tau_{1}  \}_{1} \langle \bm{r} | n
  \rangle}  ,\cr 
(-\sqrt{10})\frac{2}{N_{c}} \mathcal{K}_{1E2}(\bm{r})&=  \sum_{n \ne
 \mathrm{val} } \frac{1}{E_{n}-E_{\mathrm{val}}} {\langle\mathrm{val}
 |  \gamma^{0} \bm{\tau} | n \rangle} \cdot {\langle n | \bm{r} \rangle
 \{ \sqrt{4\pi} Y_{2}  \otimes \tau_{1}  \}_{1} \langle \bm{r} |
 \mathrm{val} \rangle} \cr  
& + \frac{1}{2} \sum_{n,m} \mathcal{R}_{5}(E_n,E_m)     {\langle n |
  \gamma^{0} \bm{\tau} | m \rangle} \cdot {\langle m | \bm{r} \rangle
  \{ \sqrt{4\pi} Y_{2}  \otimes \tau_{1}  \}_{1} \langle \bm{r} | n
                               \rangle}.
\label{eq:E2den}                               
\end{align}
The regularization functions in Eqs.~\eqref{eq:M1den}
and~\eqref{eq:E2den} are defined by
\begin{align}
&\mathcal{R}_{1}(E_{n}) = -\frac{1}{2 \sqrt{\pi}} E_{n}
  \int^{\infty}_{0} \phi(u) \frac{du}{u} e^{-u E_{n}^{2}}, \cr 
&\mathcal{R}_{2}(E_{n},E_{m}) = \frac{1}{2 \sqrt{\pi}} \int^{\infty}_{0}
  \phi(u) \frac{du}{\sqrt{u}} \frac{E_{m} e^{-u E_{m}^{2}}-E_{n} e^{-u
  E_{n}^{2}}}{E_{n} - E_{m}}, \cr 
&\mathcal{R}_{3}(E_{n},E_{m}) = \frac{1}{2 \sqrt{\pi}} \int^{\infty}_{0}
  \phi(u) \frac{du}{\sqrt{u}} \left[ \frac{ e^{-u E_{m}^{2}}- e^{-u
  E_{n}^{2}}}{u(E^{2}_{n} - E^{2}_{m})} -\frac{E_{m} e^{-u
  E_{m}^{2}}+E_{n} e^{-u E_{n}^{2}}}{E_{n} + E_{m}}  \right ], \cr 
&\mathcal{R}_{4}(E_{n},E_{m}) = \frac{1}{2 {\pi}} \int^{\infty}_{0}
  \phi(u) {du} \int^{1}_{0} d\alpha e^{-u E_{n}^{2}(1-\alpha) - u
  E^{2}_{m}\alpha}  \frac{E_{n}(1-\alpha) - \alpha
  E_{m}}{\sqrt{\alpha(1-\alpha)}}, \cr 
&\mathcal{R}_{5}(E_{n},E_{m}) =
  \frac{\mathrm{sign}(E_{n})-\mathrm{sign}(E_{m})}{2(E_{n}-E_{m})},
\end{align}
where $|\mathrm{val}\rangle$ and $|n\rangle$ denote the states of the 
valence and sea quarks with the corresponding eigenenergies
$E_{\mathrm{val}}$ and $E_n$ of the single-quark Hamiltonian $h(U_c)$,
respectively~\cite{Christov:1995vm}. 

\section{Matrix elements of the SU(3) Wigner ${D}$
  function \label{app:b}} 
The collective wavefunction of a baryon with flavor $F=(Y,T,T_3)$ and
spin $S=(Y'=-N_{c}/3,J,J_3)$ in the representation $\nu$ is expressed
in terms of a tensor with two indices, i.e. $\psi_{(\nu;\,
  F),(\overline{\nu};\,\overline{S})}$, one running over the states
$F$ in the representation $\nu$ and the other one over the 
states $\overline{S}$ in the representation $\overline{\nu}$. Here,  
$\overline{\nu}$ denotes the complex conjugate of the
$\nu$, and the complex conjugate of $S$ is written by
$\overline{S}=(N_{c}/3,\,J,\,J_3)$. Thus, the collective wavefunction
is expressed as  
\begin{align}
  \label{eq:SolitonWF1}
\psi_{(\nu;\, F),(\overline{\nu};\,\overline{S})}(R) =
  \sqrt{\mathrm{dim}(\nu)} (-1)^{Q_S} [D_{F\,S}^{(\nu)}(R)]^*,
\end{align}
where $\mathrm{dim}(\nu)$ stands for the dimension of the
representation $\nu$ and $Q_S$ a charge corresponding to the baryon
state $S$, i.e. $Q_S=J_3+Y'/2$.   
\begin{align}
&|B_{{8}_{1/2}}\rangle = |{\bm{8}}_{1/2},B\rangle + 
c^{B}_{{\overline{10}}}|{{\overline{\bm{10}}}}_{1/2},B\rangle + 
c^{B}_{{27}}|{{\bm{27}}}_{1/2},B\rangle, \cr
&|B_{{\bm10}_{3/2}}\rangle = |{\bm{10}}_{3/2},B\rangle + 
a^{B}_{{27}}|{{\bm{27}}}_{3/2},B\rangle + 
a^{B}_{{35}}|{{\bm{35}}}_{3/2},B\rangle,
\label{eq:mixedWF1}
\end{align}
with the mixing coefficients
\begin{eqnarray}
c_{{\overline{10}}}^{B}
\;=\;
c_{{\overline{10}}}\left[\begin{array}{c}
\sqrt{5}\\
0 \\
\sqrt{5} \\
0
\end{array}\right], 
& 
c_{27}^{B}
\;=\; 
c_{27}\left[\begin{array}{c}
\sqrt{6}\\
3 \\
2 \\
\sqrt{6}
\end{array}\right], \ 
a_{{27}}^{B}
\;=\;
a_{{27}}\left[\begin{array}{c}
\sqrt{15/2}\\
2 \\
\sqrt{3/2} \\
0
\end{array}\right], 
& 
a_{35}^{B}
\;=\; 
a_{35}\left[\begin{array}{c}
5/\sqrt{14}\\
2\sqrt{{5}/{7}} \\
3\sqrt{{5}/{14}} \\
2\sqrt{{5}/{7}}
\end{array}\right], 
\label{eq:pqmix}
\end{eqnarray}
respectively, in the basis
$\left[N,\;\Lambda,\;\Sigma,\;\Xi\right]$ and
$\left[\Delta,\;\Sigma^{*},\;\Xi^{*},\;\Omega\right]$. The   
parameters $c_{\overline{10}}$, $c_{27}$, $a_{{27}}$ and $a_{35}$ are given by 
\begin{align}
c_{\overline{10}} \;=\;
{\displaystyle -\frac{{I}_{2}}{15} \left ( \alpha + \frac{1}{2}\gamma
  \right)}, & 
c_{27} \;=\;
{\displaystyle -\frac{{I}_{2}}{25} \left ( \alpha - \frac{1}{6}\gamma
  \right)}, & 
a_{27} \;=\;
{\displaystyle -\frac{{I}_{2}}{8} \left ( \alpha + \frac{5}{6}\gamma
  \right)}, \ 
a_{35} \;=\; {\displaystyle -\frac{{I}_{2}}{24} \left( \alpha -
  \frac{1}{2}\gamma \right)}, 
\label{eq:pqmix2}
\end{align}
where $\alpha$ and $\gamma$ are the parameters appearing in the
  collective Hamiltonian, which are written by
\begin{align}
\alpha=\left (-\frac{{\Sigma}_{\pi N}}{3\overline{m}}+\frac{
  K_{2}}{I_{2}}{Y'}  
\right )m_{\mathrm{s}},
\;\;\; \beta=-\frac{K_{2}}{I_{2}}m_{s} ,
\;\;\;  \gamma=2\left ( \frac{K_{1}}{I_{1}}-\frac{K_{2}}{I_{2}} 
 \right ) m_{\mathrm{s}}.
\label{eq:alphaetc}  
\end{align}
Here, $\Sigma_{\pi N}$ is the well-known $\pi N$ sigma term. The
  moments of inertia ($I_{1}, I_{2}$) and anomalous moments of inertia
  ($K_{1}, K_{2}$) are given by 
\begin{align}
\frac{6}{N_{c}} I_{1} &= \sum_{n\neq\mathrm{val}}
                        \frac{1}{E_{n}-E_{\mathrm{val}}}\langle
                        \mathrm{val} | \bm{\tau} | n \rangle \cdot
                        \langle n | \bm{\tau} | \mathrm{val} \rangle +
                        \frac{1}{2}\sum_{n,m\neq n}\langle m |
                        \bm{\tau} | n \rangle \cdot \langle n |
                        \bm{\tau} | m \rangle
                        \mathcal{R}_{3}(E_n,E_m), \cr 
\frac{4}{N_{c}} I_{2} &= \sum_{n^{0}}
                        \frac{1}{E_{n^{0}}-E_{\mathrm{val}}}\langle
                        \mathrm{val} |  n^{0} \rangle  \langle n^{0}
                        | \mathrm{val} \rangle + \sum_{n^{0},m}\langle
                        m | \bm{\tau} | n^{0} \rangle \langle n^{0}  |
                        m \rangle \mathcal{R}_{3}(E_{n^{0}},E_m), \cr 
\frac{6}{N_{c}} K_{1} &= \sum_{n\neq\mathrm{val}}
                        \frac{1}{E_{n}-E_{\mathrm{val}}}\langle
                        \mathrm{val} |  \bm{\tau} | n \rangle \cdot
                        \langle n | \gamma^{0} \bm{\tau} |
                        \mathrm{val} \rangle +
                        \frac{1}{2}\sum_{n,m\neq n}\langle m |
                        \bm{\tau} | n \rangle \cdot \langle n |
                        \gamma^{0} \bm{\tau} | m \rangle
                        \mathcal{R}_{5}(E_n,E_m), \cr 
\frac{4}{N_{c}} K_{2} &= \sum_{n^{0}}
                        \frac{1}{E_{n^{0}}-E_{\mathrm{val}}}\langle
                        \mathrm{val} |  n^{0} \rangle  \langle n^{0}
                        |\gamma^{0}  | \mathrm{val} \rangle
                        +\sum_{n^{0},m}\langle m  | \bm{\tau} | n^{0}
                        \rangle \langle n^{0} |\gamma^{0} | m \rangle
                        \mathcal{R}_{5}(E_{n^{0}},E_m). 
\end{align}

We list the results of the matrix elements of the
relevant collective operators for the EM transition form factors in
Tables~\ref{tab:FF_Leading},~\ref{tab:FF_Op},~\ref{tab:FF_827},
~\ref{tab:FF_2710}. Note that all the matrix elements of the mixed
baryon wave functions arising from $\overline{\bm{10}}$ and
${\bm{35}}$ components vanish. 

   \begin{table}[htp]
  \caption{The matrix elements of the collective operators for the
    leading-order contributions and the $1/N_c$ rotational corrections
    to the electrimagnetic transition form factors.}  
  \label{tab:FF_Leading}
\begin{center}
\begin{tabular}{ c |  c c  c  c   } 
 \hline 
  \hline 
$B_{8}\gamma^{*} \to B_{10}$ & $N \gamma^{*} \to \Delta$ & $\Sigma \gamma^{*} \to \Sigma^{*}$ & $\Xi \gamma^{*} \to \Xi^{*}$ & $\Lambda \gamma^{*} \to \Sigma^{*}$ \\  
 \hline
$\langle B_{{10}} |D^{(8)}_{33} | B_{{8}}\rangle$  
& $\frac{2}{3\sqrt{5}}$& $-\frac{1}{3\sqrt{5}}T_{3}$& $-\frac{2}{3\sqrt{5}}T_{3}$ & $\frac{1}{15}$ \\  
$\langle B_{{10}} |D^{(8)}_{83} | B_{{8}}\rangle$  
& $0$ & $-\frac{1}{15}$ & $-\frac{1}{15}$& $0$ \\  
$\langle B_{{10}} |D^{(8)}_{38}J_{3}  | B_{{8}}\rangle$  
& $0$& $0$& $0$ & $0$ \\    
$\langle B_{{10}} |D^{(8)}_{88}J_{3}  | B_{{8}}\rangle$  
& $0$ & $0$ & $0$ & $0$ \\  
$\langle B_{{10}} |d_{ab3}D^{(8)}_{3a}J_{b}  | B_{{8}}\rangle$  
&$-\frac{1}{3\sqrt{5}}$ &$\frac{1}{6\sqrt{5}}T_{3}$ &$\frac{1}{3\sqrt{5}}T_{3}$ &$-\frac{1}{2\sqrt{15}}$   \\  
$\langle B_{{10}} |d_{ab3}D^{(8)}_{8a}J_{b} | B_{{8}}\rangle$
& $0$ & $\frac{1}{2\sqrt{15}}$ & $\frac{1}{2\sqrt{15}}$ & $0$  \\  
$\langle B_{{10}} |D^{(8)}_{3i}J_{i} | B_{{8}}\rangle$  
& $0$& $0$& $0$ & $0$ \\  
$\langle B_{{10}} |D^{(8)}_{8i}J_{i} | B_{{8}}\rangle$  
& $0$& $0$& $0$ & $0$ \\  
 \hline 
 \hline
\end{tabular}
\end{center}
\end{table}

\begin{table}[htp]
  \caption{The matrix elements of the collective operators for the
    $m_s$ corrections to the electromagnetic transition form factors.} 
  \label{tab:FF_Op}
\begin{center}
\begin{tabular}{ c |  c c  c  c   } 
 \hline 
  \hline 
$B_{8}\gamma^{*} \to B_{10}$ & $N \gamma^{*} \to \Delta$ & $\Sigma \gamma^{*} \to \Sigma^{*}$ & $\Xi \gamma^{*} \to \Xi^{*}$ & $\Lambda \gamma^{*} \to \Sigma^{*}$ \\  
 \hline
$\langle B_{{10}} |D^{(8)}_{88}D^{(8)}_{33} | B_{{8}}\rangle$
& $\frac{2}{9\sqrt{5}}$& $0$& $\frac{1}{18\sqrt{5}}T_{3}$ & $\frac{1}{6\sqrt{15}}$ \\  
$\langle B_{{10}} |D^{(8)}_{88}D^{(8)}_{83} | B_{{8}}\rangle$
& $0$ & $\frac{1}{6\sqrt{15}}$ & $\frac{1}{4\sqrt{15}}$ & $0$ \\  
$\langle B_{{10}} |D^{(8)}_{83}D^{(8)}_{38} | B_{{8}}\rangle$
& $\frac{1}{18\sqrt{5}}$ & $-\frac{1}{6\sqrt{5}}T_{3}$& $-\frac{\sqrt{5}}{18}T_{3}$ & $\frac{1}{6\sqrt{15}}$ \\    
$\langle B_{{10}} |D^{(8)}_{83}D^{(8)}_{88} | B_{{8}}\rangle$
& $0$ & $\frac{1}{6\sqrt{15}}$ & $\frac{1}{4\sqrt{15}}$ & $0$ \\  
$\langle B_{{10}} |d_{ab3}D^{(8)}_{8a}D^{(8)}_{8b} | B_{{8}}\rangle$  
&$0$ &$\frac{2}{9\sqrt{5}}$ &$\frac{1}{6\sqrt{5}}$ &$0$   \\  
$\langle B_{{10}} |d_{ab3}D^{(8)}_{3a}D^{(8)}_{8b}| B_{{8}}\rangle$  
& $\frac{2}{18\sqrt{15}}$ & $-\frac{1}{6\sqrt{15}}T_{3}$
& $-\frac{4}{9\sqrt{15}}T_{3}$ & $\frac{2}{9\sqrt{5}}$  \\ 
$\langle B_{{10}} |D^{(8)}_{83}D^{(8)}_{33}| B_{{8}}\rangle$  
& $-\frac{1}{18\sqrt{15}}$ & $\frac{1}{6\sqrt{15}}T_{3}$
& $-\frac{2}{9\sqrt{15}}T_{3}$ & $\frac{1}{9\sqrt{5}}$  \\  
$\langle B_{{10}} |D^{(8)}_{83}D^{(8)}_{83}| B_{{8}}\rangle$  
& $0$ & $\frac{1}{9\sqrt{5}}$
& $-\frac{1}{6\sqrt{5}}$ & $0$  \\  
$\langle B_{{10}} |D^{(8)}_{8i}D^{(8)}_{3i}| B_{{8}}\rangle$  
& $\frac{1}{18\sqrt{15}}$ & $-\frac{1}{6\sqrt{15}}T_{3}$
& $\frac{2}{9\sqrt{15}}T_{3}$ & $-\frac{1}{9\sqrt{5}}$  \\  
$\langle B_{{10}} |D^{(8)}_{8i}D^{(8)}_{8i}| B_{{8}}\rangle$  
& $0$ & $-\frac{1}{9\sqrt{5}}$
& $\frac{1}{6\sqrt{5}}$ & $0$  \\  
 \hline 
 \hline
\end{tabular}
\end{center}
\end{table}

\begin{table}[htp]
  \caption{The relevant transition matrix elements of the collective
    operators coming from the 27plet component of the baryon wave 
    functions.}  
  \label{tab:FF_2710}
\begin{center}
\begin{tabular}{ c |  c c  c  c   } 
 \hline 
  \hline 
$B_{27}\gamma^{*} \to B_{10}$ & $N \gamma^{*} \to \Delta$ & $\Sigma \gamma^{*} \to \Sigma^{*}$ & $\Xi \gamma^{*} \to \Xi^{*}$ & $\Lambda \gamma^{*} \to \Sigma^{*}$ \\  
 \hline
$\langle B_{{10}} |D^{(8)}_{33} | B_{{27}} \rangle$
& $\frac{1}{9}\sqrt{\frac{2}{15}}$ & $-\frac{1}{6\sqrt{5}}T_{3}$
&  $-\frac{7}{9\sqrt{30}}T_{3}$ &  $\frac{2}{9\sqrt{15}}$  \\  
$\langle B_{{10}}  |D^{(8)}_{83} | B_{{27}} \rangle$
& $0$ &  $\frac{1}{3\sqrt{15}}$
&$\frac{1}{6\sqrt{10}}$ & $0$  \\  
$\langle B_{{10}}  |D^{(8)}_{38}J_{3} | B_{{27}}\rangle$  
& $0$ & $0$
& $0$ & $0$   \\  
$\langle B_{{10}}  |D^{(8)}_{88}J_{3} | B_{{27}}\rangle$ 
& $0$ & $0$ 
& $0$ &$0$ \\  
$\langle B_{{10}}  |d_{ab3}D^{(8)}_{3a}J_{b} | B_{{27}}\rangle$ 
& $\frac{2}{9}\sqrt{\frac{2}{15}}$ &$-\frac{1}{3\sqrt{5}} T_{3} $ 
& $-\frac{14}{9\sqrt{30}} T_{3} $ & $\frac{4}{9\sqrt{15}}$   \\   
$\langle B_{{10}}  |d_{ab3}D^{(8)}_{8a}J_{b} | B_{{27}}\rangle$ 
& $0$ & $\frac{2}{3\sqrt{15}}$ 
&$\frac{1}{3\sqrt{10}}$ &$0$   \\   
$\langle B_{{10}} |D^{(8)}_{3i}J_{i} | B_{{27}}\rangle$  
& $0$& $0$& $0$ & $0$ \\  
$\langle B_{{10}} |D^{(8)}_{8i}J_{i} | B_{{27}}\rangle$  
& $0$& $0$& $0$ & $0$ \\  
 \hline 
 \hline
\end{tabular}
\end{center}
\end{table}

\begin{table}[htp]
  \caption{The relevant transition matrix elements of the collective
    operators coming from the 27plet component of the baryon wave 
    functions.}  
  \label{tab:FF_827}
\begin{center}
\begin{tabular}{ c |  c c  c  c   } 
 \hline 
  \hline 
$B_{8}\gamma^{*} \to B_{27}$ & $N \gamma^{*} \to \Delta$ & $\Sigma \gamma^{*} \to \Sigma^{*}$ & $\Xi \gamma^{*} \to \Xi^{*}$ & $\Lambda \gamma^{*} \to \Sigma^{*}$ \\  
 \hline
$\langle B_{{27}} |D^{(8)}_{33} | B_{{8}} \rangle$
& $\frac{2}{9}\sqrt{\frac{2}{3}}$ & $0$
&  $-\frac{2}{9}\sqrt{\frac{2}{15}}T_{3}$ &  $\frac{2}{3\sqrt{15}}$  \\  
$\langle B_{{27}}  |D^{(8)}_{83} | B_{{8}} \rangle$
& $0$ &  $\frac{2}{3\sqrt{15}}$
&$\frac{1}{3}\sqrt{\frac{2}{5}}$ & $0$  \\  
$\langle B_{{27}}  |D^{(8)}_{38}J_{3} | B_{{8}}\rangle$  
& $0$ & $0$
& $0$ & $0$   \\  
$\langle B_{{27}}  |D^{(8)}_{88}J_{3} | B_{{8}}\rangle$ 
& $0$ & $0$ 
& $0$ &$0$ \\  
$\langle B_{{27}}  |d_{ab3}D^{(8)}_{3a}J_{b} | B_{{8}}\rangle$ 
& $\frac{1}{9}\sqrt{\frac{2}{3}}$ &$0 $ 
& $-\frac{2}{9\sqrt{30}} T_{3} $ & $\frac{1}{3\sqrt{15}}$   \\   
$\langle B_{{27}}  |d_{ab3}D^{(8)}_{8a}J_{b} | B_{{8}}\rangle$ 
& $0$ & $\frac{1}{3\sqrt{15}}$ 
&$\frac{1}{3\sqrt{10}}$ &$0$   \\   
$\langle B_{{27}} |D^{(8)}_{3i}J_{i} | B_{{8}}\rangle$  
& $0$& $0$& $0$ & $0$ \\  
$\langle B_{{27}} |D^{(8)}_{8i}J_{i} | B_{{8}}\rangle$  
& $0$& $0$& $0$ & $0$ \\  
 \hline 
 \hline
\end{tabular}
\end{center}
\end{table}

\section{M1, E2 and C2 densities \label{app:c}}
The explicit expressions of three terms in Eq.~\eqref{eq:decom} are
given for the magnetic dipole form factor
\begin{align}
  \mathcal{G}^{B_8\to B_{10} (0)}_{M1}(\bm{r}) & =
   \frac{1}{6\sqrt{5}} \begin{pmatrix} 2 \\ 
 -(  \mathcal{Q}_{\Sigma \to \Sigma^*} + 1) \\
 -2 ( \mathcal{Q}_{\Xi \to \Xi^*} + 1) \\ \sqrt{3}
 ( \mathcal{Q}_{\Lambda \to \Sigma^*} + 1) \end{pmatrix}
 \left(  \mathcal{Q}_{0} (\bm{r}) + \frac{1}{I_{1}}
 \mathcal{Q}_{1} (\bm{r}) + \frac{1}{2}
 \frac{1}{I_{2}}  \mathcal{X}_{2} (\bm{r}) \right)  ,
\label{eq:M1leadingcon} \\
\mathcal{G}^{B_8\to B_{10}(\mathrm{op})}_{M1}(\bm{r}) &=
\frac{m_{8}}{36\sqrt{15}}\begin{pmatrix} 1 \\ -3
  \mathcal{Q}_{\Sigma \to \Sigma^*} + 1 \\
  -5\mathcal{Q}_{\Xi \to \Xi^*} - 1 \\ \sqrt{3}
  (\mathcal{Q}_{\Lambda \to \Sigma^*} + 1)  \end{pmatrix}
  \left(\frac{K_{1}}{I_{1}}\mathcal{X}_{1}
  (\bm{r}) -   \mathcal{M}_{1} (\bm{r})\right) \cr 
  & +\frac{2m_{8}}{108\sqrt{5}}  \begin{pmatrix} 7 \\ 
    3(  -\mathcal{Q}_{\Sigma \to \Sigma^*} + 4) \\
    8\mathcal{Q}_{\Xi \to \Xi^*} + 5 \\
    4\sqrt{3} ( \mathcal{Q}_{\Lambda \to \Sigma^*} + 1) \end{pmatrix} 
   \left(\frac{K_{2}}{I_{2}}\mathcal{X}_{2} (\bm{r})
     -    \mathcal{M}_{2} (\bm{r})\right)  \cr   
 & \hspace{0cm} - \frac{m_1}{3\sqrt{5}} \begin{pmatrix} 2 \\ 
   -(  \mathcal{Q}_{\Sigma \to \Sigma^*} + 1) \\
   -2 ( \mathcal{Q}_{\Xi \to \Xi^*} + 1) \\
   \sqrt{3} ( \mathcal{Q}_{\Lambda \to \Sigma^*} + 1)
 \end{pmatrix}
  {\mathcal{M}}_{0} (\bm{r})+  \frac{m_{8}}{36\sqrt{15}}
\begin{pmatrix} 4 \\  1 \\  \mathcal{Q}_{\Xi \to \Xi^*} + 2 \\
         \sqrt{3} ( \mathcal{Q}_{\Lambda \to \Sigma^*} + 1)
       \end{pmatrix} \mathcal{M}_{0} (\bm{r}), 
\label{eq:M1mslinear1} \\
  {\cal{G}}^{B_8\to B_{10} (\mathrm{wf})}_{M1}(\bm{r}) &=
c_{27} \frac{1}{18\sqrt{5}} \begin{pmatrix} 2 \\ 
  - 3 \mathcal{Q}_{\Sigma \to \Sigma^*} + 2 \\
  - ( 7\mathcal{Q}_{\Xi \to \Xi^*} + 2) \\
  2\sqrt{3} ( \mathcal{Q}_{\Lambda \to \Sigma^*} + 1) \end{pmatrix} 
  \left[  \left(  \mathcal{Q}_{0} (\bm{r})  + \frac{1}{I_{1}}
\mathcal{Q}_{1} (\bm{r}) \right)  - \frac{1}{2}
 \frac{1}{I_{2}}  \mathcal{X}_{2} (\bm{r}) \right ] \cr
&+ a_{27} \frac{\sqrt{5}}{45} \begin{pmatrix} 5 \\ 
  2 \\   -\mathcal{Q}_{\Xi \to \Xi^*} + 1 \\
  2\sqrt{3} ( \mathcal{Q}_{\Lambda \to \Sigma^*} + 1) \end{pmatrix} 
  \left[  \left(  \mathcal{Q}_{0} (\bm{r})  + \frac{1}{I_{1}}
 \mathcal{Q}_{1} (\bm{r}) \right)   - \frac{1}{2}
 \frac{1}{I_{2}}  \mathcal{X}_{2} (\bm{r})\right],
  \label{eq:M1final}
\end{align}
in the basis of $[N\to \Delta,\,\Sigma\to \Sigma^*,\,\Xi\to
\Xi^*,\,\Lambda\to \Sigma^*]$. $\mathcal{Q}_{B_8\to B_{10}}$ stand
for the charges of the corresponding baryons. $c_{27}$ and $a_{27}$
are the mixing coefficients in the collective baryon wave functions,
of which the explicit expressions can be found in
Appendix~\ref{app:b}. 

Similarly, the densities for the electric quadupole form factors are
written by 
\begin{align}
  \mathcal{G}^{B_8\to B_{10} f (0)}_{E2}(\bm{r}) &=
-\frac{1}{2\sqrt{5}I_{1}} \begin{pmatrix} 2 \\ 
-(  \mathcal{Q}_{\Sigma \to \Sigma^*} + 1) \\
   -2 ( \mathcal{Q}_{\Xi \to \Xi^*} + 1) \\
 \sqrt{3} ( \mathcal{Q}_{\Lambda \to \Sigma^*} + 1) \end{pmatrix} 
  \mathcal{I}_{1E2} (\bm{r}),
\label{eq:E2leading}
  \\
\mathcal{G}^{B_8\to B_{10}  (\mathrm{op})}_{E2}(\bm{r}) &=
 \frac{4}{27\sqrt{15}}m_{8}  \left(\frac{K_{1}}{I_{1}}
 \mathcal{I}_{1E2}(\bm{r}) - \mathcal{K}_{1E2}(\bm{r}) \right) 
\begin{pmatrix} -1 \\ 
  3\mathcal{Q}_{\Sigma \to \Sigma^*} + 2 \\
  - ( 4\mathcal{Q}_{\Xi \to \Xi^*} + 5) \\
  2\sqrt{3} ( \mathcal{Q}_{\Lambda \to \Sigma^*} + 1)
\end{pmatrix},
\label{eq:linearms}
  \\
  \mathcal{G}^{B_8\to B_{10} (\mathrm{wf})}_{E2}(\bm{r}) &=
-\frac{1}{6\sqrt{5}I_{1}} \bigg{
(}c_{27} \begin{pmatrix} 2 \\ 
 -3\mathcal{Q}_{\Sigma \to \Sigma^*} + 2 \\
    7\mathcal{Q}_{\Xi \to \Xi^*} + 2 \\
                 2\sqrt{3} ( \mathcal{Q}_{\Lambda \to \Sigma^*} + 1)
               \end{pmatrix} + a_{27}  \begin{pmatrix} 2 \\ 
 4 \\  2(-\mathcal{Q}_{\Xi \to \Xi^*} + 1) \\
 4\sqrt{3} ( \mathcal{Q}_{\Lambda \to \Sigma^*} + 1)
\end{pmatrix}
  \bigg{)}{\cal {I}}_{1E2} (\bm{r})
\label{eq:E2final}
\end{align}
in the basis of $[N\to \Delta,\,\Sigma\to \Sigma^*,\,\Xi\to
\Xi^*,\,\Lambda\to \Sigma^*]$. The densities for the Coulomb
quadrupole form factors are identical to those for the $E2$ form
factors, i.e. $\mathcal{G}^{B_8\to B_{10} }_{C2}(\bm{r})=
\mathcal{G}^{B_8\to B_{10}}_{E2}(\bm{r})$.   


\end{document}